\documentstyle[prd,aps,eqsecnum]{revtex}

\def\be{\begin{equation}}
\def\ee{\end{equation}}
\def\bea{\begin{eqnarray}}
\def\nn{\nonumber}
\def\eea{\end{eqnarray}}

\def\sb{\sigma_b}
\def\sg{\sigma_\gamma}
\def\sd{\sigma_\delta}

\begin{document}


\draft

\title{Stochastic semiclassical fluctuations in Minkowski spacetime}
\author{Rosario Mart\'{\i}n and Enric Verdaguer\footnote{Institut de 
                              F\'{\i}sica d'Altes Energies (IFAE)}}
\address{Departament de F\'{\i}sica Fonamental, Universitat de
        Barcelona, Av.~Diagonal 647, \mbox{08028 Barcelona}, Spain}
\date{\today}
\maketitle

\widetext

\begin{abstract}
The semiclassical Einstein-Langevin equations
which describe the dynamics of 
stochastic perturbations of the metric induced by quantum
stress-energy fluctuations of matter fields in a given state 
are considered on the background of the
ground state of semiclassical gravity, namely, 
Minkowski spacetime and a scalar field in its vacuum
state. The relevant equations are explicitly derived for massless and
massive fields arbitrarily coupled to the curvature. In doing so, some
semiclassical results, such as the expectation value of the
stress-energy tensor to linear order in the metric perturbations
and particle creation effects, are obtained.
We then solve the equations and compute the two-point correlation
functions for the linearized Einstein tensor and for the metric
perturbations. In the conformal field case, explicit results are
obtained. These results hint that gravitational fluctuations
in stochastic semiclassical gravity have a ``non-perturbative''
behavior in some characteristic correlation lengths. 
\end{abstract}



\section{Introduction}
\label{sec:introduction}


It has been pointed out that the semiclassical theory of gravity
\cite{wald94,wald77,fulling,birrell,mostepanenko}
cannot provide a correct
description of the dynamics of the gravitational field 
in situations where the
quantum stress-energy fluctuations are important
\cite{wald94,wald77,birrell,ford82,kuo93,albert1}. 
In such situations, these
fluctuations may have relevant back-reaction effects 
in the form of induced gravitational
fluctuations \cite{ford82} which, in a certain regime, are expected
to be described as classical stochastic fluctuations. 
A generalization of the semiclassical
theory is thus necessary to account for these effects.
In two previous papers, 
Refs.~\cite{letter} and \cite{paper_1},
we have shown how a stochastic semiclassical theory of
gravity can be formulated to improve the description of the 
gravitational field when stress-energy fluctuations are relevant.

In Ref.~\cite{letter}, we adopted an axiomatic approach to construct a
perturbative generalization of semiclassical gravity which
incorporates the back reaction  
of the lowest order stress-energy fluctuations in the form of 
a stochastic correction. 
We started noting that, for 
a given solution of semiclassical gravity, the lowest order matter
stress-energy fluctuations can be associated to a
classical stochastic tensor. 
We then sought a consistent equation in which this 
stochastic tensor was the source of linear perturbations of the
semiclassical metric.
The equation obtained is the so-called semiclassical 
Einstein-Langevin equation.

In Ref.~\cite{paper_1}, we followed the idea, first proposed
by Hu \cite{hu89} in the context of 
back reaction in semiclassical gravity,
of viewing the metric field as the ``system'' of interest and the
matter fields (modeled in that paper by a single scalar field)
as being part of its ``environment.''
We then showed that the semiclassical Einstein-Langevin equation
introduced in Ref.~\cite{paper_1} can be formally derived by a method  
based on the influence functional of Feynman and Vernon
\cite{feynman-vernon} (see also Ref.~\cite{decoh}).
That derivation shed light into 
the physical meaning of the semiclassical 
Langevin-type equations around specific backgrounds 
previously obtained with the same functional approach
\cite{calzettahu,humatacz,husinha,cv96,lomb-mazz,cv97,ccv97,%
campos-hu,campos-hu2,calver98}, 
since the stochastic source term was shown to be closely linked 
to the matter stress-energy fluctuations. 
We also developed a method to compute the semiclassical
Einstein-Langevin equation using dimensional regularization,
which provides an alternative and more direct way of computing this
equation with respect to previous calculations.

This paper is intended to be a first application of the full
stochastic 
semiclassical theory of gravity, where we evaluate the stochastic
gravitational fluctuations in a Minkowski background.
In order to do so, we first 
use the method developed in Ref.~\cite{paper_1} to 
derive the semiclassical 
Einstein-Langevin equation around a class of 
trivial solutions of semiclassical gravity consisting of 
Minkowski spacetime and a linear real scalar field
in its vacuum state, which may be considered the ground
state of semiclassical gravity. 
Although the Minkowski vacuum is an eigenstate of the total
four-momentum operator of a field in Minkowski spacetime, it 
is not an eigenstate of the stress-energy operator. 
Hence, even for these solutions of semiclassical gravity,
for which the expectation value of the stress-energy operator can
always be chosen to be zero, the fluctuations of this operator are
non-vanishing. This fact leads to consider the stochastic corrections
to these solutions described by the semiclassical Einstein-Langevin
equation.

We then solve the Einstein-Langevin equation for the linearized
Einstein tensor and compute the associated two-point correlation
functions. Even though, in this case, we expect to have negligibly small
values for these correlation functions at the domain of validity of
the theory, {\it i.e.}, for points separated by lengths 
larger than the Planck length, there are several reasons why we think
that it is worth carrying out this calculation.

On the one hand, these are, to our knowledge, the first 
solutions obtained to the full semiclassical Einstein-Langevin
equation. 
We are only aware of analogous solutions to a ``reduced'' version 
of this equation inspired in a ``mini-superspace'' model
\cite{ccv97}. There is also a previous attempt to obtain a solution to
the Einstein-Langevin equation in Ref.~\cite{cv96},
but, there, the non-local terms in the 
Einstein-Langevin equation were neglected.

The Einstein-Langevin equations computed in this paper are simple
enough to be explicitly solved and, at least for the case of a
conformal field, the expressions obtained for the correlation
functions can be explicitly evaluated in terms of elementary
functions. 
Thus, our calculation can serve as a testing ground for
the solutions of the Einstein-Langevin equation in more complex
situations of physical interest
(for instance, for a Robertson-Walker background and a field in a
thermal state).

On the other hand, the results of this calculation, which confirm our
expectations that gravitational fluctuations are negligible at length
scales larger than the Planck length, can be considered as a first
check that stochastic semiclassical gravity predicts reasonable
results.

In addition, we can extract
conclusions on the possible qualitative behavior of the solutions to
the Einstein-Langevin equation. Thus, it is interesting to
note that the correlation functions are characterized by
correlation lengths of the order of the Planck length;
furthermore, such correlation lengths enter in a non-analytic
way in the correlation functions. This kind of non-analytic behavior
is actually quite common in the solutions to Langevin-type equations
with dissipative terms
and hints at the possibility that correlation functions for other
solutions to the Einstein-Langevin equation are
also non-analytic in their characteristic correlation lengths.

The plan of the paper is the following. In Sec.~\ref{sec:overview},
we give a brief overview of the method developed in
Ref.~\cite{paper_1} to compute
the semiclassical Einstein-Langevin equation.
We then consider the background solutions of semiclassical gravity 
consisting of a Minkowski spacetime and a real scalar field in the
Minkowski vacuum.
In Sec.~\ref{sec:kernels}, we compute the kernels which appear in 
the Einstein-Langevin equation.
In Sec.~\ref{sec:E-L}, we derive the Einstein-Langevin equation
for metric perturbations around Minkowski spacetime. 
As a side result, we obtain some semiclassical results, 
which include
the expectation value of the stress-energy tensor 
of a scalar field with arbitrary mass and arbitrary coupling parameter
to linear order in the metric perturbations, and also some results
concerning the production of particles by metric perturbations: the
probability of particle creation and the number and energy of created
particles.  
In Sec.~\ref{sec:correlation functions}, we solve this equation for
the components of the linearized Einstein tensor and compute the
corresponding two-point correlation functions. For the case of a
conformal field and spacelike separated points, 
explicit calculations show that the correlation functions are
characterized by 
correlation lengths of the order of the Planck length. 
We conclude in Sec.~\ref{sec:discuss} with a discussion of our
results. We also include some appendices with technical details
used in the calculations.

Throughout this paper we use the $(+++)$ sign conventions and the
abstract index notation of Ref.~\cite{wald84}, and we work with units
in which $c=\hbar =1$.



\section{Overview}
\label{sec:overview}


In this section, we give a very brief summary of the main results of 
Refs.~\cite{letter} and \cite{paper_1}
which are relevant for the computations in the present paper.
One starts with a solution of semiclassical gravity 
consisting of a globally hyperbolic spacetime $({\cal M},g_{ab})$, a
linear real scalar field quantized on it and some physically
reasonable state for 
this field (we work in the Heisenberg picture).
According to the stochastic semiclassical theory of gravity 
\cite{letter,paper_1}, 
quantum fluctuations in the stress-energy tensor
of matter induce stochastic linear
perturbations $h_{ab}$ to the semiclassical metric $g_{ab}$.
The dynamics of these perturbations is described by a stochastic
equation called the semiclassical Einstein-Langevin equation.

Assuming that our semiclassical gravity solution 
allows the use of dimensional analytic continuation to define
regularized 
matrix elements of the stress-energy ``operator,'' 
we shall write the equations in dimensional regularization,
that is, assuming an arbitrary dimension $n$ of the spacetime. 
Using this regularization method, 
we use a notation in which a subindex $n$ is attached to those 
quantities that have different physical dimensions from the 
corresponding physical quantities.
The $n$-dimensional spacetime $({\cal M},g_{ab})$ has to
be a solution of the semiclassical Einstein equation
in dimensional regularization 
\be
{1\over 8 \pi G_{B}} \left( G^{ab}[g]+ \Lambda_{B} g^{ab} \right)-
\left({4\over 3}\, \alpha_{B} D^{ab} 
+2  \beta_{B} B^{ab}\right)\! [g]
= \mu^{-(n-4)} \!
\left\langle \hat{T}_{n}^{ab}\right\rangle \! [g], 
\label{semiclassical eq in n} 
\ee
where $G_B$, $\Lambda_B$, $\alpha_B$ and $\beta_B$ are bare
coupling constants and $G_{ab}$ is the Einstein tensor. The tensors 
$D^{ab}$ and $B^{ab}$ are obtained by functional
derivation with respect to the metric of the action terms
corresponding to the Lagrangian densities 
$R_{abcd}R^{abcd}-R_{ab}R^{ab}$ and $R^2$, respectively,
where $R_{abcd}$ is the
Riemann tensor, $R_{ab}$ is the Ricci tensor and $R$ is the scalar
curvature (see Ref.~\cite{paper_1} for the explicit expressions for
the tensors $D^{ab}$ and $B^{ab}$). In the last equation,
$\hat{T}_{n}^{ab}$ is the stress-energy ``operator'' in dimensional
regularization and the expectation value is taken in some state for
the scalar field in the $n$-dimensional spacetime.
Writing the bare coupling constants 
in Eq.~(\ref{semiclassical eq in n}) as
renormalized coupling constants plus some counterterms which absorb
the ultraviolet divergencies of the right hand side,
one can take the limit $n\!\rightarrow \! 4$, which leads to
the physical semiclassical Einstein equation.

Assuming that $g_{ab}$ is a solution of
Eq.~(\ref{semiclassical eq in n}), 
the semiclassical Einstein-Langevin equation can be similarly written
in dimensional regularization as 
\be
{1\over 8 \pi G_{B}}\biggl(  G^{ab}[g\!+\!h]+ 
\Lambda_{B} \left(g^{ab}\!-\!h^{ab}\right) \biggr)-
\left({4\over 3}\, \alpha_{B} D^{ab}
+ 2 \beta_{B} B^{ab} \right)\![g \!+\! h] = \mu^{-(n-4)} \!
\left\langle \hat{T}_{n}^{ab}\right\rangle \![g\!+\!h]
+ 2 \mu^{-(n-4)} \xi_n^{ab}, 
\label{Einstein-Langevin eq in n} 
\ee
where $h_{ab}$ is a linear stochastic perturbation to 
$g_{ab}$, and
$h^{ab}\!\equiv\! g^{ac}g^{bd}h_{cd}$. 
In this last equation, $\xi_n^{ab}$ is a Gaussian stochastic tensor
characterized by the correlators
\be
\langle\xi_n^{ab}(x) \rangle_{c}\!= 0, 
\hspace{10ex} 
\langle\xi_n^{ab}(x)\xi_n^{cd}(y) \rangle_{c}\!=
N_n^{abcd}(x,y),
\label{correlators in n}
\ee
where $8 N_n^{abcd}(x,y) \equiv \langle \{\hat{t}_n^{ab}(x), 
\hat{t}_n^{cd}(y)\}\rangle [g]$, 
with $\hat{t}_n^{ab}\equiv \hat{T}_{n}^{ab}-
\langle \hat{T}_{n}^{ab}\rangle$; here,
$\: \langle \hspace{1.5ex} \rangle_c$ means statistical 
average and 
$\{ \; , \:\}$ denotes an anticommutator.
As we pointed out in Ref.~\cite{paper_1}, the noise kernel 
$N_n^{abcd}(x,y)$ is free of ultraviolet divergencies in the limit 
$n \!\rightarrow \!4$.
Therefore, in the semiclassical Einstein-Langevin equation
(\ref{Einstein-Langevin eq in n}),
one can perform exactly the same renormalization procedure
as the one for the semiclassical Einstein equation 
(\ref{semiclassical eq in n}), and 
Eq.~(\ref{Einstein-Langevin eq in n}) yields the physical 
semiclassical Einstein-Langevin equation 
in four spacetime dimensions.

In Ref.~\cite{paper_1}, we used a method based on the CTP functional
technique applied to a system-environment interaction, more
specifically, on the influence action formalism of Feynman and Vernon,
to obtain an explicit expression for the expansion of 
$\langle \hat{T}_{n}^{ab} \rangle [g\!+\!h]$ up to first order in
$h_{cd}$. In this way, we can write the Einstein-Langevin 
equation (\ref{Einstein-Langevin eq in n}) in a more explicit form.
This expansion involves the kernel
$H_n^{abcd}(x,y) \equiv 
H_{\scriptscriptstyle \!{\rm S}_{\scriptstyle n}}^{abcd}(x,y)
+H_{\scriptscriptstyle \!{\rm A}_{\scriptstyle n}}^{abcd}(x,y)$,
with
\be
H_{\scriptscriptstyle \!{\rm S}_{\scriptstyle n}}^{abcd}
(x,y) \equiv 
{1\over 4}\:{\rm Im} \left\langle {\rm T}^{\displaystyle \ast}\!\! 
\left( \hat{T}_n^{ab}(x) \hat{T}_n^{cd}(y) 
\right) \right\rangle \![g], 
\hspace{3ex}    
H_{\scriptscriptstyle \!{\rm A}_{\scriptstyle n}}^{abcd}
(x,y) \equiv 
-{i\over 8}\, \left\langle  
\left[ \hat{T}_n^{ab}(x), \, \hat{T}_n^{cd}(y)
\right] \right\rangle \![g], 
\label{kernels}  
\ee
where $[ \; , \: ]$ means a commutator, 
and we use the symbol ${\rm T}^{\displaystyle \ast}$ 
to denote that
we have to time order the field operators $\hat{\Phi}_{n}$ first 
and then to apply the derivative operators which appear in each term 
of the product $T^{ab}(x) T^{cd}(y)$, where $T^{ab}$ is
the classical stress-energy tensor; see Ref.~\cite{paper_1} for more
details. 
In Eq.~(\ref{Einstein-Langevin eq in n}), all the ultraviolet
divergencies in 
the limit $n \!\rightarrow \!4$, which shall be removed by
renormalization of the coupling constants, are in some terms containing
$\langle \hat{\Phi}_{n}^2(x) \rangle$ and in
$H_{\scriptscriptstyle \!{\rm S}_{\scriptstyle n}}^{abcd}(x,y)$,
whereas the 
kernels $N_n^{abcd}(x,y)$ and
$H_{\scriptscriptstyle \!{\rm A}_{\scriptstyle n}}^{abcd}(x,y)$ are
free of ultraviolet divergencies. These two last kernels can be 
related to the real and imaginary parts of 
$\left\langle \hat{t}_n^{ab}(x)\,\hat{t}_n^{cd}(y)  \right\rangle$
by
\be
N_n^{abcd}(x,y)= 
{1\over 4}\,{\rm Re} \left\langle \hat{t}_n^{ab}(x)\,
\hat{t}_n^{cd}(y)  \right\rangle, 
\hspace{7ex}
H_{\scriptscriptstyle \!{\rm A}_{\scriptstyle n}}^{abcd}(x,y)= 
{1\over 4}\,{\rm Im} \left\langle \hat{t}_n^{ab}(x)\,
\hat{t}_n^{cd}(y)  \right\rangle. 
\label{finite kernels}
\ee

We now consider the case in which we start with a vacuum state 
$|0 \rangle$ for the field quantized in spacetime $({\cal M},g_{ab})$.
In this case, it was shown in Ref.~\cite{paper_1} that all the
expectation values entering the Einstein-Langevin equation
(\ref{Einstein-Langevin eq in n})  
can be written in terms of the Wightman
and Feynman functions, defined as
\be
G_n^+(x,y) \equiv \langle 0| \,
   \hat{\Phi}_{n}(x)  \hat{\Phi}_{n}(y) \,
   |0 \rangle [g],
\hspace{5 ex}
i G\!_{\scriptscriptstyle F_{\scriptstyle \hspace{0.1ex}  n}}
 \hspace{-0.2ex}(x,y) 
  \equiv \langle 0| \,
  {\rm T}\! \left( \hat{\Phi}_{n}(x)  \hat{\Phi}_{n}(y) \right)
  \hspace{-0.2ex}
  |0 \rangle [g].
\label{Wightman and Feynman functions}
\ee 
For instance, we can write
$\langle \hat{\Phi}_{n}^2(x) \rangle =
i G\!_{\scriptscriptstyle F_{\scriptstyle \hspace{0.1ex}  n}}
      \hspace{-0.2ex}(x,x)=G_n^+(x,x)$. The expressions for the
kernels,
which shall be used in our calculations,
can be found in Appendix \ref{app:kernels in vac}.


\subsection{Perturbations around Minkowski spacetime}
\label{subsec:Minkowski}


An interesting case to be analyzed in the framework of the
semiclassical stochastic theory of gravity is that of a 
Minkowski spacetime solution of semiclassical gravity. 
The flat metric $\eta_{ab}$ in a manifold
${\cal M} \!\equiv \!{\rm I\hspace{-0.4 ex}R}^{4}$ (topologically) and
the usual Minkowski vacuum, denoted as $|0 \rangle$, 
give the class of simplest solutions 
to the semiclassical Einstein equation [note that each possible value
of the parameters $(m^2, \xi)$ leads to a different solution],
the so called trivial solutions of semiclassical gravity
\cite{flanagan}. 
In fact, 
we can always choose a renormalization scheme in which
the renormalized expectation value 
$\langle 0|\, \hat{T}_{R}^{ab}\, |0 \rangle [\eta]=0$. Thus,
Minkowski spacetime 
$({\rm I\hspace{-0.4 ex}R}^{4},\eta_{ab})$
and the vacuum state $|0 \rangle$ are a
solution to the semiclassical Einstein equation
with renormalized cosmological constant $\Lambda\!=\!0$. 
The fact that the vacuum expectation 
value of the renormalized stress-energy operator in Minkowski
spacetime should vanish was originally proposed by Wald \cite{wald77} 
and it may be understood as a renormalization convention 
\cite{fulling,mostepanenko}. 
There are other possible renormalization prescriptions (see,
for instance, Ref.~\cite{brown}) in which such vacuum expectation
value is proportional to $\eta^{ab}$, and this would determine the
value of the cosmological constant $\Lambda$ in the semiclassical
equation. Of course, all these 
renormalization schemes give physically equivalent results: 
the total effective cosmological constant, {\it i.e.},  
the constant of proportionality in the sum of all the 
terms proportional to the metric in the semiclassical Einstein and
Einstein-Langevin equations, has to be zero.

Although the
vacuum $|0 \rangle$ is an eigenstate of the total four-momentum
operator in Minkowski spacetime, this state
is not an eigenstate of $\hat{T}^{R}_{ab}[\eta]$. Hence, even in 
these trivial solutions of semiclassical gravity, there are quantum
fluctuations in the stress-energy tensor of matter and, as a result,
the noise kernel does not vanish. 
This fact leads to consider the stochastic corrections 
to this class of trivial solutions of semiclassical
gravity.
Since, in this case, the Wightman and Feynman functions 
(\ref{Wightman and Feynman functions}), their values in the two-point
coincidence limit, and the products of derivatives of two of such
functions appearing in expressions (\ref{Wightman expression 2}) and
(\ref{Feynman expression 3}) (Appendix \ref{app:kernels in vac})
are known in dimensional regularization, 
we can compute the semiclassical Einstein-Langevin
equation using the method outlined above.

In order to perform the calculations, it is convenient to work in a
global inertial coordinate system $\{x^\mu \}$ and 
in the associated basis, in which the components of the flat metric are
simply $\eta_{\mu\nu}={\rm diag}(-1,1,\dots,1)$. 
In Minkowski spacetime, the components of the classical stress-energy
tensor functional reduce to
\be
T^{\mu\nu}[\eta,\Phi]=\partial^{\mu}\Phi
\partial^{\nu} \Phi - {1\over 2}\, \eta^{\mu\nu} \hspace{0.2ex}
\partial^{\rho}\Phi \partial_{\rho} \Phi 
-{1\over 2}\, \eta^{\mu\nu}\hspace{0.2ex} m^2 \Phi^2 
+\xi \left( \eta^{\mu\nu} \Box
-\partial^{\mu} \partial^{\nu} \right) \Phi^2,
\label{flat class s-t} 
\ee
where $\Box \!\equiv\! \partial_{\mu}\partial^{\mu}$, and the formal
expression for the 
components of the corresponding ``operator'' 
in dimensional regularization is
\be
\hat{T}_{n}^{\mu\nu}[\eta] = {1\over 2} \left\{
     \partial^{\mu}\hat{\Phi}_{n} , 
     \partial^{\nu}\hat{\Phi}_{n} \right\}
     + {\cal D}^{\mu\nu} \hat{\Phi}_{n}^2,
\label{flat regul s-t}
\ee
where ${\cal D}^{\mu\nu}$ are the differential operators
${\cal D}^{\mu\nu}_{x} \equiv (\xi-1/4) 
\eta^{\mu\nu} \Box_{x}- \xi\, \partial^{\mu}_{x} 
\partial^{\nu}_{x}$
and $\hat{\Phi}_{n}(x)$ is the field operator in the Heisenberg
picture in 
an $n$-dimensional Minkowski spacetime, which satisfies the
Klein-Gordon equation 
$( \Box -m^2 ) \hspace{0.2ex}\hat{\Phi}_n=0$.

Notice, from (\ref{flat regul s-t}),
that the stress-energy tensor depends on the coupling parameter $\xi$ of
the scalar field to the scalar curvature even in the limit of a flat
spacetime. Therefore, that tensor differs in general from the canonical 
stress-energy tensor in flat spacetime, which corresponds to the
value $\xi\!=\!0$.  
Nevertheless, it is
easy to see \cite{paper_1} that the 
$n$-momentum density components 
$\hat{T}_{n}^{0\mu}\mbox{}_{\mbox{}_{\!\!\!\scriptstyle (\xi)}}[\eta]$ 
(we temporary use this notation to indicate the dependence on the
parameter $\xi$) and 
$\hat{T}_{n}^{0\mu}\mbox{}_{\mbox{}_{\!\!\!\scriptstyle (\xi=0)}}[\eta]$ 
differ in a space divergence and, hence, dropping surface terms, they
both yield the same $n$-momentum operator: 
\be
\hat{P}^\mu \equiv \int\!  d^{n-1}{\bf x} \:
: \hspace{-0.1ex} 
\hat{T}_{n}^{0\mu}\mbox{}_{\mbox{}_{\!\!\!\scriptstyle (\xi)}}[\eta]
\hspace{-0.3ex}: \hspace{0.4ex} 
= \int\!  d^{n-1}{\bf x} \: 
: \hspace{-0.1ex} \hat{T}_{n}^{0\mu}\mbox{}
_{\mbox{}_{\!\!\!\scriptstyle (\xi=0)}}[\eta]
\hspace{-0.3ex}: \hspace{0.4ex} , 
\label{n-momentum}
\ee
where the integration is on a hypersurface $x^0\!= {\rm constant}$ 
($\hat{P}^\mu$ is actually independent of the value of $x^0$) 
and we use the notation for coordinates 
$x^\mu \equiv (x^0,{\bf x})$, {\it i.e.}, ${\bf x}$ are
space coordinates on each of the hypersurfaces 
$x^0\!= {\rm constant}$. The symbol $: \; :$ in Eq.~(\ref{n-momentum})
means normal ordering of the creation and annihilation operators on the
Fock space built on the Minkowski vacuum $|0 \rangle$
(in $n$ spacetime dimensions), which is an
eigenstate with zero eigenvalue of the operators (\ref{n-momentum}).

The Wightman and Feynman functions 
(\ref{Wightman and Feynman functions}) in Minkowski spacetime are well
known:
\bea
&&G_n^+(x,y) \equiv \langle 0| \,
   \hat{\Phi}_{n}(x)  \hat{\Phi}_{n}(y) \,
   |0 \rangle [\eta]= i \hspace{0.2ex}\Delta_n^+(x-y),
\nn   \\
&&G\!_{\scriptscriptstyle F_{\scriptstyle \hspace{0.1ex}  n}}
 \hspace{-0.2ex}(x,y) 
  \equiv -i \,\langle 0| \,
  {\rm T}\! \left( \hat{\Phi}_{n}(x)  \hat{\Phi}_{n}(y) \right)
  \hspace{-0.2ex} |0 \rangle [\eta]= 
  \Delta_{\scriptscriptstyle F_{\scriptstyle \hspace{0.1ex} n}}
  \hspace{-0.2ex}(x-y),
\label{flat Wightman and Feynman functions}
\eea
with
\bea
&&\Delta_n^+(x)=-2 \pi i \int \! {d^n k \over (2\pi)^n} \,
e^{i kx}\, \delta (k^2+m^2) \,\theta (k^0),
\nn   \\
&&\Delta_{\scriptscriptstyle F_{\scriptstyle \hspace{0.1ex} n}}
  \hspace{-0.2ex}(x)=- \int \! {d^n k \over (2\pi)^n} \, 
{e^{i kx}  \over k^2+m^2-i \epsilon} , 
\hspace{5ex} \epsilon \!\rightarrow \! 0^+,
\label{flat propagators}
\eea
where 
$k^2 \equiv \eta_{\mu\nu} k^{\mu} k^{\nu}$ and
$k x \equiv \eta_{\mu\nu} k^{\mu} x^{\nu}$. 
Note that the derivatives of these functions satisfy
$\partial_{\mu}^{x}\Delta_n^+(x-y)
= \partial_{\mu}\Delta_n^+(x-y)$ and
$\partial_{\mu}^{y}\Delta_n^+(x-y)=
 - \partial_{\mu}\Delta_n^+(x-y)$,
and similarly for the Feynman propagator 
$\Delta_{\scriptscriptstyle F_{\scriptstyle \hspace{0.1ex} n}}
 \hspace{-0.2ex}(x-y)$.

To write down the semiclassical Einstein equation 
(\ref{semiclassical eq in n}) for this case, we need to
compute the vacuum expectation value of the 
stress-energy operator components  
(\ref{flat regul s-t}). Since, from 
(\ref{flat Wightman and Feynman functions}), we have that
$\langle 0 |\hat{\Phi}_{n}^2(x)|0 \rangle=
i\Delta_{\scriptscriptstyle F_{\scriptstyle \hspace{0.1ex} n}}
\hspace{-0.2ex}(0)
=i \Delta_n^+(0)$, which is a constant (independent
of $x$), we have simply 
\be
\langle 0 |\, \hat{T}_{n}^{\mu\nu}\, |0 \rangle [\eta]=
{1\over 2} \,\langle 0 | \left\{ \partial^{\mu}\hat{\Phi}_{n} , 
     \partial^{\nu}\hat{\Phi}_{n} \right\} |0 \rangle [\eta]=
-i \left( \partial^{\mu} \partial^{\nu} 
\Delta_{\scriptscriptstyle F_{\scriptstyle \hspace{0.1ex} n}} 
\right) \!(0)=-i \int {d^n k \over (2\pi)^n} \, 
{k^{\mu} k^{\nu} \over k^2+m^2-i \epsilon}
= {\eta^{\mu\nu} \over 2} \left( {m^2 \over 4 \pi} \right)^{\! n/2} 
\! \Gamma \!\left(- {n \over 2}\right),
\label{vev}
\ee
where the integrals in dimensional regularization have been computed
in the standard way (see Appendix \ref{sec:integrals in dimensional}) 
and where $\Gamma (z)$ is the 
Euler's gamma function. The semiclassical Einstein equation 
(\ref{semiclassical eq in n}), which now reduces to
\be
{\Lambda_{B} \over 8 \pi G_{B}}\, \eta^{\mu\nu}
= \mu^{-(n-4)}
\langle 0 | \hat{T}_{n}^{\mu\nu}|0 \rangle [\eta] , 
\label{flat semiclassical eq} 
\ee
simply sets the value of the bare coupling constant 
$\Lambda_{B}/G_{B}$.
Note, from (\ref{vev}), that in order to have 
$\langle 0|\, \hat{T}_{R}^{ab}\, |0 \rangle [\eta]\!=\! 0$, 
the renormalized (and regularized) stress-energy tensor 
``operator'' for a scalar field in Minkowski spacetime
has to be defined as 
\be
\hat{T}_{R}^{ab}[\eta] =  
\mu^{-(n-4)}\, \hat{T}_{n}^{ab}[\eta]
-{ \eta^{ab} \over 2} \, {m^4 \over (4\pi)^2}  
\left( {m^2 \over 4 \pi \mu^2} \right)^{\!_{\scriptstyle n-4 \over 2}}
\! \Gamma \!\left(- {n \over 2}\right),
\label{flat renorm s-t operator}
\ee 
which corresponds to a renormalization of the cosmological constant
\be
{\Lambda_{B} \over G_{B}}={\Lambda \over G}
-{2 \over \pi} \, {m^4 \over n \hspace{0.2ex}(n\!-\!2)} 
\: \kappa_n 
+O(n-4),
\label{cosmological ct renorm 2}
\ee
where
\be
\kappa_n \equiv {1 \over (n\!-\!4)} 
\left({e^\gamma m^2 \over 4 \pi \mu^2} \right)
^{\!_{\scriptstyle n-4 \over 2}}=
{1 \over n\!-\!4}
+{1\over 2}\, 
\ln \!\left({e^\gamma m^2 \over 4 \pi \mu^2} \right)+O (n-4), 
\label{kappa}
\ee
being $\gamma$ the Euler's constant. In the case of a
massless scalar field, $m^2\!=\!0$, one simply has  
$\Lambda_{B} / G_{B}=\Lambda / G$. Introducing this renormalized
coupling constant into Eq.~(\ref{flat semiclassical eq}), we can 
take the limit $n \!\rightarrow \! 4$.
We find again that, 
for $({\rm I\hspace{-0.4 ex}R}^{4}, \eta_{ab},|0 \rangle )$ to 
satisfy the semiclassical Einstein equation, 
we must take $\Lambda\!=\!0$.

We are now in the position to write down the  
Einstein-Langevin equations for the components 
$h_{\mu\nu}$ of the stochastic metric perturbation 
in dimensional regularization. 
In our case, using $\langle 0 |\hat{\Phi}_{n}^2(x)|0 \rangle=
i\Delta_{\scriptscriptstyle F_{\scriptstyle \hspace{0.1ex} n}}
\hspace{-0.2ex}(0)$ 
and the explicit expression for 
Eq.~(\ref{Einstein-Langevin eq in n}) found in Ref.~\cite{paper_1}, 
we obtain that this equation reduces to 
\bea
&&{1\over 8 \pi G_{B}}\Biggl[
G^{{\scriptscriptstyle (1)}\hspace{0.1ex} \mu\nu} + 
\Lambda_{B} \left( h^{\mu\nu}
\!-\!{1\over 2}\, \eta^{\mu\nu} h \right) 
\Biggr](x) -
{4\over 3}\, \alpha_{B} D^{{\scriptscriptstyle
(1)}\hspace{0.1ex} \mu\nu}(x)
-2\beta_{B}B^{{\scriptscriptstyle (1)}\hspace{0.1ex} \mu\nu}(x) 
\nn   \\
&&-\, \xi\, G^{{\scriptscriptstyle (1)}\hspace{0.1ex} \mu\nu}(x)
\mu^{-(n-4)}\, 
i\Delta_{\scriptscriptstyle F_{\scriptstyle \hspace{0.1ex} n}}
\hspace{-0.2ex}(0) 
+2 \!\int\! d^ny \, \mu^{-(n-4)} 
H_n^{\mu\nu\alpha\beta}(x,y)\, h_{\alpha\beta}(y)
=2 \xi^{\mu\nu}(x), 
\label{flat Einstein-Langevin eq}
\eea 
where $\xi^{\mu\nu}$ are the components of a Gaussian stochastic
tensor of zero average and
\be
\langle \xi^{\mu\nu}(x)\xi^{\alpha\beta}(y) \rangle_c
=\mu^{-2 \hspace{0.2ex} (n-4)}  N_n^{\mu\nu\alpha\beta}(x,y),
\label{correlator}
\ee
and where indices are raised in $h_{\mu\nu}$ with the flat 
metric and $h \equiv h_{\rho}^{\rho}$.
We use a superindex ${\scriptstyle (1)}$ to denote the components of a
tensor linearized around the flat metric.
In the last expressions, $N_n^{\mu\nu\alpha\beta}(x,y)$ and 
$H_n^{\mu\nu\alpha\beta}(x,y)$ are the components of the kernels 
defined above.
In Eq.~(\ref{flat Einstein-Langevin eq}), we have made use of the
explicit expression for  
$G^{{\scriptscriptstyle (1)}\hspace{0.1ex} \mu\nu}$. This expression 
and those for  
$D^{{\scriptscriptstyle (1)}\hspace{0.1ex} \mu\nu}$ and 
$B^{{\scriptscriptstyle (1)}\hspace{0.1ex} \mu\nu}$ are given in
Appendix \ref{sec:linearized tensors}; the last two can also be
written as
\be
D^{{\scriptscriptstyle (1)}\hspace{0.1ex} \mu\nu}(x)=
{1 \over 2}\, (3 {\cal F}^{\mu\alpha}_{x} {\cal F}^{\nu\beta}_{x}
-{\cal F}^{\mu\nu}_{x} {\cal F}^{\alpha\beta}_{x})
\, h_{\alpha\beta}(x),
\hspace{7.2ex}
B^{{\scriptscriptstyle (1)}\hspace{0.1ex} \mu\nu}(x)= 
2  {\cal F}^{\mu\nu}_{x} {\cal F}^{\alpha\beta}_{x} 
h_{\alpha\beta}(x),
\label{D, B tensors}
\ee
where ${\cal F}^{\mu\nu}_{x}$ is the differential operator
${\cal F}^{\mu\nu}_{x} \equiv \eta^{\mu\nu} \Box_x
-\partial^\mu_{x} \partial^\nu_{x}$.



\section{The kernels 
for a Minkowski background}
\label{sec:kernels}


The kernels $N_n^{\mu\nu\alpha\beta}(x,y)$ and 
$H_n^{\mu\nu\alpha\beta}(x,y)=
H_{\scriptscriptstyle \!{\rm S}_{\scriptstyle n}}
^{\mu\nu\alpha\beta}(x,y)
+H_{\scriptscriptstyle \!{\rm A}_{\scriptstyle n}}
^{\mu\nu\alpha\beta}(x,y)$ 
can now be computed using 
(\ref{finite kernels}) and the expressions
(\ref{Wightman expression 2}) and 
(\ref{Feynman expression 3}).
In Ref.~\cite{paper_1}, we have shown that the kernel 
$H_{\scriptscriptstyle \!{\rm A}_{\scriptstyle n}}
^{\mu\nu\alpha\beta}(x,y)$ plays the role of a dissipation kernel,
since it is related to the noise kernel, 
$N_n^{\mu\nu\alpha\beta}(x,y)$, by a fluctuation-dissipation relation.
{}From the definitions (\ref{kernels}) 
and the fact that the Minkowski vacuum $|0 \rangle$ is
an eigenstate of the operator $\hat{P}^\mu$, given by
(\ref{n-momentum}), these kernels satisfy
\be
\int d^{n-1}{\bf x} \: N_n^{0\mu\alpha\beta}(x,y)=
\int d^{n-1}{\bf x} \: 
H_{\scriptscriptstyle \!{\rm A}_{\scriptstyle n}}
^{0\mu\alpha\beta}(x,y)=0.
\label{spatial integrals}
\ee


\subsection{The noise and dissipation kernels}
\label{subsec:noise and dissipation kernels}


Since the two kernels (\ref{finite kernels}) are free of ultraviolet
divergencies in the limit
$n\!\rightarrow \! 4$, we can deal directly with
\be
M^{\mu\nu\alpha\beta}(x-y) \equiv 
\lim_{n \rightarrow 4} \mu^{-2 \hspace{0.2ex} (n-4)} \,
\langle 0 |\, \hat{t}_n^{\mu\nu}(x)\,
\hat{t}_n^{\alpha\beta}(y) \,|0 \rangle [\eta].
\label{M}
\ee
The kernels 
$4 N^{\mu\nu\alpha\beta}(x,y)
={\rm Re}\, M^{\mu\nu\alpha\beta}(x-y)$ and 
$4 H_{\scriptscriptstyle \!{\rm A}}^{\mu\nu\alpha\beta}(x,y) 
= {\rm Im}\, M^{\mu\nu\alpha\beta}(x-y)$ 
are actually 
the components of the ``physical''
noise and dissipation kernels that will appear in the 
Einstein-Langevin equations once the renormalization procedure has
been carried out. 
Note that, in the renormalization scheme in which 
$\hat{T}_{R}^{ab}[\eta]$ is given by (\ref{flat renorm s-t operator}),
we can write 
$M^{\mu\nu\alpha\beta}(x\!-\!y)=\langle 0 |\,
\hat{T}_R^{\mu\nu}(x)\, 
\hat{T}_R^{\alpha\beta}(y) \,|0 \rangle [\eta]$, where the limit
$n \!\rightarrow \! 4$ is understood. 
This kernel can be
expressed in terms of the Wightman function in four spacetime
dimensions,
\be
\Delta^+(x)=-2 \pi i \int\! {d^4 k \over (2\pi)^4} \,
e^{i kx}\, \delta (k^2+m^2) \,\theta (k^0),
\label{Wightman function}
\ee
in the following way:
\bea
M^{\mu\nu\alpha\beta}(x)=-  2 \, &&\left[ 
\partial^\mu \partial^{( \alpha} \Delta^+(x) \,
\partial^{\beta )} \partial^\nu \Delta^+(x)
+{\cal D}^{\mu\nu} \!\left( \partial^\alpha \Delta^+(x) \,
\partial^\beta \Delta^+(x) \right)
\right.
\nn   \\
&&\left. \hspace{2ex}
+\, {\cal D}^{\alpha\beta} \!\left( \partial^\mu \Delta^+(x) \,
\partial^\nu \Delta^+(x) \right)
+{\cal D}^{\mu\nu} {\cal D}^{\alpha\beta} 
\!\left( \Delta^{+ \hspace{0.2ex} 2}(x) \right)
\right].  
\label{M 2}
\eea

The different terms in Eq.~(\ref{M 2}) can be easily computed using
the integrals
\bea
&&I(p) \equiv \int\! {d^4 k \over (2\pi)^4} \:
\delta (k^2+m^2) \,\theta (-k^0)  \,
\delta [(k-p)^2+m^2]\,\theta (k^0-p^0),
\nn   \\
&&I^{\mu_1 \dots \mu_r}(p) \equiv \int\! {d^4 k \over (2\pi)^4} \:
k^{\mu_1} \cdots k^{\mu_r} \,
\delta (k^2+m^2) \,\theta (-k^0)  \,
\delta [(k-p)^2+m^2]\,\theta (k^0-p^0),
\label{integrals}
\eea
with $r \!=\! 1, 2, 3 ,4$, given in Appendix 
\ref{sec:integrals in dimensional}; all of them can be expressed in terms
of $I(p)$. 
We obtain expressions (\ref{Wightman 1})-(\ref{Wightman 3}). 
It is convenient to separate  $I(p)$ in its even
and odd parts with respect to the variables $p^{\mu}$ as
\be
I(p)=I_{\scriptscriptstyle {\rm S}}(p)
+I_{\scriptscriptstyle {\rm A}}(p),
\label{I}
\ee
where $I_{\scriptscriptstyle {\rm S}}(-p)=
I_{\scriptscriptstyle {\rm S}}(p)$ and 
$I_{\scriptscriptstyle {\rm A}}(-p)=
-I_{\scriptscriptstyle {\rm A}}(p)$. These two functions are
explicitly given by
\bea
&&I_{\scriptscriptstyle {\rm S}}(p)={1 \over 8 \, (2 \pi)^3} \;
\theta (-p^2-4m^2) \, \sqrt{1+4 \,{m^2 \over p^2} },
\nn  \\
&&I_{\scriptscriptstyle {\rm A}}(p)={-1 \over 8 \, (2 \pi)^3} \;
{\rm sign}\,p^0 \;
\theta (-p^2-4m^2) \, \sqrt{1+4 \,{m^2 \over p^2} }.
\label{S and A parts of I}
\eea
Using the results of Appendix 
\ref{sec:integrals in dimensional}, we obtain expressions
(\ref{Wightman 4})-(\ref{Wightman 6})
and, after some calculations, we find
\bea
M^{\mu\nu\alpha\beta}(x)= &&{\pi^2 \over 45}\, 
(3 {\cal F}^{\mu (\alpha}_{x}{\cal F}^{\beta )\nu}_{x}-
{\cal F}^{\mu\nu}_{x}{\cal F}^{\alpha\beta}_{x})
\int\! {d^4 p \over (2\pi)^4} \,
e^{-i px}\hspace{0.1ex} 
\left(1+4 \,{m^2 \over p^2} \right)^2 I(p)
\nn   \\
&&+\,{8 \pi^2 \over 9 } \, 
{\cal F}^{\mu\nu}_{x}{\cal F}^{\alpha\beta}_{x}
\int\! {d^4 p \over (2\pi)^4} \,
e^{-i px}\hspace{0.1ex} 
\left(3 \hspace{0.3ex}\Delta \xi+{m^2 \over p^2} \right)^2 I(p),
\label{M 3}
\eea
where $\Delta \xi \equiv \xi - 1/6$. The real and imaginary parts of 
the last expression, which
yield the noise and dissipation kernels, are easily recognized as
the terms containing $I_{\scriptscriptstyle {\rm S}}(p)$ and
$I_{\scriptscriptstyle {\rm A}}(p)$, respectively. To write them
explicitly, it is useful to introduce the new kernels
\bea
&&N_{\rm A}(x;m^2) \equiv
{1 \over 1920 \pi} \int\! {d^4 p \over (2\pi)^4} \,
e^{i px}\,
\theta (-p^2-4m^2) \, \sqrt{1+4 \,{m^2 \over p^2} } 
\left(1+4 \,{m^2 \over p^2} \right)^2,
\nn \\
&&N_{\rm B}(x;m^2,\Delta \xi) \equiv
{1 \over 288 \pi} \int\! {d^4 p \over (2\pi)^4} \,
e^{i px}\, 
\theta (-p^2-4m^2) \, \sqrt{1+4 \,{m^2 \over p^2} } 
\left(3 \hspace{0.3ex}\Delta \xi+{m^2 \over p^2} \right)^2,
\nn \\
&&D_{\rm A}(x;m^2) \equiv
{-i \over 1920 \pi} \int\! {d^4 p \over (2\pi)^4} \,
e^{i px}\, {\rm sign}\,p^0 \;
\theta (-p^2-4m^2) \, \sqrt{1+4 \,{m^2 \over p^2} } 
\left(1+4 \,{m^2 \over p^2} \right)^2,
\nn \\
&&D_{\rm B}(x;m^2,\Delta \xi) \equiv
{-i \over 288 \pi} \int\! {d^4 p \over (2\pi)^4} \,
e^{i px}\, {\rm sign}\,p^0 \;
\theta (-p^2-4m^2) \, \sqrt{1+4 \,{m^2 \over p^2} } 
\left(3 \hspace{0.3ex}\Delta \xi+{m^2 \over p^2} \right)^2,
\label{N and D kernels}
\eea
and we finally get
\bea
&&N^{\mu\nu\alpha\beta}(x,y)=
{1 \over 6}\, 
(3 {\cal F}^{\mu (\alpha}_{x}{\cal F}^{\beta )\nu}_{x}-
{\cal F}^{\mu\nu}_{x}{\cal F}^{\alpha\beta}_{x}) \,
N_{\rm A}(x\!-\!y;m^2)
+{\cal F}^{\mu\nu}_{x}{\cal F}^{\alpha\beta}_{x} 
N_{\rm B}(x\!-\!y;m^2,\Delta \xi),
\nn \\
&&H_{\scriptscriptstyle \!{\rm A}}^{\mu\nu\alpha\beta}(x,y)=
{1 \over 6}\, 
(3 {\cal F}^{\mu (\alpha}_{x}{\cal F}^{\beta )\nu}_{x}-
{\cal F}^{\mu\nu}_{x}{\cal F}^{\alpha\beta}_{x}) \,
D_{\rm A}(x\!-\!y;m^2)
+{\cal F}^{\mu\nu}_{x}{\cal F}^{\alpha\beta}_{x} 
D_{\rm B}(x\!-\!y;m^2,\Delta \xi).
\label{noise and dissipation kernels 2}
\eea
Notice that the noise and dissipation kernels defined in 
(\ref{N and D kernels}) are actually real because, for the noise
kernels, only the $\cos px$ terms of the exponentials $e^{i px}$
contribute to the integrals, and, for the dissipation kernels, the
only contribution of such exponentials comes from the $i \sin px$
terms.

We can now evaluate the contribution of the dissipation kernel 
components 
$H_{\scriptscriptstyle \!{\rm A}}^{\mu\nu\alpha\beta}(x,y)$ 
to the 
Einstein-Langevin equations (\ref{flat Einstein-Langevin eq})
[after taking the limit $n \! \rightarrow \! 4$]. 
{}From (\ref{noise and dissipation kernels 2}), 
integrating by parts, and using (\ref{D, B tensors}) and the fact
that, in four spacetime dimensions, 
$D^{{\scriptscriptstyle (1)}\hspace{0.1ex} \mu\nu}(x)=
(3/2) \, 
A^{{\scriptscriptstyle (1)}\hspace{0.1ex} \mu\nu}(x)$ 
[the tensor $A^{ab}$ is obtained from the derivative with respect to
the metric of an action term corresponding to the Lagrangian density
$C_{abcd}C^{abcd}$, where $C_{abcd}$ is the Weyl tensor, 
see Ref.~\cite{paper_1} for details], it is easy to see that
\be
2 \!\int\! d^4y \,  
H_{\scriptscriptstyle \!{\rm A}}^{\mu\nu\alpha\beta}(x,y)\, 
h_{\alpha\beta}(y)=\int\! d^4y 
\left[ D_{\rm A}(x\!-\!y;m^2) 
A^{{\scriptscriptstyle (1)}\hspace{0.1ex} \mu\nu}(y)
+D_{\rm B}(x\!-\!y;m^2,\Delta \xi)
B^{{\scriptscriptstyle (1)}\hspace{0.1ex} \mu\nu}(y) \right].
\label{dissip kernel contribution}
\ee
These non-local terms in the semiclassical Einstein-Langevin equations
can actually be identified as being part of 
$\langle \hat{T}_R^{\mu\nu}\rangle [\eta+h]$.


\subsection{The kernel 
$H_{\scriptscriptstyle \!{\rm S}_{\scriptstyle n}}
^{\mu\nu\alpha\beta}(x,y)$ }
\label{subsec:symmetric part of H kernel}


The evaluation of the kernel components
$H_{\scriptscriptstyle \!{\rm S}_{\scriptstyle n}}
^{\mu\nu\alpha\beta}(x,y)$ is a much more cumbersome task. 
Since these quantities contain divergencies in the limit
$n\!\rightarrow \! 4$, we shall compute them using dimensional
regularization. Using Eq.~(\ref{Feynman expression 3}), 
these components can be written in terms of the Feynman propagator 
(\ref{flat propagators}) as
\be
\mu^{-(n-4)}
H_{\scriptscriptstyle \!{\rm S}_{\scriptstyle n}}
^{\mu\nu\alpha\beta}(x,y)= 
{1\over 4}\,{\rm Im}\, K^{\mu\nu\alpha\beta}(x-y),
\label{kernel H_S}
\ee
where
\bea
&& K^{\mu\nu\alpha\beta}(x) \equiv - \mu^{-(n-4)} \biggl\{
2 \partial^\mu \partial^{( \alpha} 
\Delta_{\scriptscriptstyle F_{\scriptstyle \hspace{0.1ex} n}}
   \hspace{-0.2ex}(x) \,
\partial^{\beta )} \partial^\nu 
\Delta_{\scriptscriptstyle F_{\scriptstyle \hspace{0.1ex} n}}
   \hspace{-0.2ex}(x) 
+2 {\cal D}^{\mu\nu} \!\left( \partial^\alpha 
\Delta_{\scriptscriptstyle F_{\scriptstyle \hspace{0.1ex} n}}
   \hspace{-0.2ex}(x)  \,
\partial^\beta 
\Delta_{\scriptscriptstyle F_{\scriptstyle \hspace{0.1ex} n}}
   \hspace{-0.2ex}(x)  \right)
\nn   \\ 
&& \hspace{0.85ex}
+ \, 2 {\cal D}^{\alpha\beta}  \Bigl( \partial^\mu 
\Delta_{\scriptscriptstyle F_{\scriptstyle \hspace{0.1ex} n}}
   \hspace{-0.2ex}(x) \,
\partial^\nu 
\Delta_{\scriptscriptstyle F_{\scriptstyle \hspace{0.1ex} n}}
   \hspace{-0.2ex}(x) \Bigr)
+2 {\cal D}^{\mu\nu} {\cal D}^{\alpha\beta} 
\!\left(  
\Delta_{\scriptscriptstyle F_{\scriptstyle \hspace{0.1ex} n}}^2
   \hspace{-0.2ex}(x) \right)
+\biggl[ \eta^{\mu\nu} \partial^{( \alpha} 
\Delta_{\scriptscriptstyle F_{\scriptstyle \hspace{0.1ex} n}}
   \hspace{-0.2ex}(x)  \,
\partial^{\beta )} 
+ \eta^{\alpha\beta} \partial^{( \mu} 
 \Delta_{\scriptscriptstyle F_{\scriptstyle \hspace{0.1ex} n}}
   \hspace{-0.2ex}(x)  \,
\partial^{\nu )} 
\nn   \\ 
&& \hspace{0.85ex} \left.  
+\, \Delta_{\scriptscriptstyle F_{\scriptstyle \hspace{0.1ex} n}}
   \hspace{-0.2ex}(0) \left( \eta^{\mu\nu} 
{\cal D}^{\alpha\beta}+ \eta^{\alpha\beta} 
{\cal D}^{\mu\nu}  \right)
+{1 \over 4}\, \eta^{\mu\nu} \eta^{\alpha\beta}
\left( 
\Delta_{\scriptscriptstyle F_{\scriptstyle \hspace{0.1ex} n}}
   \hspace{-0.2ex}(x) \Box 
-m^2 
\Delta_{\scriptscriptstyle F_{\scriptstyle \hspace{0.1ex} n}}
   \hspace{-0.2ex}(0)  \right) \right] \delta^n (x)
\biggr\}.  
\label{K}
\eea
Let us define the integrals 
\bea
&&J_n(p) \equiv 
\mu^{-(n-4)} \!\int\! {d^n k \over (2\pi)^n} \:
{1 \over (k^2+m^2-i \epsilon) \,
[(k-p)^2+m^2-i \epsilon] },
\nn   \\
&&J_n^{\mu_1 \dots \mu_r}(p) \equiv 
\mu^{-(n-4)} \!\int\! {d^n k \over (2\pi)^n} \:
{k^{\mu_1} \cdots k^{\mu_r} \over (k^2+m^2-i \epsilon) \,
[(k-p)^2+m^2-i \epsilon] }, 
\label{integrals in n dim}
\eea
with $r \!=\! 1, 2, 3 ,4$, and
\bea
&&I_{0_{\scriptstyle n}} \equiv 
\mu^{-(n-4)} \!\int\! {d^n k \over (2\pi)^n} \:
{1 \over (k^2+m^2-i \epsilon) },
\nn   \\
&&I_{0_{\scriptstyle n}}^{\mu_1 \dots \mu_r} \equiv 
\mu^{-(n-4)} \!\int\! {d^n k \over (2\pi)^n} \:
{k^{\mu_1} \cdots k^{\mu_r} \over (k^2+m^2-i \epsilon) }, 
\label{constant integrals in n dim}
\eea
with $r \!=\! 1, 2$, where a limit $\epsilon \!\rightarrow \! 0^+$
is understood in all these expressions. Then, the different terms in 
Eq.~(\ref{K}) can be computed using 
Eqs.~(\ref{Feynman 1})-(\ref{Feynman 6}).
The results for the expansions of the integrals 
(\ref{integrals in n dim}) and (\ref{constant integrals in n dim}) 
around $n\!=\!4$
are given in Appendix \ref{sec:integrals in dimensional}. In fact, 
$I_{0_{\scriptstyle n}}^{\mu}=0$ and the remaining integrals can be
written in terms of $I_{0_{\scriptstyle n}}$ and $J_n(p)$
given in Eqs.~(\ref{integral in dim regul 1}) and 
(\ref{integral in dim regul 4}).
Using the results of Appendix \ref{sec:integrals in dimensional}, 
we obtain Eqs.~(\ref{Feynman 7}) and (\ref{Feynman 8}) 
and, from Eqs.~(\ref{Feynman 4})-(\ref{Feynman 6}), we get
\bea
&& \mu^{-(n-4)} \left[ \eta^{\mu\nu} \partial^{( \alpha} 
\Delta_{\scriptscriptstyle F_{\scriptstyle \hspace{0.1ex} n}}
   \hspace{-0.2ex}(x)  \,
\partial^{\beta )} 
+ \eta^{\alpha\beta} \partial^{( \mu} 
 \Delta_{\scriptscriptstyle F_{\scriptstyle \hspace{0.1ex} n}}
   \hspace{-0.2ex}(x)  \,
\partial^{\nu )} \right] \delta^n (x)=
2 \hspace{0.3ex}\eta^{\mu\nu} \eta^{\alpha\beta} 
\, {m^2 \over n}\, I_{0_{\scriptstyle n}}
\hspace{0.2ex} \delta^n (x),
\nn  \\
&& \mu^{-(n-4)} \left( 
\Delta_{\scriptscriptstyle F_{\scriptstyle \hspace{0.1ex} n}}
   \hspace{-0.2ex}(x) \Box 
-m^2 
\Delta_{\scriptscriptstyle F_{\scriptstyle \hspace{0.1ex} n}}
   \hspace{-0.2ex}(0)  \right) \delta^n (x)=
- I_{0_{\scriptstyle n}} \hspace{0.2ex} \Box \delta^n (x).
\label{Feynman products 4}
\eea

We are now in the position to work out the explicit expression
for $K^{\mu\nu\alpha\beta}(x)$, defined in (\ref{K}).
We use Eqs.~(\ref{Feynman products 4}), the results
(\ref{Feynman 1}), (\ref{Feynman 4}),
(\ref{Feynman 7}) and (\ref{Feynman 8}), the
identities
$\delta^n (x)= (2\pi)^{-n} \int\! d^n p \, e^{i p x}$,
${\cal F}^{\mu\nu}_{x} \int\! d^n p \,
e^{i p x}\, f(p)
= - \!\int\! d^n p \, e^{i p x}\, f(p) \, p^2 P^{\mu\nu}$
and $\partial^\mu_{x} \partial^\nu_{x}  
\int\! d^n p \, e^{i p x}\, f(p)
= - \!\int\! d^n p \,
e^{i p x}\, f(p) \, p^\mu p^\nu$,
where $f(p)$ is an arbitrary function of $p^\mu$ and 
$P^{\mu\nu}$ is the projector orthogonal to $p^\mu$
defined as 
$p^2 P^{\mu\nu} \!\equiv \! \eta^{\mu\nu} p^2- p^\mu p^\nu$,
and the expansions in (\ref{integral in dim regul 1}) and
(\ref{integral in dim regul 4})
for $J_n(p)$ and $I_{0_{\scriptstyle n}}$.
After a rather long but straightforward calculation, we get, 
expanding around $n\!=\!4$, 
\bea
&&K^{\mu\nu\alpha\beta}(x)={i \over (4\pi)^2} \,
\Biggl\{ \kappa_n \left[ {1 \over 90} \, 
(3 {\cal F}^{\mu (\alpha}_{x}{\cal F}^{\beta )\nu}_{x}-
{\cal F}^{\mu\nu}_{x}{\cal F}^{\alpha\beta}_{x}) \, \delta^n (x)
+4 \hspace{0.3ex} \Delta \xi^2 \,
{\cal F}^{\mu\nu}_{x}{\cal F}^{\alpha\beta}_{x}
 \delta^n (x)
\right. 
\nn  \\
&& \hspace{2ex} 
+\,{2 \over 3}\, {m^2 \over (n\!-\!2)} \:
\bigr( \eta^{\mu\nu} \eta^{\alpha\beta} \Box_x
-\eta^{\mu (\alpha } \eta^{\beta )\nu} \Box_x
+\eta^{\mu (\alpha } \partial^{\beta )}_x \partial^\nu_x 
+\eta^{\nu (\alpha } \partial^{\beta )}_x \partial^\mu_x 
-\eta^{\mu\nu} \partial^\alpha_x \partial^\beta_x
-\eta^{\alpha\beta} \partial^\mu_x \partial^\nu_x \bigl)
\, \delta^n (x)
\nn  \\
&& \hspace{2ex}  
+\, {4 \hspace{0.2ex} m^4 \over n (n\!-\!2)} \:
(2 \hspace{0.2ex}\eta^{\mu (\alpha } \eta^{\beta )\nu}
\!- \eta^{\mu\nu} \eta^{\alpha\beta}) \, \delta^n (x) 
\biggr]
+{1 \over 180} \, 
(3 {\cal F}^{\mu (\alpha}_{x}{\cal F}^{\beta )\nu}_{x}-
{\cal F}^{\mu\nu}_{x}{\cal F}^{\alpha\beta}_{x})
\nn  \\
&& \hspace{2ex}
\times \!\int \! {d^n p \over (2\pi)^n} \,
e^{i p x} \left(1+4 \,{m^2 \over p^2} \right)^2 \!
\phi (p^2)
+{2 \over 9}  \,
{\cal F}^{\mu\nu}_{x}{\cal F}^{\alpha\beta}_{x}
\!\int \! {d^n p \over (2\pi)^n} \, e^{i p x} 
\left(3 \hspace{0.2ex}\Delta \xi+{m^2 \over p^2} \right)^2 \!
\phi (p^2)
\nn  \\
&& \hspace{2ex}
- \left[ {4 \over 675} \, 
(3 {\cal F}^{\mu (\alpha}_{x}{\cal F}^{\beta )\nu}_{x}-
{\cal F}^{\mu\nu}_{x}{\cal F}^{\alpha\beta}_{x}) 
+{1 \over 270} \, (60 \hspace{0.1ex}\xi \!-\! 11) \,
{\cal F}^{\mu\nu}_{x}{\cal F}^{\alpha\beta}_{x}
\right] \delta^n (x)
\nn  \\
&& \hspace{2ex}
-\, m^2 \left[ {2 \over 135} \, 
(3 {\cal F}^{\mu (\alpha}_{x}{\cal F}^{\beta )\nu}_{x}-
{\cal F}^{\mu\nu}_{x}{\cal F}^{\alpha\beta}_{x}) 
+{1 \over 27} \, {\cal F}^{\mu\nu}_{x}{\cal F}^{\alpha\beta}_{x}
\right] \Delta_n(x)
\Biggr\}+ O(n-4),
\label{result for K}
\eea
where $\kappa_n$ and $\phi (p^2)$ have been defined in (\ref{kappa})
and (\ref{phi}), and 
$\Delta_n(x)$ is given by
\be
\Delta_n(x) \equiv 
\int\! {d^n p \over (2\pi)^n} \: e^{i p x}\; {1 \over p^2}.
\label{Delta_n}
\ee
The imaginary part of (\ref{result for K}) [which,
using (\ref{kernel H_S}), gives the kernel components
$\mu^{-(n-4)}
H_{\scriptscriptstyle \!{\rm S}_{\scriptstyle n}}
^{\mu\nu\alpha\beta}(x,y)$] can be easily obtained multiplying 
this expression by $-i$ and retaining only the real part,
$\varphi (p^2)$, of the function $\phi (p^2)$. Making use of this
result, it is easy to compute the contribution of these 
kernel components
to the Einstein-Langevin equations (\ref{flat Einstein-Langevin eq}).
Integrating by parts, using Eqs.~(\ref{G tensor})-(\ref{R tensor})
and Eq.~(\ref{D, B tensors}), and taking into account
that, from Eqs.~(\ref{vev}) and (\ref{flat semiclassical eq}),
\be
{\Lambda_{B} \over 8 \pi G_{B}}= -{1 \over 4 \pi^2} \, 
{m^4 \over n (n\!-\!2)} \: \kappa_n + O(n-4),
\label{bare cosmological ct 2}
\ee
we finally find
\bea
&&2 \!\int\! d^n y \, \mu^{-(n-4)}
H_{\scriptscriptstyle \!{\rm S}_{\scriptstyle n}}
^{\mu\nu\alpha\beta}(x,y)\, h_{\alpha\beta}(y)=
-{\Lambda_{B} \over 8 \pi G_{B}} \left[ h^{\mu\nu}
\!-\!{1\over 2}\, \eta^{\mu\nu} h \right]
\!\hspace{-0.2ex} (x)
+{\kappa_n \over (4\pi)^2}\, \Biggl[
{2 \over 3} \, {m^2 \over (n\!-\!2)} \,
G^{{\scriptscriptstyle (1)}\hspace{0.1ex} \mu\nu} 
\nn \\
&& \hspace{9ex}
+\, {1 \over 90} \, 
D^{{\scriptscriptstyle (1)}\hspace{0.1ex} \mu\nu}
+\Delta \xi^2 \hspace{0.2ex}
B^{{\scriptscriptstyle (1)}\hspace{0.1ex} \mu\nu} \Biggr]
\hspace{-0.2ex} (x) 
+{1 \over 2880 \pi^2} \, \Biggl\{ 
-{16 \over 15} \, 
D^{{\scriptscriptstyle (1)}\hspace{0.1ex} \mu\nu}(x)
+\left({1 \over 6}-\! 10\hspace{0.2ex} \Delta \xi \right) \!
B^{{\scriptscriptstyle (1)}\hspace{0.1ex} \mu\nu}(x)
\nn  \\
&& \hspace{9ex}
+ \int\! d^n y 
\int\! {d^n p \over (2\pi)^n} \, e^{i p (x-y)} \,
\varphi (p^2) \,
\Biggl[\left(1+4 \,{m^2 \over p^2} \right)^2 \!
D^{{\scriptscriptstyle (1)}\hspace{0.1ex} \mu\nu}(y)
+10 \! 
\left(3 \hspace{0.2ex}\Delta \xi+{m^2 \over p^2} \right)^2 \!
B^{{\scriptscriptstyle (1)}\hspace{0.1ex} \mu\nu}(y)
\Biggr]
\nn  \\
&& \hspace{9ex}
-\, {m^2 \over 3} \!\int\! d^n y \, \Delta_n(x-y) \,
\Bigl( 8 D^{{\scriptscriptstyle (1)}\hspace{0.1ex} \mu\nu}(y)
+ 5 B^{{\scriptscriptstyle (1)}\hspace{0.1ex} \mu\nu}(y)
\Bigr)
\Biggr\} +O(n-4).
\label{symmetric kernel contribution}
\eea


\subsection{Fluctuation-dissipation relation}
\label{f-d rel}


{}From expressions (\ref{noise and dissipation kernels 2}) and 
(\ref{N and D kernels})
it is easy to check that there exists a
relation between the noise and dissipation kernels in the form of a
fluctuation-dissipation relation which was derived in
Ref.~\cite{paper_1} in a 
more general context. Introducing the Fourier transforms in the
time coordinates of these kernels as
\be
N^{\mu\nu\alpha\beta}(x,y)=
\int^{\infty}_{-\infty} {dp^0 \over 2 \pi}\, 
e^{-i p^0 (x^0-y^0)}\, 
\bar{N}^{\mu\nu\alpha\beta}(p^0;{\bf x},{\bf y}),
\ee 
and similarly for the dissipation kernel, this relation can be written
as
\be
\bar{H}_{\scriptscriptstyle \!{\rm A}}^{\mu\nu\alpha\beta}
(p^0;{\bf x},{\bf y})=
-i \, {\rm sign}\, p^0 \,
\bar{N}^{\mu\nu\alpha\beta}
(p^0;{\bf x},{\bf y}),
\label{f-d relation}
\ee
or, equivalently, as
\be
H_{\scriptscriptstyle \!{\rm A}}^{\mu\nu\alpha\beta}
(x^0,{\bf x};y^0,{\bf y})= -{1 \over \pi} 
\int^{\infty}_{-\infty} dz^0 \;
{\rm P}\!\hspace{-0.1ex} \left( {1 \over x^0\!-\!z^0} \right)
N^{\mu\nu\alpha\beta}(z^0,{\bf x};y^0,{\bf y}),
\label{f-d relation 2}
\ee
where ${\rm P} ( 1/ x^0)$ 
denotes the principal value distribution.

{}From (\ref{spatial integrals}), 
taking the limit $n \! \rightarrow \! 4$,
we see that the noise and dissipation kernels must satisfy 
\be
\int\! d^3{\bf x} \: N^{0\mu\alpha\beta}(x,y)=
\int\! d^3{\bf x} \: 
H_{\scriptscriptstyle \!{\rm A}}
^{0\mu\alpha\beta}(x,y)=0.
\label{spatial integrals 2}
\ee 
In order to check the last relations, it is useful to write the 
${\cal F}^{\mu\nu}_{x}$ derivatives in expressions
(\ref{noise and dissipation kernels 2}) using 
${\cal F}^{\mu\nu}_{x}\! \int\! d^4 p \,
e^{i p (x-y)}\, f(p)= - \int\! d^4 p \,
e^{i p (x-y)}\, f(p) \, p^2 P^{\mu\nu}$,
where $f(p)$ is any function of $p^\mu$ and
$P^{\mu\nu}$ is the projector orthogonal to $p^\mu$ defined 
above.
The identities (\ref{spatial integrals 2}) follow by noting that
$p^2 P^{00}= -p^i p_i$ and $p^2 P^{0i}=-p^0 p^i$, where we use
the index $i\!=\!1,2,3$ to denote the space components, and that  
$\int\! d^3{\bf x} \, \exp (i\hspace{0.2ex} p_i x^i)= (2 \pi)^3
\prod_{i=1}^3 \delta(p^i)$. It is also easy to check 
that the noise kernel satisfies
$\partial_{\mu}^{ \mbox{}^{\scriptscriptstyle x} } 
N^{\mu\nu\alpha\beta}(x,y) \!=\!0$ and, hence, the stochastic source
in the Einstein-Langevin equations will be conserved up to first order
in perturbation theory.



\section{The semiclassical Einstein-Langevin 
equations}
\label{sec:E-L}


The results of the previous section are now ready to be introduced
into the 
Einstein-Langevin equations (\ref{flat Einstein-Langevin eq}).
In fact, substituting expression 
(\ref{symmetric kernel contribution}) in such equations, and
using Eqs.~(\ref{Feynman 4}) 
and (\ref{integral in dim regul 1}) for the
$\mu^{-(n-4)} 
\Delta_{\scriptscriptstyle F_{\scriptstyle \hspace{0.1ex} n}}
   \hspace{-0.2ex}(0)$ term, we get
\bea
&&{1\over 8 \pi G_{B}} \, 
G^{{\scriptscriptstyle (1)}\hspace{0.1ex} \mu\nu}(x)
-{4\over 3}\, \alpha_{B} D^{{\scriptscriptstyle
(1)}\hspace{0.1ex} \mu\nu}(x)
-2\beta_{B}B^{{\scriptscriptstyle (1)}\hspace{0.1ex} \mu\nu}(x)
+{\kappa_n \over (4\pi)^2} \, \Biggl[ 
-4 \hspace{0.2ex}\Delta \xi \, {m^2 \over (n\!-\!2)} \, 
G^{{\scriptscriptstyle (1)}\hspace{0.1ex} \mu\nu}
+{1 \over 90} \, 
D^{{\scriptscriptstyle (1)}\hspace{0.1ex} \mu\nu}
\nn  \\
&&+\, \Delta \xi^2  
B^{{\scriptscriptstyle (1)}\hspace{0.1ex} \mu\nu}
\Biggr]\hspace{-0.2ex} (x)
+{1 \over 2880 \pi^2} \, \Biggl\{ 
-{16 \over 15} \, 
D^{{\scriptscriptstyle (1)}\hspace{0.1ex} \mu\nu}(x)
+\left({1 \over 6}-\! 10\hspace{0.2ex} \Delta \xi \right) \!
B^{{\scriptscriptstyle (1)}\hspace{0.1ex} \mu\nu}(x)
\nn  \\
&& 
+ \int\! d^n y \!
\int\! {d^n p \over (2\pi)^n} \, e^{i p (x-y)} \,
\varphi (p^2) \,
\Biggl[\left(1+4 \,{m^2 \over p^2} \right)^2 \!
D^{{\scriptscriptstyle (1)}\hspace{0.1ex} \mu\nu}(y)
+10 \! 
\left(3 \hspace{0.2ex}\Delta \xi+{m^2 \over p^2} \right)^2 \!
B^{{\scriptscriptstyle (1)}\hspace{0.1ex} \mu\nu}(y)
\Biggr]
\nn  \\
&& 
-\, {m^2 \over 3} \!\int\! d^n y \, \Delta_n(x\!-\!y) \,
\Bigl( 8 D^{{\scriptscriptstyle (1)}\hspace{0.1ex} \mu\nu}
+ 5 B^{{\scriptscriptstyle (1)}\hspace{0.1ex} \mu\nu}
\Bigr)\hspace{-0.2ex} (y)
\Biggr\} \!
+2 \!\int\! d^ny \, \mu^{-(n-4)} 
H_{\scriptscriptstyle \!{\rm A}_{\scriptstyle n}}
^{\mu\nu\alpha\beta}(x,y)\, h_{\alpha\beta}(y)
+O(n\!-\!4)
\nn \\
&& =2 \xi^{\mu\nu}(x). 
\label{flat Einstein-Langevin eq 2}
\eea
Notice that the terms containing the bare cosmological constant have
canceled. These equations can now be renormalized, that is,
we can now write the bare coupling constants as renormalized coupling
constants plus some suitably chosen counterterms and take 
the limit $n\!\rightarrow \! 4$. In order to carry out such a
procedure, it is convenient to distinguish between   
massive and massless scalar fields. We shall evaluate these two
cases in different subsections.


\subsection{Massive field ($m \neq 0$)}
\label{subsec:massive case}


In the case of a scalar field with mass $m \neq 0$, we can use,
as we have done in Eq.~(\ref{cosmological ct renorm 2})
for the cosmological constant, a renormalization scheme consisting on
the subtraction of terms proportional to $\kappa_n$. 
More specifically, we may introduce the renormalized coupling
constants $1/G$, $\alpha$ and $\beta$ as
\bea
&&{1 \over G_B}={1 \over G} 
+{2 \over \pi} \, \Delta \xi \, {m^2 \over (n\!-\!2)} \: \kappa_n 
+O(n-4),
\nn  \\
&&\alpha_{B}=\alpha 
+{1 \over (4 \pi)^2} \, {1 \over 120} \: \kappa_n + O(n-4),
\nn  \\
&&\beta_{B}=\beta 
+ {\Delta \xi^2 \over 32 \pi^2} \: \kappa_n + O(n-4).
\label{massive renormalization}
\eea
Note that for conformal coupling, $\Delta \xi=0$, one
has $1/G_B=1/G$ and $\beta_{B}=\beta$, that is, only
the coupling constant $\alpha$ and the cosmological constant need
renormalization. Substituting the above expressions into 
Eq.~(\ref{flat Einstein-Langevin eq 2}), we can now take the 
limit $n\!\rightarrow \! 4$, using Eqs.~(\ref{Delta_n}), 
(\ref{dissip kernel contribution}) and the fact that, for $n=4$,
$D^{{\scriptscriptstyle (1)}\hspace{0.1ex} \mu\nu}(x)=
(3/ 2) \, 
A^{{\scriptscriptstyle (1)}\hspace{0.1ex} \mu\nu}(x)$.
We obtain the semiclassical 
Einstein-Langevin equations for the physical stochastic
perturbations $h_{\mu\nu}$ in the four-dimensional manifold 
${\cal M} \!\equiv \!{\rm I\hspace{-0.4 ex}R}^{4}$. 
Introducing the two new kernels
\bea
&&H_{\rm A}(x;m^2) \equiv
{1 \over 1920 \pi^2}\! \int\! {d^4 p \over (2\pi)^4} \,
e^{i px}\,
\Biggl\{ \! \left(1+4 \,{m^2 \over p^2} \right)^{\! 2} 
\Biggl[ - i \pi \, {\rm sign}\,p^0 \;
\theta (-p^2\!-\!4m^2) \, \sqrt{1+4 \,{m^2 \over p^2} } 
\nn  \\
&& \hspace{47.8ex}
+\, \varphi(p^2) \Biggr]
-{8 \over 3} \, {m^2 \over p^2} \Biggr\},
\nn \\
&&H_{\rm B}(x;m^2,\Delta \xi) \equiv
{1 \over 288 \pi^2}\! \int\! {d^4 p \over (2\pi)^4} \,
e^{i px}\, 
\Biggl\{ \! 
\left(3 \hspace{0.3ex}\Delta \xi+{m^2 \over p^2} \right)^{\! 2} 
\Biggl[ - i \pi \, {\rm sign}\,p^0 \;
\theta (-p^2\!-\!4m^2) \, \sqrt{1+4 \,{m^2 \over p^2} } 
\nn  \\
&& \hspace{52.5ex}
+\, \varphi(p^2) \Biggr]
-{1 \over 6} \, {m^2 \over p^2} \Biggr\},
\label{H kernels}
\eea
where $\varphi(p^2)$ is given by the restriction to $n=4$ of 
expression (\ref{varphi}),
these Einstein-Langevin equations can be written as
\bea
&&{1\over 8 \pi G} \, 
G^{{\scriptscriptstyle (1)}\hspace{0.1ex} \mu\nu}(x)
\hspace{-0.2ex}-\hspace{-0.1ex} 2 \left( \alpha 
A^{{\scriptscriptstyle (1)}\hspace{0.1ex} \mu\nu}(x)
\hspace{-0.1ex}+\hspace{-0.1ex}
\beta B^{{\scriptscriptstyle (1)}\hspace{0.1ex} \mu\nu}(x)
\right) 
\hspace{-0.2ex}+\hspace{-0.1ex}
{1 \over 2880 \pi^2} \left[ -{8 \over 5} \, 
A^{{\scriptscriptstyle (1)}\hspace{0.1ex} \mu\nu}(x)
\!+\!
\left({1 \over 6}\!-\! 10\hspace{0.2ex} \Delta \xi \right) 
\! \hspace{-0.1ex}
B^{{\scriptscriptstyle (1)}\hspace{0.1ex} \mu\nu}(x) \right] 
\nn  \\
&&+ \int\! d^4y 
\left[ H_{\rm A}(x\!-\!y;m^2) 
A^{{\scriptscriptstyle (1)}\hspace{0.1ex} \mu\nu}(y)
+H_{\rm B}(x\!-\!y;m^2,\Delta \xi)
B^{{\scriptscriptstyle (1)}\hspace{0.1ex} \mu\nu}(y) \right]
=2 \xi^{\mu\nu}(x),
\label{massive Einstein-Langevin eq}
\eea
where $\xi^{\mu\nu}$ are the components of a Gaussian stochastic
tensor of vanishing mean value and 
two-point correlation function
$\langle\xi^{\mu\nu}(x)\xi^{\alpha\beta}(y) \rangle_c
=N^{\mu\nu\alpha\beta}(x,y)$, given in
(\ref{noise and dissipation kernels 2}).
Note that the two kernels defined in (\ref{H kernels})
are real and can be split into an even part 
and an odd part with respect to the variables $x^\mu$, with the 
odd terms being the dissipation kernels $D_{\rm A}(x;m^2)$ and 
$D_{\rm B}(x;m^2,\Delta \xi)$ defined in (\ref{N and D kernels}).
In spite of appearances, one can show that the Fourier transforms of
the even parts of these kernels are finite in the limit
$p^2 \! \rightarrow \! 0$ and, hence, the kernels $H_{\rm A}$ and
$H_{\rm B}$ are well defined distributions.

We should mention that, in a previous work in
Ref.~\cite{lomb-mazz}, the same Einstein-Langevin equations were
calculated using rather different methods. 
The way in which the result is written makes difficult a
direct comparison with our equations 
(\ref{massive Einstein-Langevin eq}). For instance, it is not obvious
that in those previously derived equations there is some analog of the
dissipation kernels related to the noise kernels by a
fluctuation-dissipation relation of the form 
(\ref{f-d relation}) or (\ref{f-d relation 2}).


\subsection{Massless field ($m=0$)}
\label{subsec:massless case}


In this subsection, we consider the limit 
$m \! \rightarrow \! 0$ of equations 
(\ref{flat Einstein-Langevin eq 2}).
The renormalization scheme used in the previous
subsection becomes singular in the massless limit 
because the expressions 
(\ref{massive renormalization}) for $\alpha_{B}$ and $\beta_{B}$ 
diverge when $m \! \rightarrow \! 0$. Therefore, a different
renormalization scheme is needed in this case. 
First, note that we may separate $\kappa_n$ in (\ref{kappa}) as
$\kappa_n=\tilde{\kappa}_n +{1 \over 2}\ln (m^2/\mu ^2)+O(n\!-\!4)$,
where
\be
\tilde{\kappa}_n \equiv {1 \over (n\!-\!4)} 
\left({e^\gamma \over 4 \pi} \right)
^{\!_{\scriptstyle n-4 \over 2}}=
{1 \over n\!-\!4}
+{1\over 2}\, 
\ln \!\left({e^\gamma \over 4 \pi } \right)+O (n-4), 
\label{kappa tilde}
\ee
and that [see Eq.~(\ref{varphi})]
\be
\lim_{m^2 \rightarrow 0} \left[ \varphi (p^2)+\ln (m^2/\mu ^2)
\right]=-2+\ln 
\left| \hspace{0.2ex} {p^2 \over \mu^2} \hspace{0.2ex}\right|.
\label{massless limit}
\ee
Hence, in the massless limit, equations 
(\ref{flat Einstein-Langevin eq 2}) reduce to
\bea
&&{1\over 8 \pi G_{B}} \, 
G^{{\scriptscriptstyle (1)}\hspace{0.1ex} \mu\nu}(x)
-{4\over 3}\, \alpha_{B} D^{{\scriptscriptstyle
(1)}\hspace{0.1ex} \mu\nu}(x)
-2\beta_{B}B^{{\scriptscriptstyle (1)}\hspace{0.1ex} \mu\nu}(x)
+{1 \over (4\pi)^2} \, (\tilde{\kappa}_n\!-\!1) \hspace{-0.1ex}
\left[ {1 \over 90} \, 
D^{{\scriptscriptstyle (1)}\hspace{0.1ex} \mu\nu}
+\Delta \xi^2  
B^{{\scriptscriptstyle (1)}\hspace{0.1ex} \mu\nu} \right]
\!\hspace{-0.2ex} (x)
\nn  \\
&&
+{1 \over 2880 \pi^2} \, \Biggl\{ 
-{16 \over 15} \, 
D^{{\scriptscriptstyle (1)}\hspace{0.1ex} \mu\nu}(x)
+\left({1 \over 6}-\! 10\hspace{0.2ex} \Delta \xi \right) \!
B^{{\scriptscriptstyle (1)}\hspace{0.1ex} \mu\nu}(x)
+\! \int\! d^n y \!
\int\! {d^n p \over (2\pi)^n} \, e^{i p (x-y)} \,
\ln \left| \hspace{0.2ex} {p^2 \over \mu^2} \hspace{0.2ex}\right|
\biggl[ D^{{\scriptscriptstyle (1)}\hspace{0.1ex} \mu\nu}(y)
\nn  \\
&& 
+\, 90 \hspace{0.2ex}\Delta \xi^2 
B^{{\scriptscriptstyle (1)}\hspace{0.1ex} \mu\nu}(y)
\biggr] \Biggr\} 
+\lim_{m^2 \rightarrow 0}
2 \!\int\! d^ny \, \mu^{-(n-4)} 
H_{\scriptscriptstyle \!{\rm A}_{\scriptstyle n}}
^{\mu\nu\alpha\beta}(x,y)\, h_{\alpha\beta}(y)
+O(n\!-\!4)
=2 \xi^{\mu\nu}(x).
\label{massless Einstein-Langevin eq}
\eea
These equations can be renormalized by introducing the renormalized
coupling constants $1/G$, $\alpha$ and $\beta$ as
\be
{1 \over G_B}\hspace{-0.2ex}=\hspace{-0.2ex} {1 \over G},
\hspace{4ex}
\alpha_{B} \hspace{-0.2ex} =\hspace{-0.2ex} \alpha 
\hspace{-0.1ex}+\hspace{-0.1ex}
{1 \over (4 \pi)^2} \, {1 \over 120} \, (\tilde{\kappa}_n\!-\!1)  
\hspace{-0.1ex}+\hspace{-0.1ex} O(n\!-\!4),
\hspace{4ex}
\beta_{B}\hspace{-0.2ex}=\hspace{-0.2ex} \beta 
\hspace{-0.1ex}+\hspace{-0.1ex} 
{\Delta \xi^2 \over 32 \pi^2} \, (\tilde{\kappa}_n\!-\!1)  
\hspace{-0.1ex}+\hspace{-0.1ex} O(n\!-\!4).
\label{massless renormalization}
\ee
Thus, in the massless limit, the Newtonian gravitational constant 
is not renormalized and, in the conformal coupling
case, $\Delta \xi=0$, we have again that $\beta_{B}\!=\! \beta$. 
Introducing the last expressions
into Eq.~(\ref{massless Einstein-Langevin eq}), we can take the 
limit $n \! \rightarrow \! 4$. 
Note that, by making $m \!=\!0$ in (\ref{N and D kernels}), 
the noise and dissipation kernels can be written as
\bea
&&N_{\rm A}(x;m^2\!=\!0)=N(x),
\hspace{7ex}
N_{\rm B}(x;m^2\!=\!0,\Delta \xi)
=60 \hspace{0.2ex} \Delta \xi^2 \hspace{0.2ex}  N(x),
\nn \\
&&D_{\rm A}(x;m^2\!=\!0)=D(x),
\hspace{7ex}
D_{\rm B}(x;m^2\!=\!0,\Delta \xi)
=60 \hspace{0.2ex} \Delta \xi^2 \hspace{0.2ex} D(x),
\label{massless N and D kernels}
\eea
where
\be
N(x) \equiv {1 \over 1920 \pi} \!\int \! {d^4 p \over (2\pi)^4} \,
e^{i px}\, \theta (-p^2),
\hspace{7ex}
D(x) \equiv {-i \over 1920 \pi} \!\int \! {d^4 p \over (2\pi)^4} \,
e^{i px}\, {\rm sign}\,p^0 \;
\theta (-p^2).
\label{N and D}
\ee
It is now convenient to introduce the new kernel
\bea
H(x;\mu^2) &\equiv & {1 \over 1920 \pi^2} 
\!\int \! {d^4 p \over (2\pi)^4} \, e^{i px}
\left[ 
\ln \left| \hspace{0.2ex} {p^2 \over \mu^2} \hspace{0.2ex}\right|
- i \pi \, {\rm sign}\,p^0 \; \theta (-p^2) \right]
\nn  \\
&=& {1 \over 1920 \pi^2} \lim_{\epsilon \rightarrow 0^+}
\!\int \! {d^4 p \over (2\pi)^4} \, e^{i px} \,
\ln\! \left( {-(p^0+i \epsilon)^2+p^i p_i \over \mu^2}
\right).
\label{H}
\eea
Again, this kernel is real and can be written as the
sum of an even part and an odd part in the variables $x^\mu$, 
where the odd part is the dissipation kernel $D(x)$. 
The Fourier transforms (\ref{N and D}) and (\ref{H}) can actually 
be computed
and, thus, in this case, we have explicit expressions for
the kernels in position space. For $N(x)$ and $D(x)$, we get
(see, for instance, Ref.~\cite{jones})
\be
N(x)= {1 \over 1920 \pi} \left[ {1 \over \pi^3} \, 
{\cal P}\hspace{-0.4ex}f  \hspace{-0.5ex}
\left( {1 \over (x^2)^2} \right) +
\delta^4(x) \right],
\hspace{10ex}
D(x)= {1 \over 1920 \pi^3} \: {\rm sign}\,x^0 \: 
\frac{d}{d(x^{2})}\, \delta(x^{2}),
\label{N and D 2}
\ee
where ${\cal P}\hspace{-0.4ex}f$ denotes a distribution generated by
the Hadamard finite part of a divergent integral 
(see Refs.~\cite{schwartz} for the definition of these distributions).
The expression for the kernel $H(x;\mu^2)$ can be found in 
Refs.~\cite{cmv95,horowitz} and it is given by
\bea
H(x;\mu^2) &=& {1 \over 960 \pi^2} 
\left\{ {\cal P}\hspace{-0.4ex} f \hspace{-0.5ex}
\left( \frac{1}{\pi}\,
\theta(x^{0})\,
\frac{d}{d(x^{2})}\, \delta(x^{2}) \right)+
\left( 1 \!-\!\gamma -\ln \! \mu \right) \delta^{4}(x)
\right\}
\nn    \\
&=& {1 \over 960 \pi^2} \,
\lim_{\lambda \rightarrow 0^{+}} \!\left\{ \frac{1}{\pi}\,
\theta(x^{0})\,
\theta( |{\bf x}| \!-\! \lambda )\,\frac{d}{d(x^{2})}\,
\delta(x^{2}) +
\left[ 1 \!-\!\gamma -\ln  (\mu \lambda) \right] \delta^{4}(x)
\right\}.
\eea
See Ref.~\cite{cmv95} for the details on how this last 
distribution acts on a test function. 
Finally, the semiclassical Einstein-Langevin
equations for the physical stochastic perturbations $h_{\mu\nu}$ in
the massless case are 
\bea
&&{1\over 8 \pi G} \, 
G^{{\scriptscriptstyle (1)}\hspace{0.1ex} \mu\nu}(x)
\hspace{-0.2ex}-\hspace{-0.1ex} 2 \left(\alpha 
A^{{\scriptscriptstyle (1)}\hspace{0.1ex} \mu\nu}(x)
\hspace{-0.1ex}+\hspace{-0.1ex}
\beta B^{{\scriptscriptstyle (1)}\hspace{0.1ex} \mu\nu}(x)
\right) 
\hspace{-0.2ex}+\hspace{-0.1ex}
{1 \over 2880 \pi^2} \left[ -{8 \over 5} \, 
A^{{\scriptscriptstyle (1)}\hspace{0.1ex} \mu\nu}(x)
\!+\!
\left({1 \over 6}\!-\! 10\hspace{0.2ex} \Delta \xi \right) 
\! \hspace{-0.1ex}
B^{{\scriptscriptstyle (1)}\hspace{0.1ex} \mu\nu}(x) \right] 
\nn  \\
&&+\int\! d^4y \, H(x\!-\!y;\mu^2) 
\left[ 
A^{{\scriptscriptstyle (1)}\hspace{0.1ex} \mu\nu}(y)
+60 \hspace{0.2ex}  \Delta \xi^2
B^{{\scriptscriptstyle (1)}\hspace{0.1ex} \mu\nu}(y) \right]
=2 \xi^{\mu\nu}(x),
\label{massless Einstein-Langevin eq 2}
\eea
where the Gaussian stochastic source components 
$\xi^{\mu\nu}$ have zero mean value and 
\be
\langle\xi^{\mu\nu}(x)\xi^{\alpha\beta}(y) \rangle_c
=\lim_{m \rightarrow 0}\! N^{\mu\nu\alpha\beta}(x,y)=
\left[ {1 \over 6}\, 
(3 {\cal F}^{\mu (\alpha}_{x}{\cal F}^{\beta )\nu}_{x}\!-\!
{\cal F}^{\mu\nu}_{x}{\cal F}^{\alpha\beta}_{x}) 
+ 60 \hspace{0.2ex}  \Delta \xi^2
{\cal F}^{\mu\nu}_{x}{\cal F}^{\alpha\beta}_{x} 
\right]\! N(x\!-\!y).
\label{massless noise}
\ee

It is interesting to consider the conformally coupled scalar field, 
{\it i.e.}, the case $\Delta \xi\!=\!0$, of particular interest
because of its similarities with the electromagnetic field. 
It was shown in Refs.~\cite{letter,paper_1} that, 
for this field, the
stochastic source tensor must be ``traceless'' (up to first order
in perturbation theory around semiclassical gravity), in the sense that
the stochastic variable 
$\xi^\mu_\mu \!\equiv \!\eta_{\mu\nu}\xi^{\mu\nu}$ behaves
deterministically as a vanishing scalar field. 
This can be easily checked by noticing, from 
Eq.~(\ref{massless noise}), that, when $\Delta \xi\!=\!0$, one has 
$\langle\xi^\mu_\mu(x)\xi^{\alpha\beta}(y) \rangle_c
=0$, since ${\cal F}^\mu_\mu\!=\! 3 \hspace{0.2ex}\Box $ and 
${\cal F}^{\mu \alpha}{\cal F}^\beta_\mu \!=\! 
\Box {\cal F}^{\alpha\beta}$. 
The semiclassical Einstein-Langevin equations for this 
particular case [and generalized to a spatially flat 
Robertson-Walker (RW) background]
were first obtained in Ref.~\cite{cv96}
(in this reference, the coupling constant
$\beta$ was set to zero). 
In order to compare with this
previous result, it is worth noticing that the
description of the stochastic source in terms of a symmetric and
``traceless'' tensor, with nine independent components $\xi^{\mu\nu}$,
is equivalent to a description in terms of a Gaussian 
stochastic tensor with the same symmetry properties as the 
Weyl tensor, with components $\xi_c^{\mu\nu\alpha\beta}$, defined as 
$\xi^{\mu\nu}
\!=\! -2 \partial_\alpha \partial_\beta 
\xi_c^{\mu\alpha\nu\beta}$; this tensor is used in Ref.~\cite{cv96}. 
The symmetry properties of the $\xi_c^{\mu\nu\alpha\beta}$
ensure that there are also nine independent components in 
$-2 \partial_\alpha \partial_\beta \xi_c^{\mu\alpha\nu\beta}$. 
It is easy to show that, for this combination to satisfy the
correlation relation (\ref{massless noise}) with $\Delta \xi\!=\!0$,
the relevant correlators for the new stochastic tensor must be
\be
\langle\xi_c^{\mu\nu\alpha\beta}(x)
      \xi_c^{\rho\sigma\lambda\theta}(y)\rangle_{\xi_c}
=T^{\mu\nu\alpha\beta\rho\sigma\lambda\theta} \hspace{0.1ex}
N(x-y),
\label{conformal correlation}
\ee
where $T^{\mu\nu\alpha\beta\rho\sigma\lambda\theta}$ is a linear
combination of terms like 
$\eta^{\mu\rho}\eta^{\nu\sigma}
 \eta^{\alpha\lambda}\eta^{\beta\theta}$ in such a way that it has the
same symmetries as the product of two Weyl tensor components
$C^{\mu\nu\alpha\beta} C^{\rho\sigma\lambda\theta}$, its explicit 
expression is given in Ref.~\cite{cv96}.
Thus, after a redefinition of the arbitrary mass scale $\mu$ in 
Eq.~(\ref{massless Einstein-Langevin eq 2}) to
absorb the constants of proportionality of the local terms with
$A^{{\scriptscriptstyle (1)}\hspace{0.1ex} \mu\nu}(x)$, one can see
that the resulting equations for the $\Delta \xi\!=\!0$ case are
actually equivalent to those found in Ref.~\cite{cv96}.


\subsection{Expectation value of the
stress-energy tensor}
\label{subsec:vev of stress-energy tensor}


{}From the above
equations one may extract the expectation value of the renormalized
stress-energy tensor for a scalar field in a spacetime 
$({\rm I\hspace{-0.4 ex}R}^{4}, 
\eta_{ab}+h_{ab})$,
computed up to first order in perturbation theory around the trivial
solution of semiclassical gravity. 
Such an expectation value can be obtained
by identification of Eqs.~(\ref{massive Einstein-Langevin eq})
and (\ref{massless Einstein-Langevin eq 2}) with the components of
the physical Einstein-Langevin equation, which in our particular case
simply reads
\be
{1\over 8 \pi G}\: G^{{\scriptscriptstyle (1)}\hspace{0.1ex} \mu\nu}  
-2 \left( 
\alpha A^{{\scriptscriptstyle (1)}\hspace{0.1ex} \mu\nu}
+\beta B^{{\scriptscriptstyle (1)}\hspace{0.1ex} \mu\nu}
\right)
=\left\langle
\hat{T}_R^{\mu\nu} \right\rangle \![\eta+h]
+2 \xi^{\mu\nu}.
\label{flat Einstein-Langevin eq 3}
\ee 
By comparison of Eqs.\ (\ref{massive Einstein-Langevin eq}) and
(\ref{massless Einstein-Langevin eq 2}) with the last equation, we can
identify 
\bea
&& \!\!\! \left\langle
\hat{T}_R^{\mu\nu}(x) \right\rangle \![\eta+h] =
{1 \over 2880 \pi^2} \left[ {8 \over 5} \, 
A^{{\scriptscriptstyle (1)}\hspace{0.1ex} \mu\nu}(x)
\!-\!
\left({1 \over 6}\!-\! 10\hspace{0.2ex} \Delta \xi \right) 
\! \hspace{-0.1ex}
B^{{\scriptscriptstyle (1)}\hspace{0.1ex} \mu\nu}(x) \right]
\nn  \\ 
&& \hspace{11.5ex}
- \! \int\! d^4y 
\left[ H_{\rm A}(x\!-\!y;m^2) 
A^{{\scriptscriptstyle (1)}\hspace{0.1ex} \mu\nu}(y)
+H_{\rm B}(x\!-\!y;m^2,\Delta \xi)
B^{{\scriptscriptstyle (1)}\hspace{0.1ex} \mu\nu}(y) \right]
+O(h^2),
\label{massive vev s-t}
\eea
for a massive scalar field, $m \!\neq\! 0$, and
\bea
\left\langle
\hat{T}_R^{\mu\nu}(x) \right\rangle \![\eta+h]=
&&{1 \over 2880 \pi^2} \left[ {8 \over 5} \, 
A^{{\scriptscriptstyle (1)}\hspace{0.1ex} \mu\nu}(x)
\!-\!
\left({1 \over 6}\!-\! 10\hspace{0.2ex} \Delta \xi \right) 
\! \hspace{-0.1ex}
B^{{\scriptscriptstyle (1)}\hspace{0.1ex} \mu\nu}(x) \right]
\nn  \\ 
&& 
-\! \int\! d^4y \, H(x\!-\!y;\mu^2) 
\left[ 
A^{{\scriptscriptstyle (1)}\hspace{0.1ex} \mu\nu}(y)
+60 \hspace{0.2ex}  \Delta \xi^2
B^{{\scriptscriptstyle (1)}\hspace{0.1ex} \mu\nu}(y) \right]
+O(h^2),
\label{massless vev s-t}
\eea
for a massless scalar field, $m \!=\!0$. Notice that in the massive
case we have chosen, as usual, a renormalization scheme such that 
the expectation value of the renormalized stress-energy tensor 
does not have local terms proportional to the metric and the 
Einstein tensor \cite{birrell}. 
The result (\ref{massless vev s-t}) agrees with the
general form found by Horowitz \cite{horowitz,horowitz81} 
using an axiomatic
approach and coincides with that given in Ref.~\cite{flanagan}.
The particular cases of conformal coupling, $\Delta \xi \!=\!0$, and
minimal coupling, $\Delta \xi \!=\!-1/6$, are also in agreement with
the results for this cases given in 
Refs.~\cite{horowitz,horowitz81,horowitz_wald,cv94,jordan} 
(modulo local terms proportional to 
$A^{{\scriptscriptstyle (1)}\hspace{0.1ex} \mu\nu}$ and
$B^{{\scriptscriptstyle (1)}\hspace{0.1ex} \mu\nu}$ due to different
choices of the renormalization scheme).
For the case of a massive minimally coupled scalar field,
$\Delta \xi \!=\!-1/6$, our result 
(\ref{massive vev s-t}) is equivalent to that of 
Ref.~\cite{tichy}.

As it was pointed out above, in the case of conformal coupling, both for
massive and massless scalar fields, one has $\beta_{B}\!=\!\beta$.
This means that, in these cases, the terms proportional to
$B^{{\scriptscriptstyle (1)}\hspace{0.1ex} \mu\nu}$
in the above expectation values of the stress-energy tensor are 
actually independent of the renormalization scheme 
chosen. Due to the conformal invariance of
$\int \! d^4 x \, \sqrt{- g}\, C_{cabd}C^{cabd}$, the tensor $A^{ab}$
is traceless and we have  
$A^{{\scriptscriptstyle (1)}}\mbox{}^{\mu}_\mu=0$.
Therefore, the terms with
$B^{{\scriptscriptstyle (1)}\hspace{0.1ex} \mu\nu}$ are precisely
those which give trace to the expectation value of the stress-energy 
tensor in (\ref{massive vev s-t}) and (\ref{massless vev s-t}).
In the massless conformally coupled case, $m\!=\!0$ and
$\Delta \xi \!=\!0$, such terms give the trace
anomaly \cite{birrell} up to first order in $h_{\mu\nu}$:
\be
\left\langle 
\hat{T}_{R}^{\;\;\;\mu}\mbox{}_{\!\!\!\!\mu}(x) 
\right\rangle \![\eta+h] =
-{1 \over 2880 \pi^2} \: {1 \over 6} \, 
B^{{\scriptscriptstyle (1)}}\mbox{}^{\mu}_\mu 
+O(h^2)=
{1 \over 2880 \pi^2} \, \Box R^{{\scriptscriptstyle (1)}}
+O(h^2),
\label{trace anomaly}
\ee
where we have used expression (\ref{B tensor}) 
for $B^{{\scriptscriptstyle (1)}\hspace{0.1ex} \mu\nu}$.


\subsection{Particle creation}
\label{subsec:particle creation}


We can also use the result (\ref{noise and dissipation kernels 2}) for
the noise kernel to evaluate the total probability of
particle creation and the number of created particles 
for a real scalar field in a spacetime 
$({\rm I\hspace{-0.4 ex}R}^{4}, \eta_{ab}+h_{ab})$. The metric
perturbation $h_{ab}$ (here an arbitrary
perturbation) is assumed to vanish, either in an exact way or
``asymptotically,'' in the ``remote past'' and in the 
``far future,'' so that the scalar field has well defined
``in'' and ``out'' many particle states. In that case, the 
absolute value of the logarithm of the vacuum persistence probability
$|\langle 0,{\rm out}|0,{\rm in}\rangle |^2$, where 
$ |0,{\rm in}\rangle$ and  $ |0,{\rm out}\rangle$ are, respectively,
the ``in'' and ``out'' vacua in the Heisenberg picture, 
gives a measure of the total probability of particle creation. 
On the other hand, the number of created particles can be defined as
the expectation value in the ``in'' vacuum of the number operator
for ``out'' particles. 
As it was shown in Ref.~\cite{paper_1}, the total probability
of particle creation and one half of the number of created particles
coincide to lowest non-trivial order in the metric perturbation, 
these are   
\be
P[h]=\! \int\! d^4x\, d^4y \: h_{\mu\nu}(x)\, 
N^{\mu\nu\alpha\beta}(x,y)\, h_{\alpha\beta}(y)+0(h^3),
\label{particle creation}
\ee
where $N^{\mu\nu\alpha\beta}(x,y)$ is the noise kernel given in
(\ref{noise and dissipation kernels 2}), which in the massless case
reduces to (\ref{massless noise}). The above expression
for the total probability of pair creation by metric 
perturbations about Minkowski spacetime was first derived 
in Ref.~\cite{sexl}. Using 
(\ref{noise and dissipation kernels 2}), we can write
$P[h] \!=\!P_{\rm A}[h]+P_{\rm B}[h]+0(h^3)$, where
\bea
&&P_{\rm A}[h] \equiv {1 \over 6} \int\! d^4x\, d^4y \:
(3 {\cal F}^{\mu\alpha}_{x}{\cal F}^{\nu\beta}_{x}-
{\cal F}^{\mu\nu}_{x}{\cal F}^{\alpha\beta}_{x}) \hspace{0.1ex}
N_{\rm A}(x\!-\!y;m^2) \: 
h_{\mu\nu}(x)\, h_{\alpha\beta}(y),
\nn  \\
&&P_{\rm B}[h] \equiv \! \int\! d^4x\, d^4y \:
{\cal F}^{\mu\nu}_{x}{\cal F}^{\alpha\beta}_{x} \hspace{-0.2ex}
N_{\rm B}(x\!-\!y;m^2,\Delta \xi) \: 
h_{\mu\nu}(x)\, h_{\alpha\beta}(y).
\label{P_A and P_B}
\eea
Integrating by parts (we always neglect surface terms), 
using expression (\ref{R tensor}) for 
$R^{{\scriptscriptstyle (1)}}$, which can also be written as
$R^{{\scriptscriptstyle (1)}}=-{\cal F}^{\mu\nu}h_{\mu\nu}$, we
find
\be
P_{\rm B}[h] = \int\! d^4x\, d^4y \: 
R^{{\scriptscriptstyle (1)}}(x) \,
N_{\rm B}(x\!-\!y;m^2,\Delta \xi) \,
R^{{\scriptscriptstyle (1)}}(y). 
\label{P_B}
\ee
In order to work out $P_{\rm A}[h]$,
it is useful to take into account that, using
the symmetry properties of the Weyl and Riemann tensors and the
expression (\ref{Riemann tensor}) for 
$R^{{\scriptscriptstyle (1)} \rho\sigma\lambda\tau}$, 
one can write
\be
C^{{\scriptscriptstyle (1)}}_{\rho\sigma\lambda\tau}(x)
C^{{\scriptscriptstyle (1)} \rho\sigma\lambda\tau}
\hspace{-0.1ex}(y)=
C^{{\scriptscriptstyle (1)}}_{\rho\sigma\lambda\tau}(x)
R^{{\scriptscriptstyle (1)} \rho\sigma\lambda\tau}
\hspace{-0.1ex}(y)=
-2 C^{{\scriptscriptstyle (1)} \rho\sigma\lambda\tau}
\hspace{-0.1ex}(x) \hspace{0.2ex}
\delta^\alpha_\rho \delta^\beta_\lambda \hspace{0.2ex}
\partial_\sigma \partial_\tau h_{\alpha \beta}(y).
\ee
Using the last identity, the expression (\ref{Weyl tensor}) for 
$C^{{\scriptscriptstyle (1)} \rho\sigma\lambda\tau}$  
and integrating by parts
the first expression in (\ref{P_A and P_B}) we get
\be
P_{\rm A}[h] = \int\! d^4x\, d^4y \: 
C^{{\scriptscriptstyle (1)}}_{\mu\nu\alpha\beta}(x)\,
N_{\rm A}(x\!-\!y;m^2) \,
C^{{\scriptscriptstyle (1)} \mu\nu\alpha\beta}
\hspace{-0.1ex}(y).
\label{P_A}
\ee
Thus, $P_{\rm A}[h]$ and $P_{\rm B}[h]$ depend, respectively, on the
Weyl tensor and the scalar curvature to first order in the metric
perturbation. 
The result for the massless case, $m \!=\!0$, can be
easily obtained from the above expressions, using 
Eqs.~(\ref{massless N and D kernels}). 
If, in addition, we make 
$\Delta \xi =0$, {\it i.e.}, conformal coupling, we
have $P_{\rm B}[h]\!=\!0$. 
Hence, for a conformal scalar field, particle creation is due to the
breaking of conformal flatness in the spacetime, which implies a
non-zero Weyl tensor.

In order to compare with previously obtained results, it is useful
to introduce the Fourier transform of a field $f(x)$ as
$ \tilde{f}(p) \equiv \!\int\! d^4x\, e^{-i px} f(x)$. Note that, if
$f(x)$ is real, then $\tilde{f}(-p)= \tilde{f}
^{\displaystyle \hspace{0.1ex}\ast \hspace{0.1ex} \!}(p)$.
Using the expressions (\ref{N and D kernels}) for the kernels 
$N_{\rm A}$ and $N_{\rm B}$, 
the above result for the total probability of particle 
creation and the number of particles created can also be written as 
\bea
P[h]={1 \over 1920 \pi} \!\int \! {d^4 p \over (2\pi)^4} \:
\theta (-p^2-4m^2) \, \sqrt{1+4 \,{m^2 \over p^2} } \:
&& \left[ 
\tilde{C}^{{\scriptscriptstyle (1)}}_{\mu\nu\alpha\beta}(p)\,
\tilde{C}^{{\scriptscriptstyle (1)} 
{\displaystyle \hspace{0.1ex}\ast \hspace{0.1ex}}
\mu\nu\alpha\beta}(p) \!
\left(1+4 \,{m^2 \over p^2} \right)^{\! 2}
\right.
\nn  \\
&& \hspace{-3ex}
\left. 
+\,{20 \over 3}  \left| \hspace{0.1ex} 
\tilde{R}^{{\scriptscriptstyle (1)}}(p)
\hspace{0.1ex}\right|^2 \!
\left(3 \hspace{0.3ex}\Delta \xi+{m^2 \over p^2} \right)^{\! 2}
\right]\! +O(h^3),
\label{P}
\eea
in agreement with the results of Ref.~\cite{frieman}
(except for a sign in the coefficient of the term with 
$| \hspace{0.1ex} 
\tilde{R}^{{\scriptscriptstyle (1)}}(p)
\hspace{0.1ex}|^2$).
It is also easy to see that the above result is equivalent to that
found in Ref.~\cite{cespedes} if we take into account that, for
integrals of the form 
$I \equiv \!\int\! d^4p\, \tilde{f}_{a_1 \cdots a_r}(p) \,
G(p^2) \,  \tilde{f}
^{{\displaystyle \hspace{0.1ex}\ast\hspace{0.1ex}} a_1 \cdots a_r}(p)$,
where $f_{a_1 \cdots a_r}(x)$ is any real tensor field 
in Minkowski spacetime and
$G(p^2)$ is any scalar function of $p^2$, one has that
\be
I= 2 \!\int\! d^4p\: \theta(p^0) \,
\tilde{f}_{a_1 \cdots a_r}(p) \,
G(p^2) \,  \tilde{f}
^{{\displaystyle \hspace{0.1ex}\ast\hspace{0.1ex}} a_1 \cdots
a_r}(p)
=2 \!\int\! d^4p\: \theta(-p^0) \,
\tilde{f}_{a_1 \cdots a_r}(p) \,
G(p^2) \,  \tilde{f}
^{{\displaystyle \hspace{0.1ex}\ast\hspace{0.1ex}} a_1 \cdots
a_r}(p).
\label{identity}
\ee
In the massless conformally coupled case, $m\!=\!0$ and 
$\Delta \xi\!=\!0$, the result (\ref{P}) reduces to that found in 
Ref.~\cite{zeldovich}.

The energy of the created particles, $E[h]$, defined as the
expectation value of the ``out'' energy operator in the ``in'' vacuum
can be computed using the expressions derived in Ref.~\cite{paper_1}.
We find that this energy is given by an expression like (\ref{P}), but
with a 
factor $2 p^0 \, \theta(p^0)$ inserted in the integrand
\cite{frieman,paper_1}. Since the kernels $N_{\rm A}$ and $D_{\rm A}$
are related by the fluctuation-dissipation relation 
(\ref{f-d relation}), and the same holds for $N_{\rm B}$ and 
$D_{\rm B}$, it is easy to see [similarly to (\ref{identity})] that
\be
E[h] = i \int \! {d^4 p \over (2\pi)^4 } \:
p^0 \left[ \tilde{C}^{{\scriptscriptstyle (1)}}_{\mu\nu\alpha\beta}(p)\,
\tilde{C}^{{\scriptscriptstyle (1)} 
{\displaystyle \hspace{0.1ex}\ast \hspace{0.1ex}}
\mu\nu\alpha\beta}(p) \, \tilde{D}_{\rm A}(p)+
\left| \hspace{0.1ex} 
\tilde{R}^{{\scriptscriptstyle (1)}}(p)
\hspace{0.1ex}\right|^2 \! \tilde{D}_{\rm B}(p) \right]
+O(h^3),
\label{energy of particles}
\ee
where $\tilde{D}_{\rm A}(p)$ and $\tilde{D}_{\rm B}(p)$ are the
Fourier transforms of the dissipation kernels defined in 
(\ref{N and D kernels}).
For perturbations of a spatially flat RW spacetime
({\it i.e.},  
$h_{\mu\nu} \!=\! 2 \hspace{0.2ex} \Delta a(\eta) \, \eta_{\mu\nu}$,
where $x^0 \!\equiv \! \eta$ is the conformal time and
$\Delta a(\eta)$ is the perturbation of the scale factor),
this last expression agrees with that of
Ref.~\cite{calzettahu}, see also Ref.~\cite{albert2}.

So far in this subsection the metric perturbations are arbitrary.
We may also be interested in the particles created by the back
reaction on the metric due to the stress-energy fluctuations.
Then we would have to use the solutions of the Einstein-Langevin
equations (\ref{massive Einstein-Langevin eq}) and 
(\ref{massless Einstein-Langevin eq 2}) in the above results.
However, to be consistent, one should look for solutions 
whose moments vanish asymptotically in the ``remote past''
and in the ``far future.'' These conditions are generally too strong,
since they would break the time translation invariance
in the correlation functions. In fact, the solutions that we find in
the next section do not satisfy these conditions.



\section{Correlation functions for gravitational
perturbations}
\label{sec:correlation functions}


In this section, we solve the semiclassical Einstein-Langevin
equations (\ref{massive Einstein-Langevin eq}) and 
(\ref{massless Einstein-Langevin eq 2}) for the components 
$G^{{\scriptscriptstyle (1)}\hspace{0.1ex} \mu\nu}$ 
of the linearized Einstein tensor.
In subsection \ref{subsec:Einstein}
we use these solutions to
compute the corresponding two-point correlation functions,
which give a measure of the gravitational fluctuations
predicted by the stochastic semiclassical theory of gravity in the
present case. Since the linearized Einstein tensor 
is invariant under gauge transformations 
of the metric perturbations, these two-point correlation functions are
also gauge invariant. Once we have computed the two-point correlation
functions for the linearized Einstein tensor, we find solutions for the
metric perturbations in subsection \ref{subsec:metric perturbations}
and we show how the associated two-point correlation functions can be
computed. 
This procedure to solve the Einstein-Langevin equations is similar to
the one used by Horowitz \cite{horowitz}, see also
Ref.~\cite{flanagan}, to analyze the stability of Minkowski spacetime 
in semiclassical gravity.

{}From expressions (\ref{D tensor}) and (\ref{B tensor}) 
restricted to $n\!=\!4$, it
is easy to see that 
$A^{{\scriptscriptstyle (1)}\hspace{0.1ex} \mu\nu}$ and 
$B^{{\scriptscriptstyle (1)}\hspace{0.1ex} \mu\nu}$ can
be written in terms of 
$G^{{\scriptscriptstyle (1)}\hspace{0.1ex} \mu\nu}$ as
\be
A^{{\scriptscriptstyle (1)}\hspace{0.1ex} \mu\nu} =
{2 \over 3} \, ({\cal F}^{\mu\nu} 
G^{{\scriptscriptstyle (1)}}\mbox{}^{\alpha}_\alpha
-{\cal F}^{\alpha}_\alpha 
G^{{\scriptscriptstyle (1)}\hspace{0.1ex} \mu\nu}),
\hspace{10ex}
B^{{\scriptscriptstyle (1)}\hspace{0.1ex} \mu\nu} =
2 \hspace{0.2ex} {\cal F}^{\mu\nu} 
G^{{\scriptscriptstyle (1)}}\mbox{}^{\alpha}_\alpha,
\label{A and B}
\ee
where we have used that 
$3 \hspace{0.2ex}\Box={\cal F}^{\alpha}_\alpha$.
Therefore, the Einstein-Langevin equations 
(\ref{massive Einstein-Langevin eq}) and 
(\ref{massless Einstein-Langevin eq 2}) can be seen as linear
integro-differential stochastic equations for the components
$G^{{\scriptscriptstyle (1)}\hspace{0.1ex} \mu\nu}$. 
Such equations can be
written in both cases, $m \!\neq \!0$ and $m\!=\!0$, as
\be
{1\over 8 \pi G} \, 
G^{{\scriptscriptstyle (1)}\hspace{0.1ex} \mu\nu}(x)
\hspace{-0.2ex}-\hspace{-0.2ex} 2 \left(\bar{\alpha} 
A^{{\scriptscriptstyle (1)}\hspace{0.1ex} \mu\nu}(x)
\hspace{-0.2ex}+\hspace{-0.2ex}
\bar{\beta} B^{{\scriptscriptstyle (1)}\hspace{0.1ex} \mu\nu}(x)
\right)\! 
+\!\!\int\! d^4y 
\left[ H_{\rm A}(x\!-\!y) 
A^{{\scriptscriptstyle (1)}\hspace{0.1ex} \mu\nu}(y)
\hspace{-0.2ex}+\hspace{-0.2ex}
H_{\rm B}(x\!-\!y)
B^{{\scriptscriptstyle (1)}\hspace{0.1ex} \mu\nu}(y) \right]
=2 \xi^{\mu\nu}(x),
\label{unified Einstein-Langevin eq}
\ee
where the new constants $\bar{\alpha}$ and $\bar{\beta}$, and 
the kernels $H_{\rm A}(x)$ and $H_{\rm B}(x)$ can be identified 
in each case by comparison of this last equation with 
Eqs.~(\ref{massive Einstein-Langevin eq}) and 
(\ref{massless Einstein-Langevin eq 2}). 
For instance, when $m \!=\! 0$, we have 
$H_{\rm A}(x)=H(x;\mu^2)$ and 
$H_{\rm B}(x)=60 \hspace{0.2ex} \Delta \xi^2 H(x;\mu^2)$.
In this case, we can  
use the arbitrariness of the mass scale $\mu$ to
eliminate one of the parameters $\bar{\alpha}$ or $\bar{\beta}$.

In order to find solutions to these equations, it is convenient to 
Fourier transform them. Introducing Fourier transforms as in
subsection \ref{subsec:particle creation}, 
one finds, from (\ref{A and B}), 
\be
\tilde{A}^{{\scriptscriptstyle (1)}\hspace{0.1ex} \mu\nu}(p)=
2 p^2 \tilde{G}^{{\scriptscriptstyle (1)}\hspace{0.1ex} \mu\nu}(p)
-{2 \over 3} \, p^2 P^{\mu\nu} 
\tilde{G}^{{\scriptscriptstyle (1)}}\mbox{}^{\alpha}_\alpha(p),
\hspace{10ex}
\tilde{B}^{{\scriptscriptstyle (1)}\hspace{0.1ex} \mu\nu}(p)=
-2 p^2 P^{\mu\nu} 
\tilde{G}^{{\scriptscriptstyle (1)}}\mbox{}^{\alpha}_\alpha(p).
\ee 
Using these relations, the Fourier transform of 
Eq.~(\ref{unified Einstein-Langevin eq}) reads 
\be
F^{\mu\nu}_{\hspace{2ex}\alpha\beta}(p) \,
\tilde{G}^{{\scriptscriptstyle (1)}\hspace{0.1ex} \alpha\beta}(p)=
16 \pi G \, \tilde{\xi}^{\mu\nu}(p),
\label{Fourier transf of E-L eq}
\ee
where
\be
F^{\mu\nu}_{\hspace{2ex} \alpha\beta}(p) \equiv
F_1(p) \, \delta^\mu_{( \alpha} \delta^\nu_{\beta )}+
F_2(p) \, p^2 P^{\mu\nu} \eta_{\alpha\beta},
\label{F def} 
\ee
with
\be
F_1(p) \equiv 1+16 \pi G \, p^2 
\left[ \tilde{H}_{\rm A}(p)-2 \bar{\alpha}\right],
\hspace{7ex}
F_2(p) \equiv -{16 \over 3} \, \pi G 
\left[ \tilde{H}_{\rm A}(p)+3 \tilde{H}_{\rm B}(p)
-2 \bar{\alpha}-6 \bar{\beta}\right].
\label{F_1 and F_2}
\ee
In Eq.~(\ref{Fourier transf of E-L eq}), 
$\tilde{\xi}^{\mu\nu}(p)$, the Fourier transform of 
$\xi^{\mu\nu}(x)$, is a Gaussian stochastic source of zero average and
\be
\langle \tilde{\xi}^{\mu\nu}(p) 
\tilde{\xi}^{\alpha\beta}(p^\prime)
\rangle_c = (2 \pi)^4 \, \delta^4(p+p^\prime) \,
\tilde{N}^{\mu\nu\alpha\beta}(p),
\label{Fourier transf of corr funct}
\ee
where we have introduced the Fourier transform of the noise kernel.
The explicit expression for $\tilde{N}^{\mu\nu\alpha\beta}(p)$
is found from
(\ref{noise and dissipation kernels 2}) and (\ref{N and D kernels}) 
to be
\bea
\tilde{N}^{\mu\nu\alpha\beta}(p)= 
{1 \over 2880 \pi} \: 
\theta (-p^2\!-\!4m^2) \, \sqrt{1+4 \,{m^2 \over p^2} } \;
&& \left[ {1 \over 4} \left(1+4 \,{m^2 \over p^2} \right)^{\!2}
(p^2)^2 \,
\bigl( 3 P^{\mu (\alpha}P^{\beta )\nu}-P^{\mu\nu} P^{\alpha\beta}
\bigr) \right.
\nn  \\
&&\hspace{1.8ex} \left.
+\, 10 \!
\left(3 \hspace{0.2ex}\Delta \xi+{m^2 \over p^2} \right)^{\!2}
(p^2)^2 P^{\mu\nu} P^{\alpha\beta} \right],
\label{Fourier transf of noise 2}
\eea
which in the massless case reduces to
\be
\lim_{m \rightarrow 0}\! \tilde{N}^{\mu\nu\alpha\beta}(p)=
{1 \over 1920 \pi} \: 
\theta (-p^2)  \left[ {1 \over 6} \, 
(p^2)^2 \,
\bigl(3 P^{\mu (\alpha}P^{\beta )\nu}-P^{\mu\nu} P^{\alpha\beta}
\bigr)
+60 \hspace{0.2ex} \Delta \xi^2 (p^2)^2 P^{\mu\nu} P^{\alpha\beta}
\right].
\label{Fourier transf of massless noise}
\ee


\subsection{Correlation functions for the linearized
Einstein tensor}
\label{subsec:Einstein}


In general, we can write 
$G^{{\scriptscriptstyle (1)}\hspace{0.1ex} \mu\nu}=
\langle G^{{\scriptscriptstyle (1)}\hspace{0.1ex} \mu\nu} \rangle_c
+G_{\rm f}^{{\scriptscriptstyle (1)}\hspace{0.1ex} \mu\nu}$,
where 
$G_{\rm f}^{{\scriptscriptstyle (1)}\hspace{0.1ex} \mu\nu}$
is a solution to Eqs.~(\ref{unified Einstein-Langevin eq})
[or, in the Fourier transformed version, 
(\ref{Fourier transf of E-L eq})] with zero average.
The averages 
$\langle G^{{\scriptscriptstyle (1)}\hspace{0.1ex} \mu\nu} \rangle_c$
must be a solution of the linearized semiclassical Einstein equations 
obtained by averaging Eqs.~(\ref{unified Einstein-Langevin eq})
[or (\ref{Fourier transf of E-L eq})]. 
Solutions to these equations (specially in the massless case, 
$m \!=\! 0$) have been studied by several authors
\cite{horowitz,horowitz-wald78,horowitz81,hartle_horowitz,simon,%
jordan,flanagan},
particularly in connection with the issue of the stability of the
trivial solutions of semiclassical gravity. 
The two-point correlation functions for the linearized Einstein tensor
are given by 
\be
{\cal G}^{\mu\nu\alpha\beta}(x,x^{\prime}) \equiv 
\langle G^{{\scriptscriptstyle (1)}\hspace{0.1ex} \mu\nu}(x)
G^{{\scriptscriptstyle (1)}\hspace{0.1ex} \alpha\beta}(x^{\prime}) 
\rangle_c 
-\langle G^{{\scriptscriptstyle (1)}\hspace{0.1ex} \mu\nu}(x)
\rangle_c 
\langle G^{{\scriptscriptstyle (1)}\hspace{0.1ex} \alpha\beta}
(x^{\prime})\rangle_c =
\langle G_{\rm f}^{{\scriptscriptstyle (1)}\hspace{0.1ex} \mu\nu}(x)
G_{\rm f}^{{\scriptscriptstyle (1)}\hspace{0.1ex} \alpha\beta}
(x^{\prime})\rangle_c.
\label{two-p corr funct}
\ee

Next, we shall seek the family of solutions to the Einstein-Langevin
equations which can be written as a
linear functional of the stochastic source
and whose Fourier transform, 
$\tilde{G}^{{\scriptscriptstyle (1)}\hspace{0.1ex} \mu\nu}(p)$, 
depends locally on $\tilde{\xi}^{\alpha\beta}(p)$.  
Each of such solutions is a Gaussian stochastic field
and, thus, it can be completely characterized by the 
averages 
$\langle G^{{\scriptscriptstyle (1)}\hspace{0.1ex} \mu\nu} \rangle_c$
and the two-point correlation functions 
(\ref{two-p corr funct}).
For such a family of solutions, 
$\tilde{G}_{\rm f}^{{\scriptscriptstyle (1)}\hspace{0.1ex} \mu\nu}(p)$
is the most general solution to Eq.~(\ref{Fourier transf of E-L eq})
which is linear, homogeneous and local in 
$\tilde{\xi}^{\alpha\beta}(p)$. It can be written as
\be
\tilde{G}_{\rm f}^{{\scriptscriptstyle (1)}\hspace{0.1ex} \mu\nu}(p)
= 16 \pi G \, D^{\mu\nu}_{\hspace{2ex} \alpha\beta}(p) \,
\tilde{\xi}^{\alpha\beta}(p),
\label{G_f}
\ee
where
$D^{\mu\nu}_{\hspace{2ex} \alpha\beta}(p)$ 
are the components of a Lorentz invariant tensor 
field distribution in Minkowski spacetime
[by ``Lorentz
invariant'' we mean invariant under the transformations of the
orthochronous Lorentz subgroup; see  
Ref.~\cite{horowitz} for more details on the definition 
and properties of these tensor distributions], 
symmetric under the interchanges 
$\alpha \! \leftrightarrow \!\beta$ and  
$\mu \! \leftrightarrow \!\nu$, which is the most general solution
of
\be
F^{\mu\nu}_{\hspace{2ex} \rho\sigma}(p) \,
D^{\rho\sigma}_{\hspace{2ex} \alpha\beta}(p)=
\delta^\mu_{( \alpha} \delta^\nu_{\beta )}.
\label{eq for D}
\ee
In addition, we must impose the conservation condition to 
the solutions:
$p_\nu 
\tilde{G}_{\rm f}^{{\scriptscriptstyle (1)}\hspace{0.1ex} \mu\nu}(p)
= 0$, where this zero must be understood as a stochastic variable
which behaves deterministically as a zero vector field. 
We can write
$D^{\mu\nu}_{\hspace{2ex} \alpha\beta}(p)=
D^{\mu\nu}_{p \hspace{1.2ex} \alpha\beta}(p)+
D^{\mu\nu}_{h \hspace{1.2ex} \alpha\beta}(p)$, where
$D^{\mu\nu}_{p \hspace{1.2ex} \alpha\beta}(p)$ is a particular
solution to Eq.~(\ref{eq for D}) and 
$D^{\mu\nu}_{h \hspace{1.2ex} \alpha\beta}(p)$ is the most general
solution to the corresponding homogeneous equation. 
Correspondingly [see Eq.~(\ref{G_f})], we can write
$\tilde{G}_{\rm f}^{{\scriptscriptstyle (1)}\hspace{0.1ex} \mu\nu}(p)
=\tilde{G}_p^{{\scriptscriptstyle (1)}\hspace{0.1ex} \mu\nu}(p)+
\tilde{G}_h^{{\scriptscriptstyle (1)}\hspace{0.1ex} \mu\nu}(p)$. 
To find the particular solution, we try an ansatz of the form
\be
D^{\mu\nu}_{p \hspace{1.2ex} \alpha\beta}(p)=
d_1(p) \, \delta^\mu_{( \alpha} \delta^\nu_{\beta )}
+ d_2(p) \, p^2 P^{\mu\nu} \eta_{\alpha\beta}.
\label{ansatz for D}
\ee
Substituting this ansatz into
Eqs.~(\ref{eq for D}), it is easy to see that it
solves these equations if we take
\be
d_1(p)=\left[ {1 \over F_1(p)} \right]_r,
\hspace{7ex}
d_2(p)= - \left[ {F_2(p)\over F_1(p) F_3(p)} \right]_r,
\label{d's}
\ee
with
\be
F_3(p) \equiv F_1(p) + 3 p^2 F_2(p)= 1-48 \pi G \, p^2 
\left[ \tilde{H}_{\rm B}(p)-2 \bar{\beta}\right], 
\label{F_3}
\ee
and where the notation $[ \;\; ]_r$ means that the zeros of the
denominators are regulated with appropriate prescriptions
in such a way that $d_1(p)$ and $d_2(p)$ are well defined
Lorentz invariant scalar distributions. 
This yields a particular solution to the 
Einstein-Langevin equations:
\be
\tilde{G}_p^{{\scriptscriptstyle (1)}\hspace{0.1ex} \mu\nu}(p)
= 16 \pi G \, D^{\mu\nu}_{p \hspace{1.2ex} \alpha\beta}(p) \,
\tilde{\xi}^{\alpha\beta}(p),
\label{solution}
\ee
which, since the stochastic source is conserved, satisfies the
conservation condition.
Note that, in the case of a massless scalar field, $m\!=\!0$, the
above solution has a functional form analogous to that of the
solutions of linearized semiclassical gravity found in the 
Appendix of Ref.~\cite{flanagan}.
Notice also that, for a massless conformally coupled field,
$m\!=\!0$ and $\Delta \xi\!=\!0$, the second term in the right hand
side of Eq.~(\ref{ansatz for D}) will not contribute in the
correlation functions (\ref{two-p corr funct}), since,  
as we have pointed out in
Sec.~\ref{subsec:massless case}, in this case the stochastic
source is ``traceless.''

Next, we can work out the general form for 
$D^{\mu\nu}_{h \hspace{1.2ex} \alpha\beta}(p)$, which is a linear
combination of terms consisting of a Lorentz invariant scalar 
distribution times one of the products
$\delta^\mu_{( \alpha} \delta^\nu_{\beta )}$,
$p^2 \hspace{-0.2ex} P^{\mu\nu} \eta_{\alpha\beta}$,
$\eta^{\mu\nu} \eta_{\alpha\beta}$,
$\eta^{\mu\nu} p^2 \hspace{-0.2ex} P_{\alpha\beta}$,
$\delta^{( \mu}_{( \alpha} \hspace{0.5ex} 
p^2 \hspace{-0.2ex} P^{\nu )}_{\beta )}$
and $p^2 \hspace{-0.2ex} P^{\mu\nu} \hspace{0.2ex}
p^2 \hspace{-0.2ex} P_{\alpha\beta}$.
However, taking into account that the stochastic source is conserved,
we can omit some terms in 
$D^{\mu\nu}_{h \hspace{1.2ex} \alpha\beta}(p)$
and simply write
\be
\tilde{G}_h^{{\scriptscriptstyle (1)}\hspace{0.1ex} \mu\nu}(p)
= 16 \pi G \, D^{\mu\nu}_{h \hspace{1.2ex} \alpha\beta}(p) \,
\tilde{\xi}^{\alpha\beta}(p),
\ee
with
\be
D^{\mu\nu}_{h \hspace{1.2ex} \alpha\beta}(p)=
h_1(p)\, \delta^\mu_{( \alpha} \delta^\nu_{\beta )}
+ h_2(p) \, p^2 P^{\mu\nu} \eta_{\alpha\beta}
+h_3(p) \,  \eta^{\mu\nu} \eta_{\alpha\beta},
\label{bar D}
\ee
where $h_1(p)$, $h_2(p)$ and $h_3(p)$ are Lorentz invariant scalar 
distributions. {}From the fact that 
$D^{\mu\nu}_{h \hspace{1.2ex} \alpha\beta}(p)$ must satisfy the
homogeneous equation corresponding to Eq.~(\ref{eq for D}), we find
that $h_1(p)$ and $h_3(p)$ have support on the set of points 
$\{ p^\mu \}$ for which $F_1(p) \!=\! 0$, and that
$h_2(p)$ has support on the set of points 
$\{ p^\mu \}$ for which $F_1(p) \!=\! 0$ or $F_3(p) \!=\! 0$.
Moreover, the conservation condition for 
$\tilde{G}_h^{{\scriptscriptstyle (1)}\hspace{0.1ex} \mu\nu}(p)$
implies that the term with $h_3(p)$ is only allowed in the case 
of a massless conformally coupled field, 
$m\!=\!0$ and $\Delta \xi\!=\!0$. 
{}From (\ref{Fourier transf of corr funct}), we get
\be
\langle 
\tilde{G}_h^{{\scriptscriptstyle (1)}\hspace{0.1ex} \mu\nu}(p) \,
\tilde{\xi}^{\alpha\beta}(p^\prime)
\rangle_c = (2 \pi)^4 \, 16 \pi G \, \delta^4(p+p^\prime) \,
D^{\mu\nu}_{h \hspace{1.2ex} \rho\sigma}(p) \,
\tilde{N}^{\rho\sigma\alpha\beta}(p).
\label{c f}
\ee
Note, from expressions 
(\ref{Fourier transf of noise 2}) and 
(\ref{Fourier transf of massless noise}), that the support of 
$\tilde{N}^{\mu\nu\alpha\beta}(p)$ is on the set of points 
$\{ p^\mu \}$ for which $-p^2 \!\geq \! 0$ when $m\!=\!0$,
and for which $-p^2-4 m^2 \! > \! 0$ when $m \! \neq \! 0$. 
At such points, using expressions (\ref{F_1 and F_2}),
(\ref{F_3}), (\ref{H}) and (\ref{H kernels}), 
it is easy to see that $F_1(p)$ is always 
different from zero, and that $F_3(p)$ is also always different from
zero, except for some particular values of $\Delta \xi$ and 
$\bar{\beta}$: 
\begin{itemize}
\item[a)] when $m \!=\!0$, $\Delta \xi\!=\!0$ and 
               $\bar{\beta} \! > \! 0$;
\item[b)] when $m \! \neq \! 0$, 
               $0 \! < \! \Delta \xi \! <\!  (1/12)$ and
               $\bar{\beta} \!=\! (\Delta \xi/ 32 \pi^2)
                \hspace{0.2ex}
                [ \pi/(G m^2)+1/36]$.
\end{itemize}
In the case a), $F_3(p)\!=\!0$ for the set of points $\{ p^\mu \}$
satisfying $-p^2 \!=\! 1/(96 \pi G \bar{\beta})$; in the case
b), $F_3(p)\!=\!0$ for $\{ p^\mu \}$ such that 
$-p^2 \!=\! m^2/(3 \Delta \xi)$.
Hence, except for the above cases a) and b), the intersection of the
supports of $\tilde{N}^{\mu\nu\alpha\beta}(p)$ and 
$D^{\rho\sigma}_{h \hspace{1.2ex} \lambda\gamma}(p)$ is an empty
set and, thus, the correlation function (\ref{c f}) is zero.
In the cases a) and b), we can have a contribution to 
(\ref{c f}) coming from the term with $h_2(p)$ in (\ref{bar D})
of the form
$D^{\mu\nu}_{h \hspace{1.2ex} \rho\sigma}(p) \,
\tilde{N}^{\rho\sigma\alpha\beta}(p)\!=\!
H_3(p; \{ C \}) \, p^2 P^{\mu\nu} \hspace{0.2ex} 
\tilde{N}^{\alpha\beta\rho}_{\hspace{3.3ex} \rho}(p)$,
where $H_3(p; \{ C \})$ is the most general Lorentz invariant
distribution satisfying 
$F_3(p) \hspace{0.2ex} H_3(p; \{ C \})\!=\! 0$, which depends on a set
of arbitrary parameters represented as $\{ C \}$. However,
from (\ref{Fourier transf of noise 2}), we see that
$\tilde{N}^{\alpha\beta\rho}_{\hspace{3.3ex} \rho}(p)$ is proportional
to $\theta (-p^2\!-\!4m^2) \hspace{0.2ex} 
(1+4 m^2/p^2)^{(1/2)} \hspace{0.2ex} 
(3 \Delta \xi+m^2/p^2 )^2$. Thus, in the case a), 
we have $\tilde{N}^{\alpha\beta\rho}_{\hspace{3.3ex} \rho}(p) 
\!=\! 0$ and, in the case b), the intersection of the supports of 
$\tilde{N}^{\alpha\beta\rho}_{\hspace{3.3ex} \rho}(p)$ and of
$H_3(p; \{ C \})$ is an empty set. 
Therefore, from the above analysis, we conclude that
$\tilde{G}_h^{{\scriptscriptstyle (1)}\hspace{0.1ex} \mu\nu}(p)$ gives
no contribution to the correlation functions 
(\ref{two-p corr funct}), since
$\langle 
\tilde{G}_h^{{\scriptscriptstyle (1)}\hspace{0.1ex} \mu\nu}(p) \,
\tilde{\xi}^{\alpha\beta}(p^\prime)
\rangle_c \!=\! 0$, and we have simply  
${\cal G}^{\mu\nu\alpha\beta}(x,x^{\prime}) \!=\!
\langle G_p^{{\scriptscriptstyle (1)}\hspace{0.1ex} \mu\nu}(x)
G_p^{{\scriptscriptstyle (1)}\hspace{0.1ex} \alpha\beta}
(x^{\prime})\rangle_c$, where 
$G_p^{{\scriptscriptstyle (1)}\hspace{0.1ex} \mu\nu}(x)$ is the
inverse Fourier transform of (\ref{solution}).

The correlation functions (\ref{two-p corr funct}) can then be 
computed from
\be
\langle 
\tilde{G}_p^{{\scriptscriptstyle (1)}\hspace{0.1ex} \mu\nu}(p) \,
\tilde{G}_p^{{\scriptscriptstyle (1)}\hspace{0.1ex} \alpha\beta}
(p^\prime) \rangle_c = 
64 \, (2 \pi)^6 \, G^2 \, \delta^4(p+p^\prime) \,
D^{\mu\nu}_{p \hspace{1.2ex} \rho\sigma}(p) \,
D^{\alpha\beta}_{p \hspace{1.2ex} \lambda\gamma}(-p) \,
\tilde{N}^{\rho\sigma\lambda\gamma}(p).
\ee
It is easy to see from the above analysis that the prescriptions
$[ \;\; ]_r$ in the factors $D_p$ are irrelevant in the last
expression and, thus, they can be suppressed.
Taking into account that 
$F_l(-p) \!=\! F^{\displaystyle \ast}_l(p)$, with $l \!=\! 1,2,3$, 
we get from Eqs.~(\ref{ansatz for D}) and (\ref{d's})
\bea
\langle 
\tilde{G}_p^{{\scriptscriptstyle (1)}\hspace{0.1ex} \mu\nu}(p) \,
\tilde{G}_p^{{\scriptscriptstyle (1)}\hspace{0.1ex} \alpha\beta}
(p^\prime) \rangle_c = &&
64 \, (2 \pi)^6 \, G^2 \: {\delta^4(p+p^\prime) \over 
\left| \hspace{0.1ex} F_1(p) \hspace{0.1ex}\right|^2 }
\left[ 
\tilde{N}^{\mu\nu\alpha\beta}(p)
- {F_2(p) \over F_3(p)} \: p^2 P^{\mu\nu} \hspace{0.2ex} 
\tilde{N}^{\alpha\beta\rho}_{\hspace{3.3ex} \rho}(p)
\right.     \nn  \\
&& \hspace{7ex}
\left. - \,
{F_2^{\displaystyle \ast}(p) \over F_3^{\displaystyle \ast}(p)} 
\: p^2 P^{\alpha\beta} \hspace{0.2ex} 
\tilde{N}^{\mu\nu\rho}_{\hspace{3.3ex} \rho}(p)
+ { \left| \hspace{0.1ex} F_2(p) \hspace{0.1ex}\right|^2
\over \left| \hspace{0.1ex} F_3(p) \hspace{0.1ex}\right|^2 } \:
p^2 P^{\mu\nu} \hspace{0.2ex} p^2 P^{\alpha\beta} \hspace{0.2ex}
\tilde{N}
^{\rho \hspace{0.9ex} \sigma}_{\hspace{1ex} \rho \hspace{1.1ex}
\sigma} (p) \right].   \nn  \\
\mbox{}
\eea 
This last expression is well defined as a bi-distribution
and can be easily evaluated using 
Eq.~(\ref{Fourier transf of noise 2}). We find
\bea
\langle 
\tilde{G}_p^{{\scriptscriptstyle (1)}\hspace{0.1ex} \mu\nu}(p) \,
\tilde{G}_p^{{\scriptscriptstyle (1)}\hspace{0.1ex} \alpha\beta}
(p^\prime) \rangle_c = 
{2 \over 45} && \, (2 \pi)^5 \, G^2 \: 
{\delta^4(p+p^\prime) \over 
\left| \hspace{0.1ex} F_1(p) \hspace{0.1ex}\right|^2 } \:
\theta (-p^2\!-\!4m^2) \, \sqrt{1+4 \,{m^2 \over p^2} } 
\nn   \\
&&    \times \!
\left[{1 \over 4} \left(1+4 \,{m^2 \over p^2} \right)^{\!2} \!
(p^2)^2 \,
\bigl( 3 P^{\mu (\alpha}P^{\beta )\nu} \!-\!
P^{\mu\nu} P^{\alpha\beta} \bigr)  \right.
\nn   \\
&& \hspace{3.5ex} \left.
+\, 10 \!
\left(3 \hspace{0.2ex}\Delta \xi+{m^2 \over p^2} \right)^{\!2} \!
(p^2)^2 P^{\mu\nu} P^{\alpha\beta} 
\left| 1-3 p^2 \, {F_2(p) \over F_3(p)} \right|^2
\right].   
\label{Fourier tr corr funct}
\eea
To derive the correlation functions (\ref{two-p corr funct}), we have
to take the inverse Fourier transform of the above result. 
We finally obtain
\be
{\cal G}^{\mu\nu\alpha\beta}(x,x^{\prime})=
{\pi \over 45} \, G^2 \,{\cal F}^{\mu\nu\alpha\beta}_{x} \,
{\cal G}_{\rm A} (x-x^{\prime})+
{8 \pi \over 9} \, G^2 \,
{\cal F}^{\mu\nu}_{x} {\cal F}^{\alpha\beta}_{x} \,
{\cal G}_{\rm B} (x-x^{\prime}),
\label{corr funct}
\ee
with
\bea
&&\tilde{{\cal G}}_{\rm A}(p) \equiv 
\theta (-p^2-4m^2) \, \sqrt{1+4 \,{m^2 \over p^2} } 
\left(1+4 \,{m^2 \over p^2} \right)^{\!2} \!
{1 \over 
\left| \hspace{0.1ex} F_1(p) \hspace{0.1ex}\right|^2 } \, ,
\nn   \\
&&\tilde{{\cal G}}_{\rm B}(p) \equiv 
\theta (-p^2-4m^2) \, \sqrt{1+4 \,{m^2 \over p^2} } 
\left(3 \hspace{0.2ex}\Delta \xi+{m^2 \over p^2} \right)^{\!2} \!
{1 \over 
\left| \hspace{0.1ex} F_1(p) \hspace{0.1ex}\right|^2 } \,
\left| 1-3 p^2 \, {F_2(p) \over F_3(p)} \right|^2,
\label{distri}
\eea
and 
${\cal F}^{\mu\nu\alpha\beta}_{x} \equiv
3 {\cal F}^{\mu (\alpha}_{x}{\cal F}^{\beta )\nu}_{x}-
{\cal F}^{\mu\nu}_{x}{\cal F}^{\alpha\beta}_{x}$,
and where $F_l(p)$, $l=1,2,3$, are given in (\ref{F_1 and F_2}) and
(\ref{F_3}). Notice that, for a massless field ($m \!=\! 0$), we have
\bea
&&F_1(p)= 1+16 \pi G \hspace{0.2ex} p^2 \hspace{0.2ex} 
          \tilde{H}(p;\bar{\mu}^2), 
\nn  \\
&&F_2(p)= - {16 \over 3} \, \pi G \left[ 
(1 +180 \hspace{0.2ex}\Delta \xi^2 ) \, \tilde{H}(p;\bar{\mu}^2)
-6 \Upsilon \right],
\nn  \\
&&F_3(p)= 1- 48 \pi G \hspace{0.2ex} p^2
\left[ 60 \hspace{0.2ex}\Delta \xi^2 \hspace{0.2ex}
\tilde{H}(p;\bar{\mu}^2) -2 \Upsilon \right],
\eea
with $\bar{\mu} \equiv \mu\, \exp (1920 \pi^2 \bar{\alpha})$
and 
$\Upsilon \equiv \bar{\beta} 
 -60 \hspace{0.2ex}\Delta \xi^2 \hspace{0.2ex}\bar{\alpha}$, and where
$\tilde{H}(p;\mu^2)$ is the Fourier transform of
$H(x;\mu^2)$ given in (\ref{H}).


\subsection{Conformal field case}
\label{subsec:conformal}


The above correlation functions become simpler when the scalar
field is massless and conformally coupled, {\it i.e.}, when
$m \!=\! 0$ and $\Delta \xi\!=\!0$, since in this case 
${\cal G}_{\rm B} (x) \!=\!0$ and $\tilde{{\cal G}}_{\rm A}(p)$
reduces to $\tilde{{\cal G}}_{\rm A}(p)= \theta(-p^2) 
\left|\hspace{0.2ex} F_1(p) \hspace{0.2ex} \right|^{-2}$.
Introducing the function 
$\varphi (\chi ; \lambda ) \equiv \left[ 1-\chi 
\ln \hspace{-0.2ex}\left(\lambda \chi /e\right)\right]^2
+\pi^2 \chi^2$, 
with $\chi \geq 0$ and $\lambda > 0$,
${\cal G}_{\rm A}(x)$ can be written as
\be
{\cal G}_{\rm A}(x) = {(120 \pi)^{3/2} \over 2 \pi^3 L_P^3}  \,
{1 \over |{\bf x}|} \!\int_{0}^{\infty}\!\! d|{\bf q}| \hspace{0.2ex}
|{\bf q}|
\sin \!\left[ { \sqrt{120 \pi} \over L_P}\,
|{\bf x}| \hspace{0.2ex} |{\bf q}| \right]  
\int_{0}^{\infty}\! dq^0 
\cos \!\left[ { \sqrt{120 \pi} \over L_P} \,
x^0 q^0 \right] {\theta(-q^2) \over \varphi (-q^2; \lambda )}, 
\label{G}
\ee
where $L_P \equiv \sqrt{G}$ is the Planck length,
$\lambda \equiv 120 \pi e /(L_P^2 \bar{\mu}^2)$,
and we use the notation 
$x^{\mu}=(x^0,{\bf x})$ and $q^{\mu}=(q^0,{\bf q})$.
Notice that, if we assume that
$\bar{\mu} \leq L_P^{-1}$, then $\lambda \gtrsim 10^3$.
For those values of the parameter $\lambda$ (and also for smaller
values), the function $\varphi (\chi ; \lambda )$ has a minimum
at some value of $\chi$ that we denote as $\chi_0(\lambda)$.
This can be found by solving the equation 
$\pi^2 \chi_0 = \left[1-\chi_0 \ln( \lambda \chi_0 /e)\right] 
\left[1+ \ln( \lambda \chi_0 /e) \right]$
numerically (discarding a solution 
$\chi_M(\lambda) < \chi_0(\lambda)$, 
at which the function $\varphi (\chi ; \lambda )$ has
a maximum). Since the main contribution to the integral (\ref{G}) come
from the values of $-q^2$ around $-q^2 \!=\! \chi_0(\lambda)$,
$\varphi (\chi ; \lambda )$ can be approximately replaced in this
integral by 
$\varphi_{\rm ap} (\chi ; \lambda ) \equiv 
[\hspace{0.1ex} 
1 - \kappa (\lambda ) \, \chi \hspace{0.2ex} ]^2
+ \pi^2 \chi^2 = 
[\hspace{0.1ex} \kappa^2 (\lambda )+ \pi^2 \hspace{0.2ex} ]
\, \chi^2
- 2 \kappa (\lambda ) \, \chi + 1$, 
with 
$\kappa (\lambda ) \equiv \ln \hspace{-0.2ex}
\left( \lambda \chi_0(\lambda)/e \right)$.
For $(\lambda /5) \sim 10^3 - 10^7$, we have
$\kappa \sim 10$.

Let the spacetime points $x$ and $x^{\prime}$ be 
different and spacelike separated. 
In this case, we can choose an inertial coordinate system
for which $(x-x^{\prime})^\mu=(0,{\bf x}-{\bf x}^\prime)$ and
${\cal G}^{\mu\nu\alpha\beta}(x,x^{\prime})$
will be a function of ${\bf x}-{\bf x}^\prime$ only
that can be written as 
\be
{\cal G}^{\mu\nu\alpha\beta}({\bf x}-{\bf x}^\prime) =
{\cal G}_1^{\mu\nu\alpha\beta}({\bf x}-{\bf x}^\prime)
+{\cal G}_2^{\mu\nu\alpha\beta}({\bf x}-{\bf x}^\prime)
+{\cal G}_3^{\mu\nu\alpha\beta}({\bf x}-{\bf x}^\prime),
\label{G spacelike}
\ee
with 
\be
{\cal G}_a^{\mu\nu\alpha\beta}({\bf x}) \equiv
{\pi \over 45} \, G^2 \,
{\cal F}^{\mu\nu\alpha\beta}
_{\!\! a_{\scriptstyle \bf \hspace{0.1ex} x}} \hspace{0.1ex}
I_a({\bf x}),
\label{G_a}
\ee 
$a \!=\! 1,2,3$, where
$I_1({\bf x}) \!\equiv \!\left. {\cal G}_{\rm A} (x) 
\right|_{x^\mu =(0,{\bf x})}$,
$I_2({\bf x}) \!\equiv \!\left. (\partial^0_x)^2
{\cal G}_{\rm A} (x) 
\right|_{x^\mu =(0,{\bf x})}$,
$I_3({\bf x}) \!\equiv \!\left. (\partial^0_x)^4
{\cal G}_{\rm A} (x) 
\right|_{x^\mu =(0,{\bf x})}$, and ${\cal F}^{\mu\nu\alpha\beta}
_{\!\! a_{\scriptstyle \bf \hspace{0.1ex} x}}$ are some differential
operators. Note that the terms containing an odd number of 
$\partial^0_x$ derivatives are zero. The differential operators
${\cal F}^{\mu\nu\alpha\beta}
_{\! 1_{\scriptstyle \bf \hspace{0.1ex} x}}$ are given by
${\cal F}^{\mu\nu\alpha\beta}_{\! 1} \!=
3 \hspace{0.2ex}
{\cal D}^{\mu (\alpha} {\cal D}^{\beta )\nu} -
{\cal D}^{\mu\nu} {\cal D}^{\alpha\beta}$, with
${\cal D}^{\mu\nu} \!\equiv \! (\eta^{\mu\nu} \delta^{ij}-
\delta^{\mu i} \delta^{\nu j}) \, \partial_i \partial_j$.
The non-null components of the remaining operators are
${\cal F}^{00ij}_{\! 2} \!= 
    3 \hspace{0.1ex}
    \partial^i \partial^j 
    - \delta^{ij} \hspace{-0.2ex} \bigtriangleup$,	 
${\cal F}^{0i0j}_{\! 2} \!= 
     {1 \over 2} \, (\partial^i \partial^j 
     + 3 \delta^{ij} \hspace{-0.2ex} \bigtriangleup )$,
${\cal F}^{ijkl}_{\! 3} \!= - \delta^{ij} \delta^{kl}
      + 3  \delta^{i(k} \delta^{l)j}$,
${\cal F}^{ijkl}_{\! 2} \!=
     2 \, (\delta^{ij} \delta^{kl} - 3 \delta^{i(k} \delta^{l)j})
     \hspace{-0.3ex}
     \bigtriangleup - \delta^{ij} \partial^k \partial^l
     - \delta^{kl} \partial^i \partial^j
     +3 \, ( \delta^{i(k} \partial^{l)} \partial^j +
             \delta^{j(k} \partial^{l)} \partial^i )$,
where $\bigtriangleup \!\equiv \! \delta^{ij} \partial_i \partial_j$
is the usual (Euclidean space) Laplace operator.
{}From the above expressions, we can see that 
${\cal G}^{000i}({\bf x}-{\bf x}^\prime) =
{\cal G}^{0ijk}({\bf x}-{\bf x}^\prime) = 0$, but the
remaining correlation functions 
${\cal G}^{\mu\nu\alpha\beta}({\bf x}-{\bf x}^\prime)$
are in principle non-null.

With the approximation described above, the integrals $I_a({\bf x})$
can be written as
\be
I_a({\bf x}) \simeq { (-1)^{a+1} \over 2 \pi^3} 
\left( { 120 \pi \over L_P^2 } \right)^{\!\! a+1/2} \! 
{1 \over |{\bf x}|} \!\int_{0}^{\infty}\!\! d|{\bf q}| \hspace{0.2ex}
\sin \!\left[ { \sqrt{120 \pi} \over L_P}\,
|{\bf x}| \hspace{0.2ex} |{\bf q}| \right] 
|{\bf q}| \, J_a(|{\bf q}|), 
\label{I integrals}
\ee
where
\bea
&&J_1(|{\bf q}|) \equiv \int_{|{\bf q}|}^{\infty}\! dq^0 \,
{1 \over \varphi_{\rm ap}(-q^2; \lambda )}, 
\hspace{10ex}
J_2(|{\bf q}|) \equiv \int_{|{\bf q}|}^{\infty}\! dq^0 \,
{(q^0)^2 \over \varphi_{\rm ap}(-q^2; \lambda )}, 
\nn   \\
&&J_3(|{\bf q}|) \equiv 
{- |{\bf q}| \over 
\kappa^2 (\lambda )+ \pi^2 }
+\int_{|{\bf q}|}^{\infty}\! dq^0 
\left[
{(q^0)^4 \over \varphi_{\rm ap}(-q^2; \lambda )}
- {1 \over 
 [\hspace{0.1ex} \kappa^2 (\lambda )+ \pi^2 \hspace{0.1ex}] }
\right].
\label{J integrals} 
\eea
Noting that $\varphi_{\rm ap}(-q^2; \lambda )$ has four zeros in the
complex $q^0$ plane at $\pm p(|{\bf q}|)$, 
$\pm p^{\displaystyle \ast}\hspace{-0.1ex}(|{\bf q}|)$, where $p(s)$ 
(we make $s \!\equiv \! |{\bf q}|$) is the complex function with
\be
\left. \begin{array}{c}
         {\rm Re}\, p(s)  \\
         {\rm Im}\, p(s) 
       \end{array}   \right\} =
\left[{ \sqrt{\left[(\kappa^2 + \pi^2) \, s^2 
              + \kappa \right]^2+\pi^2} 
\pm (\kappa^2 + \pi^2) \, s^2 \pm \kappa
\over 2 \hspace{0.2ex}(\kappa^2 + \pi^2) }
\right]^{1/2},
\label{p(s)}
\ee
we can decompose
\be
{1 \over \varphi_{\rm ap}(-q^2; \lambda )} =
{1 \over 4 \hspace{0.2ex}(\kappa^2 + \pi^2) } \, 
{1 \over |p|^2 \, {\rm Re}\, p } 
\left[ {q^0 + 2 {\rm Re}\, p \over (q^0)^2 
+ 2 \hspace{0.2ex} {\rm Re}\: p \: q^0 + |p|^2 }- 
{(q^0 - 2 {\rm Re}\, p) \over (q^0)^2 
- 2 \hspace{0.2ex} {\rm Re}\, p \: q^0 + |p|^2 }
\right],
\ee
and then we can perform the integrals $J_a(s)$, $a \!=\! 1,2,3$. 
The results for these integrals can be found in 
Appendix \ref{app:J's}.

Next, to carry on with the calculation, we need to introduce some
suitable approximations for the functions $J_a(s)$ in the integrals 
(\ref{I integrals}). In order to do so, we study the behavior of these
functions for small and large values of $s$.
For $s \, J_1(s)$, we find that it can be well approximated by an
$\arctan$ function. In fact, on the one hand, 
$s \, J_1(s)$ tends very quickly
to a constant limiting value 
$\lim_{s \rightarrow \infty} s \, J_1(s)= a/4$, where
$a \equiv 1+ (2/ \pi) 
\arctan (\kappa / \pi)$. On the other hand, for small values of $s$,
we have $s \, J_1(s) \simeq \bigl[ \hspace{0.1ex}
\sqrt{120 \pi} \, a/(2 \pi b) \hspace{0.1ex} \bigr] \, s
+O(s^2)$, with $b \equiv  (4 a/ \pi^2) \!
\left[15 \pi \!
\left(\sqrt{\kappa^2 +\pi^2}- \kappa \right) \right]^{1/2}$.
Hence, we can approximate 
\be
s \, J_1(s) \simeq {a \over 2 \pi} \, 
\arctan \!\left( { \sqrt{120 \pi} \over b} \: s \right). 
\ee
Performing the integral $I_1({\bf x})$ [see Eq.~(\ref{I integrals})] with
this approximation, we get, for $|{\bf x}| \!\neq \! 0$,   
\be
I_1({\bf x}) \simeq {15 \over \pi^2} \, {a \over L_P^2} \,
{1 \over |{\bf x}|^2} \: e^{-b \hspace{0.2ex} |{\bf x}|/L_P}.
\label{I_1}
\ee
The function $J_2(s)$ behaves as 
$J_2(s) \simeq (a/4) \, s + O(s^{-1} \ln s)$ for large values of
$s$, and as  $J_2(s) \simeq 
(a/4 ) \, (120 \pi)^{-1/2} \, \gamma
+O(s^2)$, with 
$\gamma \equiv 240 \, (\kappa^2 +\pi^2)^{-1/2}
\, b^{-1} $, for small values of $s$.
This function can be well approximated by
\be
J_2(s) \simeq {a \over 4}  
\left[ s^2+ {\gamma^2 \over 120 \pi} \right]^{1/2}, 
\ee
and, substituting the last expression in the integral
$I_2({\bf x})$ [see (\ref{I integrals})], 
we obtain, for $|{\bf x}| \!\neq \! 0$, 
\be
I_2({\bf x}) \simeq {15 \over \pi^2} \, {a \over L_P^4} \,
{\gamma^2 \over |{\bf x}|^2} \: 
K_2 \bigl( \gamma \hspace{0.2ex} |{\bf x}|/L_P \bigr),
\label{I_2}
\ee
where $K_{\nu}(z)$ denote the modified Bessel functions of the second
kind. 
For $J_3(s)$, we find that 
$J_3(s) \simeq (a/4) \, s^3 + O(s \ln s)$ for large values of
$s$, and that $J_3(s) \simeq 
(a/4 ) \, (120 \pi)^{-3/2} \, \delta^3
+O(s)$, with 
$\delta \equiv 4 \, (\kappa^2 +\pi^2)^{-1/2}
\left[450 \pi \, b^{-1} \!
\left(2 \kappa - \sqrt{\kappa^2 +\pi^2} \right) \right]^{1/3}$,
for $s$ small. With the approximation
\be
J_3(s) \simeq {a \over 4}  
\left[ s^2+ {\delta^2 \over 120 \pi} \right]^{3/2}, 
\ee
we can compute the integral $I_3({\bf x})$ 
[see (\ref{I integrals})] for $|{\bf x}| \!\neq \! 0$,
and we find 
\be
I_3({\bf x}) \simeq 
{45 \over \pi^2} \, {a \over L_P^5} \,
{\delta^3 \over |{\bf x}|^3} \: 
K_3 \bigl( \delta \hspace{0.2ex} |{\bf x}|/L_P \bigr).
\label{I_3}
\ee
Numerical calculations confirm that the above approximations are 
reasonable. For $\kappa \sim 10$, we have $a, b, \delta \sim 1$ and
$\gamma \sim 10$.

The results (\ref{I_1}), (\ref{I_2}) and (\ref{I_3}) are now ready to
be substituted into (\ref{G_a}), from where we can compute the
different contributions to the correlation functions 
(\ref{G spacelike}). Using the relation
$(d/ dz) \bigl[ \hspace{0.2ex}
z^{- \nu} K_{\nu}(z) \hspace{0.2ex} \bigr] \!=\!
- z^{- \nu} K_{\nu+1}(z)$, 
and defining
$\sigma_b \!\equiv \! b \, |{\bf x}|/L_P$,
$\sigma_\gamma \!\equiv \! \gamma \, |{\bf x}|/L_P$,
$\sigma_\delta \!\equiv \! \delta \, |{\bf x}|/L_P$,
we get, after a rather long but straightforward calculation, the
following results for the non-zero components of 
${\cal G}_a^{\mu\nu\alpha\beta}({\bf x})$ 
[with $|{\bf x}| \!\neq \! 0$]:
\bea
&&{\cal G}_1^{0000}({\bf x}) \simeq
{2 \over 3 \pi} \, {a \hspace{0.2ex} b^6 \over L_P^4} \, 
{e^{-\sb} \over \sb^2}   
\left[1+{4 \over \sb}+ {12 \over \sb^2}+
{24 \over \sb^3}+{24 \over \sb^4} \right],
\nn   \\
&&{\cal G}_1^{00ij}({\bf x}) \simeq
{1 \over 3 \pi} \, {a \hspace{0.2ex} b^6 \over L_P^4} \, 
{e^{-\sb}  \over \sb^2} 
\left[ \delta^{ij} 
\left( 1+{5 \over \sb}+ {16 \over \sb^2}+
{32 \over \sb^3}+{32 \over \sb^4}
\right) \!
- {x^i x^j \over |{\bf x}|^2 } 
\left( 1+{7 \over \sb}+ {24 \over \sb^2}+
{48 \over \sb^3}+{48 \over \sb^4}
\right) \right],
\nn   \\
&&{\cal G}_1^{0i0j}({\bf x})=- {3 \over 2} \: 
{\cal G}_1^{00ij}({\bf x}),
\nn   \\
&&{\cal G}_1^{ijkl}({\bf x}) \simeq
{1 \over 3 \pi} \, {a \hspace{0.2ex} b^6 \over L_P^4} \, 
{e^{-\sb}  \over \sb^2} 
\left[ - \left(\delta^{ij} \delta^{kl}
\!-\! 3 \hspace{0.2ex}\delta^{i(k} \delta^{l)j} 
\right) \! \left( 1+{6 \over \sb}+ {18 \over \sb^2}+
{30 \over \sb^3}+{24 \over \sb^4} \right) \!
\right.
\nn   \\
&& \hspace{17.2ex}
+\,10 \hspace{0.2ex} \delta^{i(k} \delta^{l)j} \!
\left( {1 \over \sb^2}+{5 \over \sb^3}+{8 \over \sb^4} \right)\!
\nn   \\
&& \hspace{17.2ex}
+\, {1 \over |{\bf x}|^2 } 
\left( \delta^{ij} x^k x^l \!+\! \delta^{kl} x^i x^j
\!-\! 3 \hspace{0.2ex} \delta^{i(k} x^{l)} x^j 
\!-\! 3 \hspace{0.2ex} \delta^{j(k} x^{l)} x^i\right) \!
\left( 1+{5 \over \sb}+ {6 \over \sb^2}-
{18 \over \sb^3}-{48 \over \sb^4} \right)\!
\nn   \\
&&  \hspace{17.2ex}
-\,{10 \over |{\bf x}|^2 } \left( \delta^{i(k} x^{l)} x^j
+ \delta^{j(k} x^{l)} x^i\right) \!
\left( {1 \over \sb}+ {9 \over \sb^2}+
{33 \over \sb^3}+{48 \over \sb^4} \right)\!
\nn   \\
&& \left. \hspace{17.2ex}
+\,{2 \over |{\bf x}|^4 } \, x^i x^j x^k x^l
\left( 1+{14 \over \sb}+ {87 \over \sb^2}+
{279 \over \sb^3}+{384 \over \sb^4} \right)
\right],
\nn   \\
&&{\cal G}_2^{00ij}({\bf x}) \simeq
{1 \over 3 \pi} \, {a \hspace{0.2ex} \gamma^6 \over L_P^4} \, 
{ K_4(\sg)  \over \sg^2} 
\left( 3 \, {x^i x^j \over |{\bf x}|^2 }- \delta^{ij} \right), 
\nn   \\
&&{\cal G}_2^{0i0j}({\bf x}) \simeq
{1 \over 6 \pi} \, {a \hspace{0.2ex} \gamma^6 \over L_P^4} \, 
{ K_4(\sg)  \over \sg^2} 
\left( {x^i x^j \over |{\bf x}|^2 }
+ 3 \hspace{0.2ex} \delta^{ij} \right) \!
-{5 \over 3 \pi} \, {a \hspace{0.2ex} \gamma^6 \over L_P^4} \, 
{ K_3(\sg)  \over \sg^3}  \, \delta^{ij} ,
\nn   \\
&&{\cal G}_2^{ijkl}({\bf x}) \simeq
{1 \over 3 \pi} \, {a \hspace{0.2ex} \gamma^6 \over L_P^4} \, 
{ K_4(\sg)  \over \sg^2} 
\left[ 2 \left( \delta^{ij} \delta^{kl}
\!-\! 3 \hspace{0.2ex}\delta^{i(k} \delta^{l)j} 
\right) \! 
-{1 \over |{\bf x}|^2 } 
\left( \delta^{ij} x^k x^l 
\!+\! \delta^{kl} x^i x^j
\!-\! 3 \hspace{0.2ex} \delta^{i(k} x^{l)} x^j 
\right. \right.
\nn   \\
&& \hspace{39.9ex} \left.
- 3 \hspace{0.2ex} \delta^{j(k} x^{l)} x^i\right) 
\Biggr]
-{4 \over 3 \pi} \, {a \hspace{0.2ex} \gamma^6 \over L_P^4} \, 
{ K_3(\sg)  \over \sg^3} 
\left(\delta^{ij} \delta^{kl}
\!-\! 3 \hspace{0.2ex}\delta^{i(k} \delta^{l)j} 
\right),
\nn   \\
&&{\cal G}_3^{ijkl}({\bf x}) \simeq
-{1 \over \pi} \, {a \hspace{0.2ex} \delta^6 \over L_P^4} \, 
{ K_3(\sd)  \over \sd^3} 
\left(\delta^{ij} \delta^{kl}
\!-\! 3 \hspace{0.2ex}\delta^{i(k} \delta^{l)j} 
\right).
\label{results for G_a}
\eea
Note that, for $\sigma \!\gg \! 1$, we have the following asymptotic
expansions for the modified Bessel functions in the above expressions:
\be
K_4(\sigma) \simeq \sqrt{ { \pi \over 2 \sigma} } \:
e^{-\sigma} \left[ 1 + {63 \over 8} \, {1 \over \sigma}+
O \!\left({1 \over \sigma^2} \right)
\right],
\hspace{9ex}
K_3(\sigma) \simeq \sqrt{ { \pi \over 2 \sigma} } \:
e^{-\sigma} \left[ 1 + {35 \over 8} \, {1 \over \sigma}+
O \!\left({1 \over \sigma^2} \right)
\right].
\ee  


\subsection{Correlation functions for the metric
perturbations}
\label{subsec:metric perturbations}


Starting from the solutions found for the linearized Einstein tensor, 
which are characterized by the two-point correlation functions
(\ref{corr funct}) [or, in terms of Fourier transforms, 
(\ref{Fourier tr corr funct})], we can now solve the equations for the
metric perturbations. Working in the harmonic gauge, 
$\partial_{\nu} \bar{h}^{\mu\nu} \!=\! 0$ (this zero must be
understood in the same statistical sense as above), where
$\bar{h}_{\mu\nu} \!\equiv \! h_{\mu\nu} 
\!-\! (1/2)\hspace{0.2ex} \eta_{\mu\nu} \hspace{0.2ex}h$, 
and using Eqs.~(\ref{D, B tensors}) and (\ref{G tensor}), these
equations reduce to 
$\Box \bar{h}^{\mu\nu}(x) \!=\! -2 
G^{{\scriptscriptstyle (1)}\hspace{0.1ex} \mu\nu}(x)$, or, in terms of
Fourier transforms, 
$p^2 
\tilde{\bar{h}}^{\mbox{}_{\mbox{}_{\mbox{}_{\mbox{}_{\mbox{}
_{\scriptstyle \mu\nu}}}}}}\hspace{-0.5ex} (p) 
\!=\! 2 \tilde{G}^{{\scriptscriptstyle (1)}\hspace{0.1ex} \mu\nu}(p)$.
As above, we can write 
$\bar{h}^{\mu\nu} \!=\! \langle \bar{h}^{\mu\nu} \rangle_c
\!+\! \bar{h}^{\mu\nu}_{\rm f}$, where $\bar{h}^{\mu\nu}_{\rm f}$ is a
solution to these equations with zero average, and the two-point
correlation functions are given by
\be
{\cal H}^{\mu\nu\alpha\beta}(x,x^{\prime}) \equiv 
\langle \bar{h}^{\mu\nu}(x) \bar{h}^{\alpha\beta}(x^{\prime})
\rangle_c - \langle \bar{h}^{\mu\nu}(x) \rangle_c 
\langle \bar{h}^{\alpha\beta}(x^{\prime}) \rangle_c =
\langle \bar{h}^{\mu\nu}_{\rm f}(x) 
\bar{h}^{\alpha\beta}_{\rm f}(x^{\prime}) \rangle_c.
\label{co fu}
\ee
We can now seek solutions of the form 
$\tilde{\bar{h}}^{\mbox{}_{\mbox{}_{\mbox{}_{\mbox{}_{\mbox{}
_{\scriptstyle \mu\nu}}}}}}_{\rm f}\hspace{-0.5ex} (p) 
\!=\! 2 D(p) 
\tilde{G}^{{\scriptscriptstyle (1)}\hspace{0.1ex} \mu\nu}_{\rm f}(p)$,
where $D(p)$ is a Lorentz invariant scalar distribution in Minkowski
spacetime, which is the most general solution of $p^2 D(p) \!=\! 1$.
Note that, since the linearized Einstein tensor is conserved, 
solutions of this form automatically satisfy the harmonic
gauge condition. As above, we can write $D(p) \!=\! [ 1/p^2 ]_r 
\!+\! D_h(p)$, where $D_h(p)$ is the most general solution to the
associated homogeneous equation and, correspondingly, we have
$\tilde{\bar{h}}^{\mbox{}_{\mbox{}_{\mbox{}_{\mbox{}_{\mbox{}
_{\scriptstyle \mu\nu}}}}}}_{\rm f}\hspace{-0.5ex} (p) 
\!=\! 
\tilde{\bar{h}}^{\mbox{}_{\mbox{}_{\mbox{}_{\mbox{}_{\mbox{}
_{\scriptstyle \mu\nu}}}}}}_p\hspace{-0.5ex} (p) +
 \tilde{\bar{h}}^{\mbox{}_{\mbox{}_{\mbox{}_{\mbox{}_{\mbox{}
_{\scriptstyle \mu\nu}}}}}}_h \hspace{-0.5ex} (p)$.
However, since $D_h(p)$ has support on the set of points for which
$p^2 \!=\! 0$, it is easy to see from 
Eq.~(\ref{Fourier tr corr funct})
[from the factor $\theta (-p^2-4 m^2)$] that
$\langle \tilde{\bar{h}}^{\mbox{}_{\mbox{}_{\mbox{}_{\mbox{}_{\mbox{}
_{\scriptstyle \mu\nu}}}}}}_h \hspace{-0.5ex} (p)
\tilde{G}^{{\scriptscriptstyle (1)}\hspace{0.1ex} \alpha\beta}
_{\rm f}(p^{\prime})
\rangle_c \!=\! 0$ and, thus, 
the two-point correlation functions (\ref{co fu})
can be computed from
$\langle 
\tilde{\bar{h}}^{\mbox{}_{\mbox{}_{\mbox{}_{\mbox{}_{\mbox{}
_{\scriptstyle \mu\nu}}}}}}_{\rm f}\hspace{-0.5ex} (p)
\tilde{\bar{h}}^{\mbox{}_{\mbox{}_{\mbox{}_{\mbox{}_{\mbox{}
_{\scriptstyle \alpha\beta}}}}}}_{\rm f}\hspace{-0.5ex}(p^{\prime})
\rangle_c \!=\! 
\langle 
\tilde{\bar{h}}^{\mbox{}_{\mbox{}_{\mbox{}_{\mbox{}_{\mbox{}
_{\scriptstyle \mu\nu}}}}}}_p\hspace{-0.5ex} (p)
\tilde{\bar{h}}^{\mbox{}_{\mbox{}_{\mbox{}_{\mbox{}_{\mbox{}
_{\scriptstyle \alpha\beta}}}}}}_p\hspace{-0.5ex}(p^{\prime})
\rangle_c$. {}From Eq.~(\ref{Fourier tr corr funct})
and due to the factor $\theta (-p^2-4 m^2)$, it is also easy to see
that the prescription $[ \;\; ]_r$ is irrelevant in this correlation
function and we obtain
\be
\langle 
\tilde{\bar{h}}^{\mbox{}_{\mbox{}_{\mbox{}_{\mbox{}_{\mbox{}
_{\scriptstyle \mu\nu}}}}}}_p\hspace{-0.5ex} (p)
\tilde{\bar{h}}^{\mbox{}_{\mbox{}_{\mbox{}_{\mbox{}_{\mbox{}
_{\scriptstyle \alpha\beta}}}}}}_p\hspace{-0.5ex}(p^{\prime})
\rangle_c = {4 \over (p^2)^2} \,
\langle 
\tilde{G}_p^{{\scriptscriptstyle (1)}\hspace{0.1ex} \mu\nu}(p) \,
\tilde{G}_p^{{\scriptscriptstyle (1)}\hspace{0.1ex} \alpha\beta}
(p^\prime) \rangle_c,   
\ee
where $\langle 
\tilde{G}_p^{{\scriptscriptstyle (1)}\hspace{0.1ex} \mu\nu}(p) \,
\tilde{G}_p^{{\scriptscriptstyle (1)}\hspace{0.1ex} \alpha\beta}
(p^\prime) \rangle_c$ is given in (\ref{Fourier tr corr funct}).
The right hand side of this equation is a well defined
bi-distribution, at least for $m \!\neq \! 0$ (the $\theta$ function
provides the suitable cutoff).
In the massless field case, since the noise kernel 
is obtained as the limit $m \!\rightarrow \!0$ of the noise kernel
for a massive field, it seems that the natural prescription to avoid
the divergencies on the lightcone $p^2 \!=\! 0$ is a Hadamard finite
part (see Refs.~\cite{schwartz} for its definition).
Taking this prescription, we also get a well defined bi-distribution
for the massless limit of the last expression.
Finally, we find the result
\be
{\cal H}^{\mu\nu\alpha\beta}(x,x^{\prime})=
{4 \pi \over 45} \, G^2 \,{\cal F}^{\mu\nu\alpha\beta}_{x} \,
{\cal H}_{\rm A} (x-x^{\prime})+
{32 \pi \over 9} \, G^2 \,
{\cal F}^{\mu\nu}_{x} {\cal F}^{\alpha\beta}_{x} \,
{\cal H}_{\rm B} (x-x^{\prime}),
\label{corr funct 2}
\ee
where $\tilde{{\cal H}}_{\rm A}(p) \!\equiv \! 
[1/(p^2)^2]\, \tilde{{\cal G}}_{\rm A}(p)$ and 
$\tilde{{\cal H}}_{\rm B}(p) \!\equiv \! 
[1/(p^2)^2]\, \tilde{{\cal G}}_{\rm B}(p)$, with 
$\tilde{{\cal G}}_{\rm A}(p)$ and  
$\tilde{{\cal G}}_{\rm B}(p)$ given by (\ref{distri}).
The two-point correlation functions for the metric perturbations 
can be easily obtained using $h_{\mu\nu} \!=\! 
\bar{h}_{\mu\nu} 
\!-\! (1/2) \hspace{0.2ex}\eta_{\mu\nu} 
\hspace{0.2ex}\bar{h}^{\alpha}_{\alpha}$.



\section{Discussion}
\label{sec:discuss}


Our main results for the correlation functions are (\ref{corr funct})
and (\ref{corr funct 2}). In the case of a conformal field, the 
correlation functions of the linearized Einstein
tensor have been explicitly evaluated and the results are given
in (\ref{results for G_a}).
{}From the exponential factors $e^{-\sigma}$ in these results,
we see that the correlation functions of the linearized Einstein
tensor are in this case characterized by correlation lengths of the
order of the Planck length. A similar behavior is expected for the
correlation functions of the metric perturbations.
Hence, as expected in this case, the 
correlation functions are negligibly small for points separated by
distances large compared to the Planck length.
At such scales, the dynamics of gravitational perturbations of
Minkowski spacetime can be simply described by semiclassical gravity
\cite{horowitz,horowitz-wald78,horowitz81,hartle_horowitz,simon,%
jordan,flanagan}.
Deviations from semiclassical gravity are only important for points
separated by Planckian or sub-Planckian scales. 
However, for such scales, our results (\ref{results for G_a}) are not
reliable, since we expect that gravitational fluctuations of genuine
quantum nature to be relevant and, thus, the classical description
breaks down. 
It is interesting to note, however, that these results for correlation
functions are non-analytic in their characteristic correlation
lengths. This kind of non-analytic behavior is actually quite typical
of the solutions of Langevin-type equations with dissipative
terms. An example in the context of a reduced version of the
semiclassical Einstein-Langevin equation is given in
Ref.~\cite{ccv97}).

For background solutions of semiclassical gravity with other scales
present apart from the Planck scales (for instance, for matter fields
in a thermal state), stress-energy fluctuations may be important at
larger scales. For such backgrounds, stochastic
semiclassical gravity might predict correlation functions with
characteristic correlation lengths much larger than the Planck
scales, so as to be relevant and reliable on a certain range of
scales. It seems quite plausible, nevertheless, that these correlation
functions would remain non-analytic in their characteristic
correlation lengths. 
This would imply that these correlation functions could not be
obtained from a calculation involving a perturbative expansion in the
characteristic correlation lengths. In particular, if these  
correlation lengths are proportional to the Planck constant 
$\hbar$, the 
gravitational correlation functions could not be obtained from an
expansion in $\hbar$. Hence, stochastic semiclassical
gravity might predict a behavior for gravitational
correlation functions different from that of the analogous functions
in perturbative quantum gravity \cite{donoghue}. 
This is not necessarily inconsistent with having neglected action
terms of higher order in $\hbar$ when considering semiclassical
gravity as an effective theory \cite{flanagan}.

We conclude this section with some comments about a technical point  
on the obtained solutions of stochastic semiclassical
gravity. It concerns the issue that the Einstein-Langevin
equations, as well as the semiclassical Einstein equations, contain
derivatives of order higher than two. Due to this fact, these
equations can have some ``pathological'' solutions 
({\it e.g.}, ``runaway'' solutions) which are presumably unphysical
\cite{simon90,simon,parker-simon,flanagan}. Thus, one needs to apply
some criterion to discern the ``physical''  
from the unphysical solutions. However, as it is discussed in
Ref.~\cite{flanagan} (see also Refs.~\cite{amendola}), 
even in the context of ``pure'' (non-stochastic)
semiclassical gravity, this is still an open problem. Two main
proposals, both based in the works by Simon
\cite{simon90,simon,parker-simon}, have been made concerning this
issue: the ``perturbative expandability'' (in $\hbar$) criterion
\cite{simon90,simon,parker-simon} and the ``reduction of order''
procedure \cite{flanagan}.

The first proposal consists in identifying a subclass
of ``physical'' solutions which are
analytic in the Planck constant $\hbar$. This proposal has been
successful in eliminating the instability of Minkowski spacetime found
by Horowitz~\cite{horowitz,horowitz81}. 
However, on the one hand, 
this proposal seems to be too restrictive
since, as it has been pointed out in Ref.~\cite{flanagan}, 
one could not describe effects such as the continuous mass loss of a
black hole due to Hawking radiation. On the other hand, there can be
situations in which the formal series obtained when seeking
approximate perturbative solutions (to a finite order in $\hbar$) does
not converge to a solution to the semiclassical equations 
\cite{flanagan}. In our case, if we had tried to find 
solutions to Eq.~(\ref{unified Einstein-Langevin eq}) as a 
Taylor expansion in $\hbar$, we would have obtained a series for 
$\tilde{G}^{\scriptscriptstyle (1)}_{\mu\nu}(p)$
which, as the above solutions, would be linear and local in 
$\tilde{\xi}_{\alpha\beta}(p)$, but whose corresponding 
two-point correlation functions for the conformal
field case would not converge to (\ref{corr funct}).

The ``reduction of order'' procedure provides in some cases a
reasonable way to modify the semiclassical equations in order to
eliminate spurious solutions. 
But, as it has been emphasized in Ref.~\cite{flanagan}, it is
not clear at all whether a reduction of order procedure can always be
applied to the semiclassical Einstein equation (and how this procedure
should be applied). For the Einstein-Langevin equation, this
issue has not been, to our knowledge, properly addressed. A naive
application of the prescription to 
Eq.~(\ref{unified Einstein-Langevin eq})
seems to downplay the role
of the dissipative terms with respect to the
noise source. In fact, to lowest order, we obtain
$G^{{\scriptscriptstyle (1)}\hspace{0.1ex} \mu\nu} \!=\! 
16 \pi G \xi^{\mu\nu}$, where there is no contribution of the
dissipation kernel. 
From this equation, we get the well-known result
$\langle G^{{\scriptscriptstyle (1)}\hspace{0.1ex} \mu\nu} \rangle_c
\!=\! 0$ \cite{flanagan,simon},
and also
${\cal G}^{\mu\nu\alpha\beta}(x,x^{\prime}) \!=\!
(16 \pi)^2 L_P^4 N^{\mu\nu\alpha\beta}(x,x^{\prime})$.
For a massless field, using 
Eqs.~(\ref{noise and dissipation kernels 2}),
(\ref{massless N and D kernels}) and (\ref{N and D 2}), this gives
${\cal G}^{\mu\nu\alpha\beta}(x,x^{\prime}) \!=\!
(2/15) (L_P^4/ \pi^2) \, \Bigl[ (1/6) {\cal F}^{\mu\nu\alpha\beta}_{x}
+ 60 \hspace{0.2ex}\Delta \xi^2 
{\cal F}^{\mu\nu}_{x} {\cal F}^{\alpha\beta}_{x}
\Bigr] \Bigl[ {\cal P}\hspace{-0.4ex}f 
\Bigl[1/\bigl( (x-x^{\prime})^2 \bigr)^2 \Bigr] + 
\pi^3 \delta^4(x-x^{\prime})
\Bigr]$. For the two-point correlation functions (\ref{co fu}),
we get, in the harmonic gauge, 
${\cal H}^{\mu\nu\alpha\beta}(x,x^{\prime}) =
(4 \pi/45) L_P^4 \,{\cal F}^{\mu\nu\alpha\beta}_{x} \,
{\cal I}_{\rm A} (x-x^{\prime})+
(32 \pi/9) L_P^4 \, 
{\cal F}^{\mu\nu}_{x} {\cal F}^{\alpha\beta}_{x} \,
{\cal I}_{\rm B} (x-x^{\prime})$,
with
$ \tilde{{\cal I}}_{\rm A}(p) \!\equiv \! 
\theta (-p^2\!-\!4m^2) \, (p^2)^{-2} \, 
\sqrt{1\!+\!4 m^2/ p^2} \, \bigl(1\!+\!4 m^2/ p^2\bigr)^2$ and
$ \tilde{{\cal I}}_{\rm B}(p) \!\equiv \! 
\theta (-p^2\!-\!4m^2) \, (p^2)^{-2} \, 
\sqrt{1\!+\!4 m^2/ p^2} \, 
\bigl(3 \hspace{0.2ex}\Delta \xi\!+\! m^2/ p^2\bigr)^2$. 
Comparing the last results for the massless case 
with the ones obtained in 
Sec.~\ref{sec:correlation functions}, we note that the main
qualitative feature is the absence of the exponential factors
$e^{-\sigma}$, which make the
two-point correlation functions to decay much more slowly
with the distance, {\it i.e.}, like a power instead of an exponential
law. This fact is due to the lack of dissipative terms in
the reduced order equations. 
The conclusion is that one should probably implement a more
sophisticated 
version of the reduction of order procedure so as to keep
some contribution of the dissipation kernel in the reduced order
equations.

For these reasons, in our work we have not attempted any of these
procedures and we have simply sought some solutions to
the full equations (\ref{unified Einstein-Langevin eq}). 
Our solutions for the conformal field case
have the physically reasonable
feature of having negligible two-point functions for points separated
by scales larger than the Planck length.


\acknowledgments

We are grateful to Esteban Calzetta, Jaume Garriga,
Bei-Lok Hu, Ted Jacobson and Albert Roura
for very helpful suggestions and discussions. 
This work has been partially supported by the 
CICYT Research Project number
\mbox{AEN98-0431}, and the European Project number
\mbox{CI1-CT94-0004}.



\appendix



\section{The kernels for a vacuum 
state}
\label{app:kernels in vac}


The kernels for a vacuum state can be computed 
in terms of the Wightman and Feynman functions defined in 
Eq.~(\ref{Wightman and Feynman functions})
using 
\bea
&&\langle 0| \,  \hat{t}_n^{ab}(x)\, \hat{t}_n^{cd}(y)\, |0 \rangle
= 4 \left( N_n^{abcd}(x,y)+ i 
H_{\scriptscriptstyle \!{\rm A}_{\scriptstyle n}}^{abcd}(x,y) \right)
=\bigtriangledown^{a}_{\!\!\! \mbox{}_{x}}\!
 \bigtriangledown^{c}_{\!\!\! \mbox{}_{y}}\! G_n^+(x,y) 
 \bigtriangledown^{b}_{\!\!\! \mbox{}_{x}}\!
 \bigtriangledown^{d}_{\!\!\! \mbox{}_{y}} G_n^+(x,y)
\nn  \\
&&\hspace{25ex}+\bigtriangledown^{a}_{\!\!\! \mbox{}_{x}}\!
 \bigtriangledown^{d}_{\!\!\! \mbox{}_{y}}\! G_n^+(x,y) 
 \bigtriangledown^{b}_{\!\!\! \mbox{}_{x}}\!
 \bigtriangledown^{c}_{\!\!\! \mbox{}_{y}} G_n^+(x,y) 
+2\, {\cal D}^{ab}_{\!\! \scriptscriptstyle x} \! \left(
  \bigtriangledown^{c}_{\!\!\! \mbox{}_{y}} G_n^+(x,y)
  \bigtriangledown^{d}_{\!\!\! \mbox{}_{y}}\! G_n^+(x,y)  \right)
\nn   \\
&&\hspace{25ex}+\,
2\, {\cal D}^{cd}_{\!\! \scriptscriptstyle y} \! \left(
  \bigtriangledown^{a}_{\!\!\! \mbox{}_{x}} G_n^+(x,y)
  \bigtriangledown^{b}_{\!\!\! \mbox{}_{x}}\! G_n^+(x,y)  \right)
+2\, {\cal D}^{ab}_{\!\! \scriptscriptstyle x} 
   {\cal D}^{cd}_{\!\! \scriptscriptstyle y} \! \left(
 G_n^{+ 2}(x,y)  \right)\!,  
\label{Wightman expression 2}
\eea
where 
${\cal D}^{ab}$ is the differential operator
\be
{\cal D}^{ab}_{\!\! \scriptscriptstyle x} 
\equiv \left(\xi-{1\over 4}\right) g^{ab}(x) 
\Box_{\! \scriptscriptstyle x}+ \xi
\left( R^{ab}(x)- \bigtriangledown^{a}_{\!\! \scriptscriptstyle x}
\bigtriangledown^{b}_{\!\! \scriptscriptstyle x} \right),
\ee
and
\bea
&&H_{\scriptscriptstyle \!{\rm S}_{\scriptstyle n}}^{abcd}(x,y)=
- {1 \over 4} \, {\rm Im} \Biggl[ 
 \bigtriangledown^{a}_{\!\!\! \mbox{}_{x}}\!
 \bigtriangledown^{c}_{\!\!\! \mbox{}_{y}}\!
     G\!_{\scriptscriptstyle F_{\scriptstyle \hspace{0.1ex}  n}}
     \hspace{-0.2ex}(x,y)
 \bigtriangledown^{b}_{\!\!\! \mbox{}_{x}}\!
 \bigtriangledown^{d}_{\!\!\! \mbox{}_{y}}
     G\!_{\scriptscriptstyle F_{\scriptstyle \hspace{0.1ex}  n}}
     \hspace{-0.2ex}(x,y)
+\bigtriangledown^{a}_{\!\!\! \mbox{}_{x}}\!
 \bigtriangledown^{d}_{\!\!\! \mbox{}_{y}}\! 
     G\!_{\scriptscriptstyle F_{\scriptstyle \hspace{0.1ex}  n}}
     \hspace{-0.2ex}(x,y)
 \bigtriangledown^{b}_{\!\!\! \mbox{}_{x}}\!
 \bigtriangledown^{c}_{\!\!\! \mbox{}_{y}}
     G\!_{\scriptscriptstyle F_{\scriptstyle \hspace{0.1ex}  n}}
     \hspace{-0.2ex}(x,y)   \nn \\
&&
+\,2\, {\cal D}^{ab}_{\!\! \scriptscriptstyle x} \! \left(
  \bigtriangledown^{c}_{\!\!\! \mbox{}_{y}}
      G\!_{\scriptscriptstyle F_{\scriptstyle \hspace{0.1ex}  n}}
      \hspace{-0.2ex}(x,y)
  \bigtriangledown^{d}_{\!\!\! \mbox{}_{y}}\!
      G\!_{\scriptscriptstyle F_{\scriptstyle \hspace{0.1ex}  n}}
      \hspace{-0.2ex}(x,y)  \right)
\hspace{-0.2ex}
+2\, {\cal D}^{cd}_{\!\! \scriptscriptstyle y} \! \left(
  \bigtriangledown^{a}_{\!\!\! \mbox{}_{x}} 
      G\!_{\scriptscriptstyle F_{\scriptstyle \hspace{0.1ex}  n}}
      \hspace{-0.2ex}(x,y)  
\bigtriangledown^{b}_{\!\!\! \mbox{}_{x}}\!
      G\!_{\scriptscriptstyle F_{\scriptstyle \hspace{0.1ex}  n}}
      \hspace{-0.2ex}(x,y)  \right) \nn  \\
&&
+2\, {\cal D}^{ab}_{\!\! \scriptscriptstyle x} 
   {\cal D}^{cd}_{\!\! \scriptscriptstyle y} \! \left(
      G\!_{\scriptscriptstyle F_{\scriptstyle \hspace{0.1ex}  n}}
      ^{\;\: 2} \hspace{-0.2ex}(x,y)  \right) 
+\,{1 \over 2}\left[ g^{ab}(x)\hspace{-0.2ex} \left(
 \bigtriangledown^{c}_{\!\!\! \mbox{}_{y}}
      G\!_{\scriptscriptstyle F_{\scriptstyle \hspace{0.1ex}  n}}
      \hspace{-0.2ex}(x,y) \!
 \bigtriangledown^{d}_{\!\!\! \mbox{}_{y}}\!
+\bigtriangledown^{d}_{\!\!\! \mbox{}_{y}}\!
      G\!_{\scriptscriptstyle F_{\scriptstyle \hspace{0.1ex}  n}}
      \hspace{-0.2ex}(x,y)
 \bigtriangledown^{c}_{\!\!\! \mbox{}_{y}}  \right) \right.
    \nn  \\
&& \left.
+\,g^{cd}(y)\hspace{-0.2ex} \left(
 \bigtriangledown^{a}_{\!\!\! \mbox{}_{x}}
      G\!_{\scriptscriptstyle F_{\scriptstyle \hspace{0.1ex}  n}}
      \hspace{-0.2ex}(x,y) 
 \bigtriangledown^{b}_{\!\!\! \mbox{}_{x}}\! 
+\bigtriangledown^{b}_{\!\!\! \mbox{}_{x}}\!
      G\!_{\scriptscriptstyle F_{\scriptstyle \hspace{0.1ex}  n}}
      \hspace{-0.2ex}(x,y)
 \bigtriangledown^{a}_{\!\!\! \mbox{}_{x}}  \right)
   \right]  {\delta^n(x\!-\!y) \over \sqrt{- g(x)}}  
+ \left( g^{ab}(x){\cal D}^{cd}_{\!\! \scriptscriptstyle y}
\right.  \nn  \\
&& \left.
+\,g^{cd}(y){\cal D}^{ab}_{\!\! \scriptscriptstyle x} \right) \!\!
    \left( {\delta^n(x\!-\!y) \over \sqrt{- g(x)}} \,
       G\!_{\scriptscriptstyle F_{\scriptstyle \hspace{0.1ex}  n}}
      \hspace{-0.2ex}(x,y)   \right) \!
+{1 \over 4}\, g^{ab}(x) g^{cd}(y) 
      G\!_{\scriptscriptstyle F_{\scriptstyle \hspace{0.1ex}  n}}
      \hspace{-0.2ex}(x,y)
\left( \Box_{\! \scriptscriptstyle x} -m^2- \xi R(x) \right)
{\delta^n(x\!-\!y) \over \sqrt{- g(x)}}\,
\Biggr].  \nn \\
\mbox{}
\label{Feynman expression 3}
\eea


\section{Momentum integrals}
\label{sec:integrals in dimensional}


Some useful expressions for the momentum integrals in dimensional
regularization defined in (\ref{integrals in n dim}) and 
(\ref{constant integrals in n dim})
are:
\bea
&&I_{0_{\scriptstyle n}} =
{i \over (4\pi)^2 }\, m^2 
\left({ m^2 \over 4 \pi \mu^2} \right)
^{\!_{\scriptstyle n-4 \over 2}} 
\! \Gamma \!\left(1 \!-\! {n \over 2}\right)
={i \over (4\pi)^2 }\, {4 m^2 \over (n\!-\!2)} \,
\kappa_n+O (n \!-\! 4), 
\label{integral in dim regul 1}
    \\
&&I_{0_{\scriptstyle n}}^{\mu} = 0,
   \\
&&I_{0_{\scriptstyle n}}^{\mu \nu} 
= - m^2 \, \eta^{\mu\nu} \, {I_{0_{\scriptstyle n}} \over n} , 
    \\
&&J_n(p) 
={-i \over (4\pi)^2 } \left[2 \kappa_n+\phi (p^2) \right.
+O (n\!-\!4) \Bigr],
\label{integral in dim regul 4}  \\
&&J_n^{\mu }(p) 
= {J_n(p) \over 2} \, p^\mu, 
  \\
&&J_n^{\mu \nu}(p) 
={J_n(p) \over 4} \left[ p^\mu p^\nu 
- \biggl(1+4 \,{m^2 \over p^2} \biggr) \, 
{p^2 P^{\mu\nu} \over (n\!-\!1) } \right]
+ {I_{0_{\scriptstyle n}} \over 2} \, {1 \over p^2}
\left[ p^\mu p^\nu + {p^2 P^{\mu\nu} \over n\!-\!1 }
\right]\!,
   \\
&&J_n^{\mu \nu \alpha}(p) 
={J_n(p) \over 8} \left[ p^\mu p^\nu p^\alpha
- \biggl(1+4 \,{m^2 \over p^2} \biggr) \, 
{p^2 \over (n\!-\!1) } \, \bigl( P^{\mu\nu} p^\alpha 
+ P^{\mu\alpha} p^\nu +P^{\alpha\nu} p^\mu
\bigr) \right]
\nn  \\
&& \hspace{10.7ex}
+\, {I_{0_{\scriptstyle n}} \over 4} \, {1 \over p^2}
\left[ 3 p^\mu p^\nu p^\alpha+ {p^2 \over (n\!-\!1)} \,
\bigl( P^{\mu\nu} p^\alpha 
+ P^{\mu\alpha} p^\nu +P^{\alpha\nu} p^\mu \bigr) 
\right] \!,
   \\
&&J_n^{\mu \nu \alpha \beta}(p) 
={J_n(p) \over 16} \left[ p^\mu p^\nu p^\alpha p^\beta
- \biggl(1+4 \,{m^2 \over p^2} \biggr) \, 
{p^2 \over (n\!-\!1) } \, \bigl( P^{\mu\nu} p^\alpha p^\beta
+ P^{\nu\alpha} p^\mu p^\beta+P^{\nu\beta} p^\mu p^\alpha
\right.
\nn  \\
&& \hspace{19ex} 
+\, P^{\mu\alpha} p^\nu p^\beta+P^{\mu\beta} p^\nu p^\alpha
+P^{\alpha\beta} p^\mu p^\nu \bigr) 
+\biggl(1+4 \,{m^2 \over p^2} \biggr)^{\! 2} \hspace{0.1ex}  
{(p^2)^2 \over (n^2\!-\!1) } \, \bigl( P^{\mu\nu} P^{\alpha\beta} 
\nn  \\
&& \hspace{19ex} 
+\, P^{\mu\alpha} P^{\nu\beta} + P^{\mu\beta} P^{\nu\alpha} \bigr)
\Biggr]
\nn  \\
&& \hspace{11.7ex}
+\, {I_{0_{\scriptstyle n}} \over 8} \, {1 \over p^2}
\left[ \biggl(7-{12 \over n}\,{m^2 \over p^2} \biggr) \,
p^\mu p^\nu p^\alpha p^\beta 
+\biggl({1 \over n\!-\!1}-{4 \over n}\,{m^2 \over p^2} \biggr) \,
p^2\, \bigl( P^{\mu\nu} p^\alpha p^\beta
+ P^{\nu\alpha} p^\mu p^\beta
\right.
\nn  \\
&& \hspace{21.7ex} 
+\, P^{\nu\beta} p^\mu p^\alpha
+P^{\mu\alpha} p^\nu p^\beta+P^{\mu\beta} p^\nu p^\alpha
+P^{\alpha\beta} p^\mu p^\nu \bigr) 
\nn  \\
&& \hspace{21.7ex} \left.
-\, \biggl({1 \over n^2\!-\!1}
-{4 \hspace{0.3ex} (2n\!-\!1) \over n \hspace{0.3ex}(n^2\!-\!1)}
\,{m^2 \over p^2} \biggr) \hspace{0.2ex} (p^2)^2 \hspace{0.2ex}
 \bigl( P^{\mu\nu} P^{\alpha\beta} +
P^{\mu\alpha} P^{\nu\beta} + P^{\mu\beta} P^{\nu\alpha} \bigr)
\right]\!, 
\nn  \\
&&\mbox{}
\eea
where
$p^2 P^{\mu\nu} \equiv \eta^{\mu\nu} p^2- p^\mu p^\nu$,
$\kappa_n$ is defined in (\ref{kappa}),
\be
\phi (p^2) \equiv \int_0^1 d\alpha \: \ln \biggl(1+{p^2 \over m^2}
\, \alpha (1\!-\!\alpha)-i \epsilon \biggr)
= -i \pi \,\theta (-p^2-4m^2) \, \sqrt{1+4 \,{m^2 \over p^2} }
+\varphi (p^2),
\label{phi}
\ee
with $\epsilon \!\rightarrow \! 0^+$, and
\bea
\varphi (p^2) \equiv && \int_0^1 d\alpha \: \ln \hspace{0.2ex} 
\biggl| \hspace{0.1ex} 1+{p^2 \over m^2}
\, \alpha (1\!-\!\alpha) \hspace{0.1ex}\biggr|
=-2+ \sqrt{1+4 \,{m^2 \over p^2} }\, \ln \left| \hspace{0.1ex}
{ \sqrt{1+4 \,{m^2 \over p^2} }+1 \over 
\sqrt{1+4 \,{m^2 \over p^2} }-1 }\hspace{0.1ex} \right| 
\theta \biggl(1+4 \,{m^2 \over p^2} \biggr)
\nn  \\
&& \hspace{16ex}
+\, 2 \, \sqrt{-1-4 \,{m^2 \over p^2} } \:
{\rm arccotan} \hspace{0.2ex}
\Biggl( \sqrt{-1-4 \,{m^2 \over p^2} } \Biggr)
\: \theta \biggl(-1-4 \,{m^2 \over p^2} \biggr).
\label{varphi}
\eea
We can also write $\phi (p^2)$ in a more compact way as
\be
\phi (p^2)=-2+
\sqrt{1+4 \,{m^2 \over p^2} }\: \ln \! \left( \hspace{0.1ex}
{ \sqrt{1+4 \, (m^2-i \epsilon )/p^2 }+1 \over 
\sqrt{1+4 \, (m^2-i \epsilon )/p^2 }-1 }\hspace{0.1ex} \right).
\label{phi 2}
\ee

Other useful integrals in momentum space 
defined in (\ref{integrals}) are
\bea
&&I(p) 
={1 \over 4 \, (2 \pi)^3} \; \theta (-p^0) \,
\theta (-p^2-4m^2) \, \sqrt{1+4 \,{m^2 \over p^2} }\, ,
\label{I(P)}   \\
&&I^{\mu}(p) 
={I(p) \over 2} \, p^\mu, 
   \\
&&I^{\mu \nu}(p) 
={I(p) \over 4} \left[ p^\mu p^\nu 
- \biggl(1+4 \,{m^2 \over p^2} \biggr) \, 
{p^2 P^{\mu\nu} \over 3 } \right] \!,
  \\
&&I^{\mu \nu \alpha}(p) 
={I(p) \over 8} \left[ p^\mu p^\nu p^\alpha
- \biggl(1+4 \,{m^2 \over p^2} \biggr) \, 
{p^2 \over 3} \, \bigl( P^{\mu\nu} p^\alpha 
+ P^{\mu\alpha} p^\nu +P^{\alpha\nu} p^\mu
\bigr) \right]\!,
  \\
&&I^{\mu \nu \alpha \beta}(p) 
={I(p) \over 16} \left[ p^\mu p^\nu p^\alpha p^\beta
- \biggl(1+4 \,{m^2 \over p^2} \biggr) \, 
{p^2 \over 3} \, \bigl( P^{\mu\nu} p^\alpha p^\beta
+ P^{\nu\alpha} p^\mu p^\beta+P^{\nu\beta} p^\mu p^\alpha
+P^{\mu\alpha} p^\nu p^\beta
\right.
\nn   \\
&& \hspace{14.4ex} \left.
+\, P^{\mu\beta} p^\nu p^\alpha
+P^{\alpha\beta} p^\mu p^\nu \bigr) 
+\biggl(1+4 \,{m^2 \over p^2} \biggr)^{\! 2} \hspace{0.1ex}  
{(p^2)^2 \over 15} \, \bigl( P^{\mu\nu} P^{\alpha\beta}
+P^{\mu\alpha} P^{\nu\beta} + P^{\mu\beta} P^{\nu\alpha} \bigr)
\right] \!, 
\nn  \\
\mbox{} \label{I mu-nu-alpha-beta(P)}
\eea


\section{Products of Wightman functions}
\label{sec:Wightman products}


For the products of derivatives of Wightman functions involved in the
calculations of Sec.~\ref{subsec:noise and dissipation kernels}, we
obtain the following expressions:
\bea
&&\Delta^{+ \hspace{0.2ex} 2}(x)=
- (2\pi)^2 \!\int\! {d^4 p \over (2\pi)^4} \,
e^{-i px}\, I(p), \label{Wightman 1}
  \\
&&\partial^\mu \Delta^+(x) \, \partial^\nu \Delta^+(x)=
(2\pi)^2 \!\int\! {d^4 p \over (2\pi)^4} \,
e^{-i px}\hspace{0.1ex} 
\left[ I^\mu(p)\hspace{0.2ex} p^\nu-I^{\mu\nu}(p) \right],
\label{Wightman 2}   \\
&&\partial^\mu \partial^\nu \Delta^+(x) \,
\partial^\alpha \partial^\beta \Delta^+(x)=
- (2\pi)^2 \!\int\! {d^4 p \over (2\pi)^4} \,
e^{-i px}\hspace{0.1ex}
\left[ I^{\mu\nu}(p) \hspace{0.2ex}p^\alpha  p^\beta
-2 I^{\mu\nu (\alpha}(p) \hspace{0.2ex} p^{\beta )}
+I^{\mu\nu\alpha\beta}(p)
\right], \nn   \\
\mbox{}
\label{Wightman 3}
\eea
with $I(p)$, $I^\mu(p)$, $I^{\mu\nu}(p)$, $I^{\mu\nu\alpha}(p)$
and $I^{\mu\nu\alpha\beta}(p)$
given by Eqs.~(\ref{I(P)})-(\ref{I mu-nu-alpha-beta(P)}).
{}From these expressions, using the results of Appendix 
\ref{sec:integrals in dimensional}, we obtain
\bea
&&\partial^\mu \Delta^+(x) \, \partial^\nu \Delta^+(x)=
- \pi^2 \hspace{0.2ex}\partial^\mu_{x} \partial^\nu_{x}\! 
\int\! {d^4 p \over (2\pi)^4} \,
e^{-i px}\, I(p)
-{\pi^2 \over 3} \, {\cal F}^{\mu\nu}_{x}
\!\int\! {d^4 p \over (2\pi)^4} \,
e^{-i px}\hspace{0.1ex} 
\left(1+4 \,{m^2 \over p^2} \right) I(p),
\nn \\
\mbox{} \label{Wightman 4}  \\
&&\partial^\mu \partial^{(\alpha} \Delta^+(x) \,
\partial^{\beta )} \partial^\nu \Delta^+(x)=
-{\pi^2 \over 4}\, \partial^\mu_{x} \partial^\nu_{x}
\partial^\alpha_{x} \partial^\beta_{x}
\!\int\! {d^4 p \over (2\pi)^4} \,
e^{-i px}\, I(p)
\label{Wightman 5} \\
&&\hspace{27ex}-\,{\pi^2 \over 12} \, ({\cal F}^{\mu\nu}_{x} 
\partial^\alpha_{x} \partial^\beta_{x}
+{\cal F}^{\alpha\beta}_{x} \partial^\mu_{x} \partial^\nu_{x} )
\!\int\! {d^4 p \over (2\pi)^4} \,
e^{-i px}\hspace{0.1ex} 
\left(1+4 \,{m^2 \over p^2} \right) I(p)
\nn   \\
&&\hspace{27ex}-\,{\pi^2 \over 60} \, 
({\cal F}^{\mu\nu}_{x}{\cal F}^{\alpha\beta}_{x}
+2 {\cal F}^{\mu (\alpha}_{x}{\cal F}^{\beta )\nu}_{x})
\!\int\! {d^4 p \over (2\pi)^4} \,
e^{-i px}\hspace{0.1ex} 
\left(1+4 \,{m^2 \over p^2} \right)^2 I(p).
\nn   \\
\mbox{}
\label{Wightman 6}
\eea


\section{Products of Feynman functions}
\label{sec:Feynman products}


For the products of derivatives of Feynman functions that we need for
the calculations of Sec.~\ref{subsec:symmetric part of H kernel},
we obtain the following results:
\bea
&&\mu^{-(n-4)}
\Delta_{\scriptscriptstyle F_{\scriptstyle \hspace{0.1ex} n}}^2
   \hspace{-0.2ex}(x)=
\int\! {d^n p \over (2\pi)^n} \,
e^{i px}\, J_n(p),
\label{Feynman 1}   \\
&&\mu^{-(n-4)}
\partial^\mu 
\Delta_{\scriptscriptstyle F_{\scriptstyle \hspace{0.1ex} n}}
   \hspace{-0.2ex}(x) \,
\partial^\nu 
\Delta_{\scriptscriptstyle F_{\scriptstyle \hspace{0.1ex} n}}
   \hspace{-0.2ex}(x) =
- \int\! {d^n p \over (2\pi)^n} \,
e^{i px} \hspace{0.1ex} 
\left[J_n ^\mu(p)\hspace{0.2ex} p^\nu-J_n^{\mu\nu}(p) \right],
\label{Feynman 2}   \\
&&\mu^{-(n-4)}
\partial^\mu \partial^\nu 
\Delta_{\scriptscriptstyle F_{\scriptstyle \hspace{0.1ex} n}}
   \hspace{-0.2ex}(x) \,
\partial^\alpha \partial^\beta 
\Delta_{\scriptscriptstyle F_{\scriptstyle \hspace{0.1ex} n}}
   \hspace{-0.2ex}(x) =
\int\! {d^n p \over (2\pi)^n} \,
e^{i px} \hspace{0.1ex} 
\left[J_n^{\mu\nu}(p) \hspace{0.2ex}p^\alpha  p^\beta
-2 J_n^{\mu\nu (\alpha}(p) \hspace{0.2ex} p^{\beta )}
+J_n^{\mu\nu\alpha\beta}(p)
\right], \nn   \\
\mbox{}
\label{Feynman 3}  \\
&&\mu^{-(n-4)} 
\Delta_{\scriptscriptstyle F_{\scriptstyle \hspace{0.1ex} n}}
   \hspace{-0.2ex}(0) = -I_{0_{\scriptstyle n}},
\label{Feynman 4}   \\
&&\mu^{-(n-4)} 
\partial^\mu 
\Delta_{\scriptscriptstyle F_{\scriptstyle \hspace{0.1ex} n}}
   \hspace{-0.2ex}(x) \,
\partial^\nu \delta^n(x)= 
\int\! {d^n p \over (2\pi)^n} \,
e^{i px} \hspace{0.1ex} 
\left(I_{0_{\scriptstyle n}}^\mu\hspace{0.2ex} p^\nu
-I_{0_{\scriptstyle n}}^{\mu\nu} \right),
\label{Feynman 5}   \\
&&\mu^{-(n-4)}
\Delta_{\scriptscriptstyle F_{\scriptstyle \hspace{0.1ex} n}}
   \hspace{-0.2ex}(x) \, \Box \delta^n(x)= 
\int\! {d^n p \over (2\pi)^n} \,
e^{i px} \hspace{0.1ex} 
\left( p^2  I_{0_{\scriptstyle n}} 
+ 2 \hspace{0.1ex} p_\mu I_{0_{\scriptstyle n}}^\mu 
+I_{0_{\scriptstyle n}}^{\mu}\mbox{}_{\mu} \right).
\label{Feynman 6}
\eea
Using the results of Appendix \ref{sec:integrals in dimensional}, we
find from the above expressions
\bea
&&\mu^{-(n-4)}
\partial^\mu  \!
\Delta_{\scriptscriptstyle F_{\scriptstyle \hspace{0.1ex} n}}
   \hspace{-0.2ex}(x) \,
\partial^\nu  \!
\Delta_{\scriptscriptstyle F_{\scriptstyle \hspace{0.1ex} n}}
   \hspace{-0.2ex}(x) =
{1 \over 4}\, \partial^\mu_{x} \partial^\nu_{x} \! 
\int \!\hspace{-0.2ex} {d^n p \over (2\pi)^n} \,
e^{i px}  J_n(p)
+{1 \over 12} \, {\cal F}^{\mu\nu}_{x} \!\!
\int \!\hspace{-0.2ex} {d^n p \over (2\pi)^n} \,
e^{i px} \!
\left(1+4 \,{m^2 \over p^2} \right)\!  J_n(p)
\nn  \\
&& \hspace{15ex}
+\, {1 \over 2} 
\int \!\hspace{-0.2ex} {d^n p \over (2\pi)^n} \,
e^{i px} \! \left[ I_{0_{\scriptstyle n}} \!
\left( {p^\mu p^\nu \over p^2}+{1 \over 3}\, P^{\mu\nu} \right)
-{i \over (4\pi)^2}\, {1 \over 9}\, 
(p^2+6 m^2 )\hspace{0.2ex} P^{\mu\nu} \right]
+O(n-4), \nn  \\
\mbox{} \label{Feynman 7}  \\
&&\mu^{-(n-4)}
\partial^\mu \partial^{( \alpha} \!
\Delta_{\scriptscriptstyle F_{\scriptstyle \hspace{0.1ex} n}}
   \hspace{-0.2ex}(x) \,
\partial^{\beta )} \partial^\nu \!
\Delta_{\scriptscriptstyle F_{\scriptstyle \hspace{0.1ex} n}}
   \hspace{-0.2ex}(x) \hspace{-0.18ex}=\hspace{-0.18ex}
{1 \over 16}\, \partial^\mu_{x} \partial^\nu_{x}
\partial^\alpha_{x} \partial^\beta_{x}\! 
\int \!\hspace{-0.2ex} {d^n p \over (2\pi)^n} \,
e^{i px}  J_n(p)
\nn  \\
&& \hspace{2ex}
+ \,
{1 \over 48} \, ({\cal F}^{\mu\nu}_{x} 
\partial^\alpha_{x} \partial^\beta_{x}
\!+\!
{\cal F}^{\alpha\beta}_{x} \partial^\mu_{x} \partial^\nu_{x} )
\!  \int \!\hspace{-0.2ex} {d^n p \over (2\pi)^n} \,
e^{i px} \!
\left(1+4 \,{m^2 \over p^2} \right)\!  J_n(p)
\nn  \\
&& \hspace{2ex}
+ \,
{1 \over 240} \, ({\cal F}^{\mu\nu}_{x}{\cal F}^{\alpha\beta}_{x}
+2 {\cal F}^{\mu (\alpha}_{x}{\cal F}^{\beta )\nu}_{x})
\! \int \!\hspace{-0.2ex} {d^n p \over (2\pi)^n} \,
e^{i px} \!
\left(1+4 \,{m^2 \over p^2} \right)^2 \!\!  J_n(p)
\nn  \\
&& \hspace{2ex}
- \,
{1 \over 8} \hspace{-0.3ex}
\int \!\hspace{-0.2ex} {d^n p \over (2\pi)^n} \,
e^{i px} \! \left\{ \hspace{-0.2ex} 
I_{0_{\scriptstyle n}} \hspace{-0.3ex}
\left[ {1 \over p^2} \hspace{-0.1ex} 
\left(1\!+\!{12 \over n} \,{m^2 \over p^2} \right) 
\! p^\mu p^\nu p^\alpha p^\beta 
\hspace{-0.1ex}+\hspace{-0.1ex} {1 \over 3} \, 
(P^{\mu\nu} p^\alpha p^\beta 
+P^{\alpha\beta} p^\mu p^\nu )
\right. \right. 
\nn  \\
&& \hspace{6.7ex} 
+\, 
{4 \over n}\, {m^2 \over p^2} \, 
\bigl( P^{\mu\nu} p^\alpha p^\beta 
+P^{\alpha\beta} p^\mu p^\nu+2 P^{\mu (\alpha} p^{\beta )} p^\nu
+2 P^{\nu (\alpha} p^{\beta )} p^\mu \bigr)
\nn  \\
&& \hspace{6.7ex} 
+\,
{1 \over 15} \left(p^2\!+\!{28 \over n} \,m^2  \right) \!
(P^{\mu\nu} P^{\alpha\beta}
+2 P^{\mu (\alpha} P^{\beta ) \nu})  \Biggr] 
\hspace{-0.1ex}
- {i \over (4\pi)^2}\, {1 \over 9} \,
(p^2+6 m^2 )  \hspace{0.2ex}
(P^{\mu\nu} p^\alpha p^\beta\!+\!P^{\alpha\beta} p^\mu p^\nu)
\nn  \\
&& \hspace{6.7ex}
-\,
{i \over 4 \pi^2}\, {1 \over 225} \,
\left(2 \hspace{0.2ex} (p^2)^2
+20 \hspace{0.2ex} m^2 p^2+45\hspace{0.2ex}m^4 \right)
\hspace{-0.2ex}
\bigl( P^{\mu\nu} P^{\alpha\beta}
+2 P^{\mu (\alpha} P^{\beta ) \nu} \bigr)
\Biggr\}+O(n-4),
\label{Feynman 8}
\eea
where $P^{\mu\nu}$ is the projector orthogonal to $p^\mu$
defined above.


\section{Linearized tensors around 
flat spacetime}
\label{sec:linearized tensors}


Some curvature tensors linearized around flat spacetime are
given by the following expressions:
\bea
G^{{\scriptscriptstyle (1)}\hspace{0.1ex} \mu\nu} &=&
R^{{\scriptscriptstyle (1)}\hspace{0.1ex} \mu\nu}
-{1\over 2}\, \eta^{\mu\nu} R^{{\scriptscriptstyle (1)}},
\label{G tensor}  \\
D^{{\scriptscriptstyle (1)}\hspace{0.1ex} \mu\nu} &=&
\partial^\mu \partial^\nu \hspace{-0.2ex} 
R^{{\scriptscriptstyle (1)}}
+{1\over 2}\, \eta^{\mu\nu} \hspace{0.2ex} 
\Box R^{{\scriptscriptstyle (1)}}
-3 \, \Box R^{{\scriptscriptstyle (1)}\hspace{0.1ex} \mu\nu},
\label{D tensor}  \\
B^{{\scriptscriptstyle (1)}\hspace{0.1ex} \mu\nu} &=& 
2\: ( \partial^\mu \partial^\nu \hspace{-0.2ex} 
R^{{\scriptscriptstyle (1)}}
-\eta^{\mu\nu} \hspace{0.2ex} \Box R^{{\scriptscriptstyle (1)}} ),
\label{B tensor}
\eea
with
\bea
&&R^{{\scriptscriptstyle (1)}\hspace{0.1ex} \mu\nu}=
{1\over 2} \; ( \partial_{\alpha} \partial^{\mu} h^{\nu \alpha}
+\partial_{\alpha} \partial^{\nu} h^{\mu \alpha}
-\Box h^{\mu \nu} -\partial^\mu \partial^\nu h),
\label{Ricci tensor}  \\
&&R^{{\scriptscriptstyle (1)}}= \eta_{\alpha\beta}
R^{{\scriptscriptstyle (1)}\hspace{0.1ex} \alpha\beta}=
\partial^{\alpha} \partial^{\beta} h_{\alpha\beta}
-\Box h, 
\label{R tensor}
\eea
and
\be
R^{{\scriptscriptstyle (1)} \mu\nu\alpha\beta}=
{1 \over 2} \hspace{0.6ex}
\bigl( \partial^\mu \partial^\beta h^{\nu \alpha}
+\partial^\nu \partial^\alpha h^{\mu \beta}
-\partial^\mu \partial^\alpha h^{\nu \beta}
-\partial^\nu \partial^\beta h^{\mu \alpha} \bigr).
\label{Riemann tensor}
\ee
In four spacetime dimensions, the linearized Weyl tensor is given by
\bea
&&C^{{\scriptscriptstyle (1)} \mu\nu\alpha\beta}=
{1 \over 12} \left[ 6 \, \bigl( 
\eta^{\nu \rho} \eta^{\alpha \sigma} \partial^\mu \partial^\beta
+\eta^{\mu \rho} \eta^{\beta \sigma} \partial^\nu \partial^\alpha
-\eta^{\nu \rho} \eta^{\beta \sigma} \partial^\mu \partial^\alpha
-\eta^{\mu \rho} \eta^{\alpha \sigma} \partial^\nu \partial^\beta
\bigr)
+ 3 \, \bigl( 
\eta^{\mu \alpha} \eta^{\rho \sigma} \partial^\nu \partial^\beta
\right.
\nn   \\
&& \hspace{15ex}
+\, \eta^{\mu \alpha} \eta^{\nu \rho} \eta^{\beta \sigma} \Box
-\eta^{\mu \alpha} \eta^{\nu \rho} \partial^\beta \partial^\sigma
-\eta^{\mu \alpha} \eta^{\beta \sigma} \partial^\nu \partial^\rho
+\eta^{\nu \beta}  \eta^{\rho \sigma} \partial^\mu \partial^\alpha
+\eta^{\nu \beta} \eta^{\mu \rho} \eta^{\alpha \sigma} \Box
\nn   \\
&& \hspace{15ex}
-\, \eta^{\nu \beta} \eta^{\mu \rho} \partial^\alpha \partial^\sigma
-\eta^{\nu \beta} \eta^{\alpha \sigma} \partial^\mu \partial^\rho
-\eta^{\nu \alpha} \eta^{\rho \sigma} \partial^\mu \partial^\beta
- \eta^{\nu \alpha} \eta^{\mu \rho} \eta^{\beta \sigma} \Box
+\eta^{\nu \alpha} \eta^{\mu \rho} \partial^\beta \partial^\sigma
\nn   \\
&& \hspace{15ex}
+\, \eta^{\nu \alpha} \eta^{\beta \sigma} \partial^\mu \partial^\rho
-\eta^{\mu \beta} \eta^{\rho \sigma}  \partial^\nu \partial^\alpha 
-\eta^{\mu \beta} \eta^{\nu \rho} \eta^{\alpha \sigma} \Box
+\eta^{\mu \beta} \eta^{\nu \rho} \partial^\alpha \partial^\sigma
+\eta^{\mu \beta} \eta^{\alpha \sigma} \partial^\nu \partial^\rho
\bigr)
\nn   \\
&& \left. \hspace{15ex}
+\, 2 \, \bigl( \eta^{\mu \alpha} \eta^{\nu \beta}
-\eta^{\nu \alpha} \eta^{\mu \beta} \bigr) \, 
\bigl( \partial^\rho \partial^\sigma - \eta^{\rho \sigma} \Box
\bigr)
\right] h_{\rho \sigma}.
\label{Weyl tensor}
\eea


\section{The integrals 
$J_{\lowercase{a}}\lowercase{(s)}$}
\label{app:J's}


For the integrals $J_a(s)$, $a \!=\! 1,2,3$, defined in 
(\ref{J integrals}), we find the following results
\bea
&&J_1(s) = {1 \over 4 \hspace{0.2ex}(\kappa^2 + \pi^2) 
  \hspace{0.2ex} |p|^2 } \left\{
{1 \over 2 \hspace{0.2ex} {\rm Re}\, p} \, 
\ln \!\left[ {s^2 -2 \hspace{0.2ex} {\rm Re}\, p \: s + |p|^2
    \over s^2 +2 \hspace{0.2ex} {\rm Re}\, p \: s + |p|^2} \right]
\right.
\nn    \\
&&\hspace{24.7ex} \left.
+\, {1 \over {\rm Im}\, p} 
\left[ \pi
- \arctan \!\left( {s+ {\rm Re}\, p \over {\rm Im}\, p} \right)
- \arctan \!\left( {s- {\rm Re}\, p \over {\rm Im}\, p} \right)
\right]  \right\},
   \\
&&J_2(s) = {1 \over 4 \hspace{0.2ex}(\kappa^2 + \pi^2)}
\left\{
{1 \over 2 \hspace{0.2ex} {\rm Re}\, p} \, 
\ln \!\left[ {s^2 +2 \hspace{0.2ex} {\rm Re}\, p \: s + |p|^2
    \over s^2 -2 \hspace{0.2ex} {\rm Re}\, p \: s + |p|^2} \right]
\right.
\nn   \\
&&\hspace{21.2ex} \left.
+\, {1 \over {\rm Im}\, p} 
\left[ \pi
- \arctan \!\left( {s+ {\rm Re}\, p \over {\rm Im}\, p} \right)
- \arctan \!\left( {s- {\rm Re}\, p \over {\rm Im}\, p} \right)
\right]  \right\},
    \\
&&J_3(s) = {1 \over 4 \hspace{0.2ex}(\kappa^2 + \pi^2)}
\left\{ -4 s+
{1 \over 2 \hspace{0.2ex} {\rm Re}\, p} \, 
\Bigl[ 3 \hspace{0.2ex} ({\rm Re}\, p )^2 -({\rm Im}\, p )^2
\Bigr] \, \ln \!\left[ {s^2 +2 \hspace{0.2ex} {\rm Re}\, p \: s + |p|^2
    \over s^2 -2 \hspace{0.2ex} {\rm Re}\, p \: s + |p|^2} \right]
\right.
\nn    \\
&&\hspace{15.5ex} \left.
+\, {1 \over {\rm Im}\, p} 
\Bigl[ ({\rm Re}\, p )^2 -3 \hspace{0.2ex} ({\rm Im}\, p )^2
\Bigr] 
\left[ \pi
- \arctan \!\left( {s+ {\rm Re}\, p \over {\rm Im}\, p} \right)
- \arctan \!\left( {s- {\rm Re}\, p \over {\rm Im}\, p} \right)
\right]  \right\}\! ,
\nn    \\
\mbox{}
\eea
where $p$ is a function of $s$ given by expressions (\ref{p(s)}),
which give $|p|^2 = \left[ \left[(\kappa^2 + \pi^2) \, s^2 
   + \kappa \right]^2+\pi^2\right]^{1/2} 
 \hspace{-0.2ex}\!/(\kappa^2 + \pi^2)$.


\end{document}